%Paper: astro-ph/9305005
%From: p165bie@mpifr-bonn.mpg.de (Peter Biermann)
%Date: Thu, 6 May 93 16:31:54 +0200

\newcount\eqnum\eqnum=0%
\def\simle{\lower 2pt \hbox {$\buildrel < \over {\scriptstyle \sim }$}}
\def\simge{\lower 2pt \hbox {$\buildrel > \over {\scriptstyle \sim }$}}
\def\autnum{\global\advance\eqnum by 1{\rm(\the\eqnum)}}
\hyphenation{mono - chro - matic  sour - ces  Wein - berg
chan - ges}
\headline={\ifnum\pageno=1\firstheadline\else
\ifodd\pageno\rightheadline \else\leftheadline\fi\fi}
\def\firstheadline{\hfil}
\def\rightheadline{\hfil}
\def\leftheadline{\hfil}
        \footline={\ifnum\pageno=1\firstfootline\else\otherfootline\fi}
\def\firstfootline{\rm\hss\folio\hss}
\def\otherfootline{\hfil}

\font\tenrm=cmr10
\font\tenit=cmti10
\font\elevenbf=cmbx10 scaled\magstep 1
\font\elevenrm=cmr10 scaled\magstep 1
\font\elevenit=cmti10 scaled\magstep 1
\font\ninebf=cmbx9
\font\ninerm=cmr9

\font\sevenrm=cmr7

\newskip\ttglue
\catcode`@=11

\font\twelverm=cmr12    \font\twelvebf=cmbx12    \font\twelvemi=cmmi12
\font\twelvesl=cmsl12   \font\twelveit=cmti12

\font\ninerm=cmr9       \font\ninebf=cmbx9       \font\ninemi=cmmi9
             
\font\ninesy=cmsy9

\font\sevenrm=cmr7      \font\sevenbf=cmbx7      \font\sevenmi=cmmi7
     \font\sevensy=cmsy7

\font\sixrm=cmr6

\def\twelvepoint{\def\rm{\fam0\twelverm}
    \textfont0=\twelverm \scriptfont0=\ninerm \scriptscriptfont0=\sevenrm
   \textfont1=\twelvemi \scriptfont1=\ninemi \scriptscriptfont1=\sevenmi
  \textfont2=\tensy    \scriptfont2=\ninesy \scriptscriptfont2=\sevensy
    \textfont3=\tenex    \scriptfont3=\tenex  \scriptscriptfont3=\tenex
    \textfont\itfam=\twelveit  \def\it{\fam\itfam\twelveit}
    \textfont\slfam=\twelvesl  \def\sl{\fam\slfam\twelvesl}
    \textfont\ttfam=\tentt     \def\tt{\fam\ttfam\tentt}
    \textfont\bffam=\twelvebf  \def\bf{\fam\bffam\twelvebf}
    \scriptfont\bffam=\ninebf
    \scriptscriptfont\bffam=\sevenbf
    \tt\ttglue=.5em plus.25em minus.15em
    \normalbaselineskip=14pt
    \setbox\strutbox=\hbox{\vrule height7pt depth2pt width0pt}
    \let\sc=\sixrm
    \normalbaselines\rm}

\nopagenumbers
\line{\hfil }
\vglue 1cm
\hsize=6.0truein
\vsize=8.5truein
\parindent=3pc
\baselineskip=10pt
\centerline{\elevenbf AGN AND GALACTIC SITES OF COSMIC RAY ORIGIN}
\vglue 1.0cm
\centerline{\tenrm PETER L. BIERMANN }
\baselineskip=13pt
\centerline{\tenit Max-Planck-Institut f\"ur Radioastronomie}
\baselineskip=12pt
\centerline{\tenit Auf dem H\"ugel 69}
\baselineskip=12pt\centerline{\tenit D-53 Bonn 1, Germany 53010}

\vglue 0.8cm
\centerline{\tenrm ABSTRACT}
\vglue 0.3cm
{\rightskip=3pc
\leftskip=3pc
\tenrm\baselineskip=12pt%
\noindent}

In the following we review recent progress in our understanding of the
physics of energetic particles in our Galaxy, in active galaxies such as
starburst galaxies, in active galactic nuclei and in the jets and radio hot
spots of powerful radio galaxies and radioloud quasars.  We propose that
cosmic rays originate mainly in three sites, a) normal supernova explosions
into the approximately homogeneous interstellar medium, b) supernova
explosions into stellar winds, and c) hot spots of powerful radio galaxies.
The predictions based on this proposition have been tested successfully
against airshower data over the particle energy range from 10 TeV to
100 EeV. The energy input from supernova explosions in starburst galaxies
is so large, as to produce breakouts from the galactic disks, observable
most prominently in X-ray and nonthermal radio emission. Normal galaxies
are also likely to have winds.  The low energy cosmic ray particles carried
out by such galactic winds to the intergalactic medium may ionize and heat
this medium consistent with all known constraints. Furthermore we propose
that radioloud quasars accelerate nuclei to high energies in shocks  in
the inner sections of their jets, and thus initiate hadronic cascades which
produce the  $\gamma$-ray emission observed with GRO.  Radio observations
of the Galactic center compact source suggest that there is a jet similar
to that in quasars, but very much weaker in its emission; simple modelling
suggests that the power carried by this jet is of order similar to the
accretion disk luminosity.  By analogy we speculate that also radioweak
quasars have jets which carry a large proportion of the accretion power,
and which accelerate particles to high energy.  These energetic particles
might serve to heat the inner regions of the host galaxies to produce the
strong farinfrared dust emission, and could also initiate hadronic and
electromagnetic cascades in the inner accretion disk close to the presumed
black hole in the center, to produce a large part of the X-ray emission
observed from radioweak AGN.

\baselineskip=14pt
\elevenrm

\twelvepoint

\vglue 0.8cm
\line{\elevenbf 1. Introduction \hfil}
\vglue 0.4cm

Energetic particles can have an observable effect in many astronomical
objects, from heating and ionization of interstellar material to TeV
$\gamma$-rays from Active Galactic Nuclei (AGN).  Key to all such arguments
and an important test for our physical insight is the origin of the
observed cosmic ray particles, their energy range, spectrum and chemical
composition.  There has been recent progress in these areas and in this
review we will describe some of it.

\bigskip
\line{\elevenit 1.1. Galactic Cosmic Rays \hfil}
\smallskip

There are few well accepted arguments about the origins of cosmic rays (for
recent reviews of cosmic ray physics, see, {\elevenit e.g.} Drury (1983),
Hillas (1984), Blandford \& Eichler (1987), Berezhko \& Krymsky (1988),
and Jones \& Ellison (1991):  a)  The cosmic rays below about $10^4 \;
\rm GeV$ are predominantly due to the explosion of stars into the normal
interstellar medium (Lagage \& Cesarsky 1983).  b) The cosmic rays from
near $10^4 \;\rm GeV$ up to the knee, at $5 \; 10^6 \; \rm GeV$, are very
likely predominantly due to explosions of massive stars into their former
stellar wind (V\"olk \& Biermann, 1988).  The consequences of this concept
have been checked by calculating the cosmic ray abundances and comparing
them with observations (Silberberg  {\elevenit et al.} 1990); the comparison
suggests that up to the highest energy where abundances are known, this
concept successfully explains the data. c) For the energies beyond the knee
there has been no consensus; Jokipii \& Morfill (1987) argue that a galactic
wind termination shock might be able to provide those particles, while
Protheroe \& Szabo (1992) argue for an extragalactic origin, although in
either case the matching of the flux at the knee from two different source
populations remains somewhat difficult.  Reacceleration of the existing
cosmic ray population has been investigated by various researchers, most
recently by Wandel {\elevenit et al.} (1987), Wandel (1988), and Ip \& Axford
(1992); in such models the chemical abundances are modified at high energy
from the interstellar medium abundances in favour of heavy nuclei.  It
would be an interesting task to test these models against air shower data.
d) For the cosmic rays beyond the ankle at about $3 \; 10^9 \; \rm GeV$ an
extragalactic origin is required because of the extremely large gyroradii
of such particles.

In all such arguments (also, {\elevenit e.g.}, Bogdan \& V\"olk, 1983, Drury
{\elevenit et al.} 1989, Markiewicz {\elevenit et al.} 1990) the spectrum
of the cosmic rays remains unexplained.  And yet the observations of the
cosmic rays themselves, and of the nonthermal radioemission from our Galaxy
as well as from all other well observed galaxies (Golla 1989) strongly
suggests, that in all galactic environments studied carefully, the cosmic
rays have a universal spectrum of about $E^{-2.7}$ below the knee at $5
\,10^6\;\rm GeV$ (electrons only $< 10 \;\rm GeV$); direct air shower
experiments show the spectrum beyond the knee to well approximated by
$E^{-3}$ (Stanev 1992).  This overall spectrum is clearly  influenced by
propagation effects, since particles at different energies have different
chances to escape from the disk of the galaxy.  It appears to be a reasonable
hypothesis to approximate the interstellar turbulence spectrum by a Kolmogorov
law, which leads to an interstellar diffusion coefficient proportional to
$E^{1/3}$.  Such an energy dependence then requires a source spectrum of
cosmic rays of approximately $E^{-2.4 \pm 0.1}$ below the knee, and
approximately $E^{-2.8 \pm 0.1}$ above the knee.  We note, that Ormes \&
Freier (1978) have argued already for a source spectrum of $E^{-2.3
\pm 0.1}$.

Recently we proposed a fully developed physical concept to account for
the cosmic ray particle energies, spectrum and chemical abundances (Biermann
1993, paper CR I; Biermann \& Cassinelli 1993, paper CR II; Biermann \&
Strom 1993, paper CR III; Rachen \& Biermann 1993, paper UHE CR I).  We
suggest, that the  cosmic rays observed near earth arise from three sites:

1)  Supernova explosions into the nearly homogeneous interstellar medium;
these cosmic rays reach particle energies of about $100 \,Z \;\rm TeV$,
and have an  injection spectrum of near $E^{-2.4}$, where $Z$ is the charge
of the nucleus considered.

2)  Supernova explosions into a former stellar wind produce a slightly
flatter spectrum of near $E^{-2.3}$ at injection below a bend near $700 \, Z
\;\rm TeV$, beyond which the spectrum steepens to near $E^{-2.7}$ up to a
maximum energy of about $100 \, Z \;\rm PeV$.  This means that heavy nuclei
dominate at large particle energies.  Since here the acceleration is in a
shock that runs through the wind of an evolved star, the heavy elements are
enriched (Silberberg {\elevenit et al.} 1990).   This contribution may
dominate even at low energies for some elements.  Air shower data have been
successfully fitted with this model by Stanev {\elevenit et al.} (1993,
paper CR IV).

3)  The hot spots of powerful radio galaxies provide the third site
of origin, to be discussed in the next subsection.

We emphasize that the proposal rests on a plausible but nevertheless
speculative assumption about the nature of the transport of energetic
particles in perpendicular shock waves, namely that there are large fast
convective motions across the shock interface; this notion is, however,
supported by radio polarization observations. The model has {\elevenit
predictive power}.  The predictions for the galactic cosmic rays underwent
their most severe tests with the airshower data in paper CR IV, and the
predictions for the extragalactic component were tested with airshower data
in Rachen {\elevenit et al.} (1993, paper UHE CR II).  Cosmic ray
acceleration and propagation is also discussed by Jokipii in chapter 5 of
this book.

We will describe the essential features of the theory for the Galactic
contribution in sections 2.1 and 2.2.  In section 2.3 we will apply it to
starburst and normal galaxies.

\bigskip
\line{\elevenit 1.2. Extragalactic Cosmic Rays \hfil}
\smallskip

The third site of cosmic ray production is the hot spots of powerful radio
galaxies, as proposed already in 1990 (Biermann 1993) at the celebration
meeting for M.M. Shapiro in Washington, DC. More detailed models were
presented at various meetings, at Kofu in 1990 (Biermann 1991), and at
Bartol in 1991 (Rachen \& Biermann 1992).  The extensive model prediction
is in Rachen (1992) and Rachen \& Biermann (1993, paper UHE CR I).  The
full tests against the new airshower data from Fly's Eye (Gaisser
{\elevenit et  al.} 1993) are presented in Rachen  {\elevenit et al.} (1993,
paper UHE CR II).

The relevant observations are discussed in chapter 6 of this book by Sokolsky.
We will discuss the theory and its test in section 3.

\bigskip
\line{\elevenit 1.3. Radioloud quasars \hfil}
\smallskip

Radioloud quasars have long been recognized to provide a unique laboratory
for studying high energy particle interactions, since their emission across
nearly the entire electromagnetic spectrum could readily be interpreted as
synchrotron and inverse Compton emission from particles presumably
accelerated in shock waves in relativistic jets.  The spatial configuration
of these shocks appears to be a double helix in both radio and HST images
at high spatial resolution (see, {\elevenit e.g.}, Macchetto {\elevenit et
al.} 1991); this appearance is even suggested in radiojets with apparent
superluminal motion of the various features ({\elevenit e.g.} Krichbaum
{\elevenit et al.} 1990).  Clearly, this calls into question whether we
really have shockwaves or not some other mechanism to enhance radio emission
at the location of a double helix throughout a jet.  Radio quasars with a
flat radio spectrum ({\elevenit i.e.} with flux density $S_{\nu} \, \sim \,
\nu^{\alpha}$, and the spectral index $\alpha$ being near zero) have been
recognized to be a most interesting observationally defined class of radio
quasars; in a complete sample of flat spectrum radio quasars all objects have
been demonstrated to show evidence for bulk relativistic motion (Witzel
{\elevenit et al.} 1988).

The new GRO observations of flat spectrum radio quasars demonstrate
(Hartman {\elevenit et al.} 1992a, Hermsen {\elevenit et al.} 1992, Montigny
{\elevenit et al.} 1992, Bertsch {\elevenit et al.} 1993, and many IAU
circulars:  Kanbach {\elevenit et al.} 1992, Fichtel {\elevenit et al.} 1992,
Michelson {\elevenit et al.} 1992, Hartman {\elevenit et al.} 1992b, 1992c,
Hunter {\elevenit et al.} 1992, Dingus {\elevenit et al.} 1993), that they
have higher luminosities in the $\gamma$-ray range than at any other
wavelength, with one object even showing emission near TeV photon energies
(Whipple collaboration: Weekes {\elevenit et al.} 1992, Punch {\elevenit
et al.} 1992).  In chapters 1  and 4 of this book Fichtel \& Thompson and
Fegan review $\gamma$-ray data and possible interpretations.

We will review the hadronic interaction concept in relativistic jets to
explain these observations in section 4.  Neutrino observations provide an
important test for such concepts and are discussed by Stanev in
chapter 7 of this book.

\bigskip
\line{\elevenit 1.4. The Galactic Center and Radioweak Quasars \hfil}
\smallskip

The Galactic center has been demonstrated by recent mm-VLBI observations to
show elongated radio emission akin to a compact jet (Krichbaum {\elevenit
et al.} 1993); in simple models this jet  can be shown to carry nearly as
much power as is in the accretion disk luminosity (Falcke {\elevenit et al.}
1992, 1993).  These arguments will be summarized in section 5.1.  For
radioweak quasars it has only recently become clear (Chini {\elevenit et al.}
1989a) that their farinfrared emission is due to dust, and is thus thermal
radiation.  The origin of this  energy is not clear; several possibilities
exist, and one of these possibilities is the hypothesis, that the active
nucleus provides a large proportion of its power in the form of energetic
particles - {\elevenit e.g.} arising from shocks in a putative jet thus
following the physical interpretation of the Galactic center made here -
which in turn heat the molecular clouds which contain the dust.  Various
tests of this hypothesis will be described in section 5.2, including some
consequences for the emission from the inner accretion disk (X-rays and
neutrinos) as well as the farinfrared emission spectra of quasars.

\bigskip
\line{\elevenit 1.5. Background \hfil}
\smallskip

The following review relies heavily on the recent work of my graduate students,
postdocs and collaborators elsewhere, especially the following papers:  1.
Biermann 1993, paper CR I, 2. Biermann \& Cassinelli 1993, paper CR II, 3.
Biermann \& Strom 1993, paper CR III, 4. Stanev, Biermann, Gaisser 1993, paper
CR IV, 5. Rachen \& Biermann 1993, paper UHE CR I, 6. Rachen, Stanev, Biermann
1993, paper UHE CR II, 7. Mannheim 1993a, Proton Blazar, 8. Mannheim 1993b,
3C273, 9. Nellen, Mannheim, Biermann 1993, RQQ.Neutrinos, 10. Falcke
{\elevenit et al.} 1993 and Falcke, Mannheim, Biermann 1992, Sgr A*, 11.
Niemeyer \& Biermann 1993, RQQ.FIR, 12. Linden, Duschl, Biermann 1993, and
Biermann, Duschl, Linden 1993, G.C. Mol.clouds, 13. Nath \& Biermann 1993, IGM,
as well as further work.

\bigskip
\line{\elevenit 1.6. History and next steps \hfil}
\smallskip

Many of the essential conceptual elements used in the following have a long
history:  After the discovery of cosmic rays by Hess (1912) and Kohlh\"orster
(1913), Baade \& Zwicky (1934) already proposed that supernova explosions
produce cosmic rays; this viewpoint is upheld in the work reported here up
to particle energies of about $3 \, 10^9 \; \rm GeV$.  Alfv\'en (1939)
argued early for a local origin in our Galaxy, a viewpoint which is
confirmed by the age determinations of the cosmic rays (Garcia-Munoz
{\elevenit et al.} 1977).  Fermi (1949, 1954) proposed the basic concept of
acceleration still being used.  Shklovsky (1953) and Ginzburg (1953) made a
convincing case for particle acceleration in supernova remnants, again a
notion being used in the following, in somewhat modified form.  Ginzburg
(1953) already emphasized the interesting role of novae; and indeed, the
nova GK Per is an important test case.  Hayakawa (1956) proposed that stellar
evolution gives rise to an enhancement of heavy elements and pointed out
the importance of spallation in the interstellar medium; and again, the
enrichment found in the cosmic ray contribution from supernova explosions
into winds is due to just this enrichment.  And finally, Cocconi (1956)
already argued convincingly that the most energetic cosmic ray particles
are from an extragalactic origin, an argument made quantitative here.

There are many obvious tasks to be done next; we will outline 12 specific next
steps that appear feasible in the near future.  Additional work may take
much longer to follow through with checking the proposal made here in all
details and also to produce comprehensive alternatives.

\vglue 0.6cm
\line{\elevenbf 2. Galactic Cosmic Rays \hfil}
\vglue 0.4cm

When discussing galactic cosmic rays, the first question to ask is:  What
is the effect of the leakage out from the galactic disk on the spectrum?
There is ample evidence from a variety of sources that the turbulence
spectrum in the interstellar medium is of a Kolmogorov character.  As a
consequence the transport coefficient for cosmic ray particles out of the
galactic disk has a $E^{1/3}$ dependence.

First, if one assumes that the turbulence in the interstellar medium
responsible for scattering cosmic rays has a single powerlaw over the entire
range of energies observed, then we have the limit that even at the highest
particle energies, the transit time through the disk should be larger than
the simple light crossing time, because we observe no anisotropy.  This means
that the powerlaw in leakage time vs. particle energy dependence is weaker
than $0.4$ in the exponent.

Second, the Chicago data already by themselves suggest that over a more
limited particle energy range the exponent is close to $1/3$ (Swordy
{\elevenit et al.} 1990).

Third, all evidence from the interstellar medium itself ({\elevenit e.g.}
Larson 1979, 1981, Jokipii 1988, Narayan 1988, Rickett 1990) suggests that
the turbulence is of a Kolmogorov character under the important proviso
that the interstellar medium can be thought of as various turbulent phases
each with a power law turbulence spectrum:  Here, the most definitive
measurements come from Very Long Baseline Interferometry (VLBI) at
radiowavelength; with this technique one can observe water masers, pulsars
and quasars (Blandford {\elevenit et al.} 1986, Gwinn {\elevenit et al.}
1988a,  1988b; Spangler \& Gwinn 1990; Britzen 1993) and always finds
properties of the interstellar medium consistent with the Kolmogorov spectrum.

Fourth, and finally, detailed plasma simulations demonstrate that in the
regime where the magnetic field is of a strength similar to the thermal
pressure the natural turbulence scaling is indeed Kolmogorov (Matthaeus \&
Zhou 1989), and only in the limit of strongly dominating magnetic field do
we get a different behaviour (the Kraichnan law).

Summarizing, it appears clear that a large body of evidence points to a
Kolmogorov law.

However, there is a major difficulty:  The secondary products of nucleus-
nucleus collisions have a spectrum compared to the primaries which in a
simple stationary one zone leaky box model for the galactic disk suggests
a law for the effective turbulence which is different.  Such an analysis
suggests that the energy dependence of the leakage time is closer to $2/3$
rather than $1/3$.  However, the data on giant molecular clouds, where
surely a large part of the cosmic ray interaction happens, suggest that
simple stationary one zone models are not adequate.  Large molecular clouds
assemble out of bits of small clouds in spiral arms and then slowly disperse
again; the time scales for assembly and dispersal are comparable to the time
scale of cosmic ray leakage from the galactic disk, making it obvious that
the interaction of cosmic rays with clouds in their structure and their slow
dispersal and diffusion out of clouds has to be investigated in order to
answer the question what the secondary to primary nuclei ratio ought to be
as a function of particle energy.  This problem is not solved at present,
and remains to be worked out; this constitutes a specific next step.

We will assume in the following that the correction from observed cosmic
ray spectrum to source spectrum is a change in spectral slope by exactly
$1/3$, so that we are looking at source spectra in the range $E^{-2.4 \pm
0.1}$ below the knee, and $E^{-2.8 \pm 0.1}$ above the knee.

In the following we will first discuss the explosions into the interstellar
medium in section 2.1 and then the explosions into a former stellar wind, in
section 2.2.  Generalizations for a discussion of starburst and normal
galaxies follow in section 2.3.

\bigskip
\line{\elevenit 2.1. Supernova explosions into the interstellar medium \hfil}
\smallskip

Already for some time it has been argued that normal supernova explosions
provide the lower particle energy range of the cosmic rays observed
({\elevenit e.g.} Lagage \& Cesarsky 1983).  However, there were several
open questions:

1)  The standard diffusive particle acceleration at shocks (Drury 1983)
works only in the approximation that the shocks are parallel, {\elevenit
i.e.} that the shock normal is parallel to the prevailing direction of the
magnetic field.  This is, however, normally not true, and already Jokipii
(1987) argued that this at the very least gives an acceleration time very
much shorter than the standard approach.  The problem of perpendicular shocks
remained.

2)  The spectrum of the cosmic rays at source could not be derived reliably
from physical arguments; this problem stems partially from the lack of
understanding of the spectral correction between source and observer (here
we use $1/3$, see above).

3)  What is the maximum particle energy that can be reached?

4)  Why do we have at low particle energies such a large discrepancy
between the Hydrogen and Helium to heavy nuclei ratio in the cosmic rays
compared to the interstellar medium, out of which these particles are
presumed to be accelerated?

We propose to discuss here the acceleration of particles in the blastwaves
caused by supernovae in the normal interstellar medium.

Observational arguments suggest that the simple Sedov solution is adequate
(Biermann \& Strom 1993, paper CR III); we will use the standard Sedov
solution:  The Sedov similarity solution for an  explosion into a homogeneous
medium can be written as a function of shock radius $r$ and shock speed $U_1$;
as a function of time $t_4$, in units of $10^4 \;\rm years$; explosion energy
$E_{51}$, in units of $10^{51} \;\rm ergs$; and density of the interstellar
medium $n_o$ in units of ${\rm cm}^{-3}$. This gives (Cox 1972)

$$r\;=\;13.7 \,t_4^{2/5}\,(E_{51}/n_o)^{1/5} \,{\rm pc} \eqno\autnum $$
\noindent and

$$\eqalign{U_1/c\;&=\;1.8\,10^{-3}\,t_4^{-3/5}\,(E_{51}/n_o)^{1/5}\;\cr &=
\;0.090\,({r \over {\rm pc}})^{-3/2}\,(E_{51}/n_o)^{1/2} \cr}. \eqno\autnum $$

\bigskip
\line{\elevenit 2.1.1. Particle Acceleration \hfil}
\smallskip

Shock acceleration in its standard form just takes the Lorentz transformation
for a energetic particle at a velocity $v$ much larger than the shock
velocity  as it goes between scatterings in a weakly turbulent magnetic field
from downstream  to upstream and back.  In practice we will assume $v$ to be
close to $c$.  One such cycle gives an momentum $p$ gain of

$$\left({\Delta p \over p}\right)_{LT}\;=\;{4 \over 3} {U_1 \over v} (1 -
{U_2 \over U_1}). \eqno\autnum $$

We assume the shock to be subrelativistic and so the phase space distribution
of the particles to be nearly isotropic.  Then downstream (see Drury 1983
for the exact derivation) the particles have a finite chance to escape.
They cross the shock with with the flux $n v/4$, where $n$ is the particle
density,  $v$ their velocity, and the factor $1/4$ comes from $1/2$ of all
particles moving in one direction with a velocity component on average $1/2$
of their absolute velocity parallel to the shock normal.  Far downstream the
particle flux is, however, $n \, U_2$, where $U_2$ is the downstream flow
velocity in  the frame of the shock. The ratio $4 \, U_2/v \,=\, U_1/v$ is
the chance to escape from the shock region.  This probability is per cycle.
The combination yields a powerlaw for the distribution $p^2 \, f(p)$, where
$f(p)$ is the particle distribution function in phase space, with the
powerlaw index

$$- {{1 + 2 U_2/U_1} \over {1 - U_2/U_1}} . \eqno\autnum $$

However, this assumes that there are no other energy losses or energy gains
during a cycle of a particle going back and forth between upstream and
downstream.  In a configuration, where there is a component of the magnetic
field perpendicular to the shock normal, there can be energy gains by drifts
parallel to the electric field seen by a particle moving with the shock
system, and also losses due to adiabatic expansion in the expansion of a
curved shock.  Since drift acceleration is a rate, the resulting particle
energy gain is proportional to the time spent on either side of the shock.
Furthermore, there can be spectral changes due to the fact, that particles
were injected at a different rate in the past, when some particle under
consideration now was injected; again, this is likely to happen in a
spherical shock.

Thus, the key here is to know what the time scale of a cycle actually is, or
in other words, what the transport or diffusion coefficient is for the
particles to move from one side of the shock to the other.  Both energy
losses and energy gains (over and above the Lorentz transformation) are
proportional to the time scale of a cycle.  This we will discuss therefore
in a separate section next.

\bigskip
\line{\elevenit 2.1.2. The basic Hypothesis \hfil}
\smallskip

The {\elevenit radio observations} of young supernova remnants can be a
guide here.  The  explosions into the interstellar medium are explosions
where the unperturbed magnetic field is not far from perpendicular to the
shock direction over most of $4 \pi$ steradians. Radio polarization
observations of supernova remnants clearly indicate what the typical local
structure of these shocked plasmas is.  The observational evidence (Milne
1971, Downs \& Thompson 1972, Reynolds \&  Gilmore 1986, Milne 1987, Dickel
{\elevenit et al.} 1988) has been summarized by Dickel {\elevenit et al.}
 (1991) in the statement that all shell type supernova remnants less than
$1000$ years old show dominant radial structure in their magnetic fields
near their boundaries.  There are several possible ways to explain this:
Rayleigh-Taylor instabilities between ejected and swept up material can
lead to locally radial differential motion and so produce a locally radial
magnetic field (Gull 1973).  It could be due to strong radial velocity
gradients of various clumps of ejecta, or result from clouds being overrun
that now evaporate and cool the surrounding material (all of which are
mentioned by Chevalier [1977]). It may seem paradoxical that we will be
considering perpendicular shocks where the observations show a parallel
magnetic field,  but it must be remembered that the observed field is a
superposition of the configurations of all emitting regions in the telescope
beam. The degree of polarization in all of the objects referred to above
is low (typically 10\% or less), much less than that expected from a
perfectly uniform field (generally 65\%), which is consistent with there
being both parallel and perpendicular components present.

The radial magnetic field configuration is also found in a few shell remnants
somewhat older than $1000 \;\rm yr$, such as RCW 86 (SN 185) and Pup A (with
a kinematic age of some $3700 \;\rm yr$, Winkler {\elevenit et al.} 1988).
However, in most ``mature" remnants the field is tangential, or complex
(Milne 1987). While a radially-directed magnetic field provides the most
compelling evidence for large-scale differential motions parallel to the
shock direction, there are other indications as well. It is generally
believed that the oxygen-rich fast moving knots (fmk's) in Cas A are ejecta
from the progenitor (Kamper \& van den Bergh 1976). Oxygen-rich knots have
also been seen in Pup A (Winkler {\elevenit et al.} 1988) and the similarly
aged G 292.0+1.8 (Braun {\elevenit et al.} 1986), and may also be present
in a number of extragalactic SNRs (Winkler and Kirshner 1985). Radially-
directed radio structures have been mapped in Cas A, and their motion has
also been determined (Braun {\elevenit et al.} 1987). Recent Rosat
observations of the older ($10 000 \;\rm yr$) Vela SNR show similar features
in the X-ray emission which have apparently penetrated the main SNR shell
(seminar by Tr\"umper at the Texas conference December 1992), and are
believed to result from fmk-like ejecta. (Their very soft X-ray spectra
would make similar features difficult to detect in more distant remnants.)

These examples are for supernova explosions into the interstellar medium;
there is also an observation demonstrating the same effect for an explosion
into a wind:  Seaquist {\elevenit et al.} (1989) find for the spatially
resolved shell of the nova GK Per a radially oriented magnetic field in the
shell, while the overall dependence of the magnetic field on radial distance
$r$ is deduced to be $1/r$ just as expected for the tightly wound up magnetic
field in a wind.  The interpretation given is that the shell is the higher
density material behind a shock wave caused by the nova explosion in 1902
and now travelling through a  wind.  Seaquist {\elevenit et al.} (1989) note
the similarity to young supernova remnants.

The important conclusion for us here is that there appear to be strong
radial differential motions in perpendicular shocks which provide the
possibility that particles get {\elevenit convected} parallel to the shock
direction.  We assume this to be a diffusive process, and note that others
have also pointed out that this may be a key to shock acceleration
({\elevenit e.g.} Falle 1990). Our task here will be to derive a natural
velocity and a natural length scale, which can be combined to yield a
transport coefficient. A classical prescription is the method of Prandtl
(1925):  In Prandtl's argument an analogy to kinetic gas theory is used to
derive a diffusion coefficient from a natural scale and a natural velocity
of the system.  Despite many weaknesses of this generalization Prandtl's
theory has held up remarkably well in many areas of science far beyond the
original intent.  We will use a similar prescription here.

Consider the structure of a layer shocked by a supernova explosion
into a homogeneous medium in the case that the adiabatic index of the gas
is $5/3$ and the shock is strong.  Then there is an inherent scale in the
system, namely the thickness of the shocked layer, in the spherical case
for a strong shock $r/12$.  In a spherical strong shock into a wind this
natural length scale is $r/4$.  There is also a natural velocity scale,
namely the velocity difference of the flow with respect to the two sides
of the shock.  Both are the {\elevenbf smallest dominant scale} in velocity
and in length; we will use the assumption that the smallest dominant scale
is the relevant scale several times subsequently to derive diffusion
coefficients and other scalings.  This choice can be thought of also as
follows:  The smallest dominant scale in a transport coefficient provides
the longest time scale for the transport mechanism, and it is well
understood that in a combination of transport processes the slowest one
always wins.

Our basic conjecture, {\elevenbf postulate 1}, {\elevenit based on
observational evidence}, is that the {\elevenit convective random walk} of
energetic particles perpendicular to the {\elevenit unperturbed} magnetic
field can be described  by a diffusive process with a downstream diffusion
coefficient  $\kappa_{rr,2}$ which is given by the thickness of the shocked
layer and the velocity difference across the shock, and is independent of
energy:

$$\kappa_{rr,2} \;=\;{1 \over 9}\, {U_2 \over U_1}\, r \, (U_1 \, - \,U_2)
\eqno\autnum $$

The upstream diffusion coefficient can be derived in a similar way, but with
a larger scale.  We make here the second critical assumption, {\elevenbf
postulate 2}, namely that the upstream length scale is just $U_1/U_2$ times
larger, and so is $r/3$.  This, obviously, is the same ratio as the mass
density and the ratio of the gyroradii of the same particle energy.  Since
the  magnetic field is lower by a factor of $U_1/U_2$ upstream, that means
that the upstream gyroradius of the maximum energy particle that could be
contained in the shocked layer, is also $r/3$.  Hence $r/3$ is the natural
scale.  And so the upstream diffusion coefficient is

$$\kappa_{rr,1} \;=\; {1 \over 9}\, r \, (U_1 \, - \, U_2) \eqno\autnum $$

For these diffusion coefficients, it follows that the residence times
(Drury 1983) on both sides of the shock are equal and are

$${4 \, \kappa_{rr,1} \over U_1 c}\;=\;{4 \,\kappa_{rr,2} \over U_2 c}\;=\;{4
\over 9}\, {r \over c} \, (1 \, - \, {U_2 \over U_1}).  \eqno\autnum $$

Adiabatic losses then cannot limit the energy
reached by any particle since their time scale is proportional to the
acceleration time, both being independent of energy.  So the limiting
size of the shocked layer limits the energy that can be reached to that
energy for which the gyroradius just equals the thickness of the shocked
layer,  provided the particles can reach this energy.  We assume here that
the  average of the magnetic field $\langle B \rangle$ is not changed very
much by all this convective motion, but leave the possibility open that
the root mean square magnetic field ${\langle B^2 \rangle}^{1/2}$ is
increased; this implies that a magnetic dynamo does not work as fast as
the timescales given by the shock (see, {\elevenit e.g.}, Galloway \&
Proctor 1992 for arguments on the shortest possible dynamo time scale).
This then leads to a possible maximum energy of

$$E_{max}\;=\; {1 \over 3}\,{U_2 \over U_1}\, Z e r B_2 \;=\; {1 \over 3}
\,Z e r B_1 \eqno\autnum $$

\noindent where $Z e$ is the particle charge and $B_{1,2}$ is the magnetic
field strength on the two sides of the shock.  This means, once again, that
the energy reached corresponds to the maximum gyroradius the system will
allow.  It also means that we push the diffusive picture right up to its
limit where the diffusive scale becomes equal to the mean free path and the
gyroradius of the most energetic particles.   For supernova explosions into
stellar winds the corresponding argument produces the limiting condition,
while for supernova explosions into the interstellar medium we will derive
a more restrictive condition below.

\bigskip
\line{\elevenit 2.1.3. The Particle Spectrum \hfil}
\smallskip

Consider particles which are either upstream of the shock, or
downstream; as long as the gyrocenter is upstream we will consider the
particle to be there, and similarly downstream.

In general, the energy gain of the particles will be governed primarily by
their adiabatic motion in the electric and magnetic fields.  The expression
for the energy gain is well known and given in, {\elevenit e.g.}, Northrop
(1963, equation 1.79), for an isotropic angular distribution where a first
term arises from the drifts and a second from the induced electric field.
The expression is valid in any coordinate frame.  We explicitly work in the
shock frame,  separate the two terms and consider the drift term first.  The
second term is accounted for further below.

The $\theta$-drift can be understood as arising from the asymmetric
component of the diffusion tensor, the $\theta r$-component.  The natural
scales there are the gyroradius and the speed of light, and so we note that
for (Forman {\elevenit et al.} 1974)

$$\kappa_{\theta r}\;=\;{1 \over 3} \,r_g \,c , \eqno\autnum $$

\noindent the exact limiting form derived from ensemble averaging, we obtain
the drift velocity by taking the proper covariant divergence (Jokipii
{\elevenit et al.} 1977); this is not simply (spherical coordinates) the
$r$-derivative of  $\kappa_{\theta r}$.  The general drift velocity is given
by (see, {\elevenit e.g.}, Jokipii 1987)

$$V_{d,\theta}\;=\;c \,{E \over {3 Z e}}\,{\rm curl}_{\theta} {{\bf B} \over
B^2} . \eqno\autnum $$

The $\theta$-drift velocity is thus zero upstream and only finite downstream
due to curvature:

$$V_{d,\theta}\;=\;{1 \over 3} \, {{c \, r_g}  \over r}  \eqno\autnum $$

\noindent where $r_g$ is now taken to be positive.

It must be remembered that there is a lot of convective turbulence which
increases the curvature:  The characteristic scale of the turbulence is
$r/12$ for strong shocks, and thus the curvature is $12/r$ maximum.  Taking
half the maximum as average we obtain then for the curvature $6/r$ which is
six times the curvature without any turbulence; this increases the curvature
term by a factor of six thus changing its contribution from $1/3$ to $2$ in
the numerical factor in the expression above.  Hence the total drift
velocity is thus

$$V_{d,\theta}\;=\;{1 \over 2} \,({U_1 \over {U_2}})\, {{c \, r_g} \over r} ,
\eqno\autnum $$

\noindent now written for arbitrary shock strength.  It is easily verified
that the factor in front is two for strong shocks where $U_1/U_2=4$. With
$\Delta E_1\;=\;0 $ we have then downstream

$${\Delta E_2 \over E}\;=\;{2 \over 9}{U_1 \over c}\, (1-{U_2 \over U_1})
\eqno\autnum $$

\noindent which is the total energy gain from drifts.

Next, we consider the influence of adiabatic losses and the injection history
on the particle spectrum.

Let us consider then one full cycle of a particle
remaining near the shock and cycling back and forth from upstream to
downstream and back. Adding the energy gain in the relativistic
approximation due to drifts to the energy gain from the simple Lorentz
transformation we obtain

$${\Delta E \over E}\;=\;{4 \over3} {U_1 \over c} (1 -{U_2 \over U_1})
x \eqno\autnum $$

\noindent where

$$x\;=\;1+{1 \over 6}. \eqno\autnum $$

Consider how long it takes a particle to reach a certain energy:

$${dt \over dE}\;=\;\lbrace 8 \;{\kappa_{rr,1} \over U_1 c} \rbrace /
\lbrace {4 \over 3} {U_1 \over c} (1-{U_2 \over U_1}) x E \rbrace.
\eqno\autnum $$

Here we have used $\kappa_{rr,1} / U_1 \;=\; \kappa_{rr,2} / U_2$. Since we
have

$$r\;=\;{5 \over 2} \,U_1 t \eqno\autnum $$

\noindent this leads to

$${dt \over t}\;=\;{dE \over E} {{3 U_1} \over {U_1 - U_2}} {2 \over x}
{\kappa_{rr,1} \over {r U_1}}\,{5 \over 2} \eqno\autnum $$

\noindent and so to a dependence of

$$t(E)\;=\;t_o \; ({E \over E_o})^{\beta} \eqno\autnum $$

\noindent with

$$\beta\;=\;{{{3 U_1} \over {U_1 - U_2}} {2 \over x} {\kappa_{rr,1} \over
{r U_1}}}\,{5 \over 2}. \eqno\autnum $$

Particles that were injected some time $t$ ago
were injected at a different rate, say, proportional to $r^b$.
Also, in a $d$-dimensional space, particles have
$\sim \, r^d$ more space available to them than when they were injected,
and so we have a correction factor which is

$$({E \over E_o})^{-{2 \over 5}(b+d) \beta}. \eqno\autnum $$

The combined effect is a spectral change by

$${{3 U_1} \over {U_1 - U_2}} {2 \over x} (d+b) {\kappa_{rr,1} \over {r U_1}}.
 \eqno\autnum $$

We note that the factor $2/5$ from the Sedov expansion drops out again.
For a detailed argument on the sign conventions used here, see paper CR I.
This expression can be compared with a limiting expansion derived by Drury
(1983; eq. 3.58), who also allowed for a velocity field;  Drury (1983)
generalized earlier work on spherical shocks by Krymskii \& Petukhov (1980)
and Prishchep \& Ptuskin (1981).  Drurys expression agrees with the more
generally derived expression given here for $x=1$.  The comparison with
Drurys work clarifies that for $\kappa \,\sim \,r$ the inherent time
dependence drops out except, obviously, for the highest energy particles,
discussed further below; the same comparison shows that the statistics of
the process are properly taken into account in our simplified treatment.
The $r^d$-term describes the adiabatic losses from the {\elevenit linear}
expansion of the shock layer in their effect on the spectrum, and thus
accounts for the second term in Northrops expression mentioned above.
This also describes then the adiabatic losses, due to the general expansion
of the shock layer. Hence the total spectral difference, as compared with
the plane parallel case, is given by

$${{3 U_1} \over {U_1 - U_2}}\;\lbrace {U_2 \over U_1} ({1 \over x} -1) \;+\;
{2 \over x} (b+d) {\kappa_{rr,1} \over {r U_1}} \rbrace. \eqno\autnum $$

\noindent If this expression is positive the spectral index is steeper.

Furthermore, we have to discuss the effect of the expansion of the material
in the shell, in this context already introduced by Drury (1983).  This
expansion is only that {\elevenit beyond the linear} expansion of the overall
flow already taken into account.  We define for the flow field $U(\xi r)$ as
function of radius $\xi r$ interior to the shock of radius $r$ the velocity
by

$$U(\xi r) \;=\;V(\xi) \, \xi \, U_1.  \eqno\autnum $$

As shown by Drury (1983), this velocity field  introduces a steepening
by

$${{3 \,U_1} \over {U_1 - U_2}}\,{{\kappa_{rr,2}} \over {r U_2}}\,
{V'(1) \over V(1)}, \eqno\autnum $$

\noindent neglecting drifts and in the limit of very small diffusion
coefficient. However, in our context where we consider a finite shell, we
have to use a non-local approximation (see paper CR III).  From a detailed
discussion we have an extra term from this postshock adiabatic loss due to
the shell expansion of

$${{3 \,U_1} \over {U_1 - U_2}}\,{1 \over x}\,{{\kappa_{rr,2}} \over {r U_2}}
\, {V'(1) \over V(1)} \, 1.970 \;. \eqno\autnum $$

\noindent which is then slightly less than double the effect which Drury
discusses.   For a strong shock then this particular effect adds $0.563$
to the spectral index.

For strong shocks in stellar winds, for which the natural length scale is
$r/4$, the linear part of the expansion is three times higher and so the
remaining difference to the postshock speed of sound is only
$2.25 - \sqrt{5}$, which gives an effect of only $2 \,10^{-3}$ in the
spectral index, justifying our neglect of this effect in the case of a shock
travelling through a stellar wind (see papers CR I and CR II and below).

The exponent $b$ describes the injection as a power of the radius; in a
Sedov solution the injection, assumed to be proportional to $\rho U_1^2$, is
given by

$$\rho U_1^2\;\sim\;r^{-3}. \eqno\autnum $$

Hence we have $b = -3$, so that $b+d = 0$.
Thus we have for a pure Sedov solution, using the velocity field term
introduced above, a spectral difference to the plane parallel case of
$0.420$ and thus an injection spectrum of

$$E^{-2.420}. \eqno\autnum $$

However, we note that this spectrum is relevant for electrons obviously
only if they are injected into the acceleration process at all. This
spectrum derived here agrees with the electron spectrum deduced from the
radio synchrotron emission in the radio shell of the supernova remnant
Cas A.  Adding  the term from diffusive losses from the Galaxy the final
observable spectrum expected is

$$E^{-2.753} ,\eqno\autnum $$

\noindent very close to what is observed, both in Galactic cosmic rays
(protons and nuclei), as through the nonthermal radioemission of other
galaxies (Golla 1989, electrons).

Clearly, there is uncertainty in our method to treat the non-linear flow
field.  Further non-linear corrections, such as also averaging the non-linear
flow field itself, lead to spectral indices within $0.04$ of the answer given
above.

\bigskip
\line{\elevenit 2.1.4. Observational Tests \hfil}
\smallskip

Calculating now the radio emission contribution from the shell of
$r\,U_2/(3\,U_1)$, we note that in the Sedov solution the ram pressure is
proportional to $U_1^2$, which in turn is proportional to $r^{-3}$.  The
total radio emission from the shell is then constant with time, if the
efficiency of injecting electrons $\eta$ is a constant.  The luminosity
is then given by

$$L_{\nu}\;=\;5.4 \,10^{23}\,B_{-5.3}^{1.710}\, \eta_{-1} \,E_{51} \,
{\nu_{9.0}}^{-0.710}\,{\rm erg/sec/Hz}  \eqno\autnum $$

\noindent and so does not depend on the interstellar medium density; since
the adiabatic expansion is a condition of constant energy, all such
dependencies drop out.  We have used here as reference $0.1$ for
$\eta$, $5 \, \rm \mu Gauss$ for the interstellar unperturbed magnetic
field strength, and $1 \;\rm GHz$ for the emission frequency observed.
We note, that the emission here is coming from a shell of thickness $r/12$;
additional emission further inside might arise if there is a reverse shock
resulting from the transition of free expansion to adiabatic expansion.
Furthermore, there is observational evidence that the synchrotron emission
systematically samples the higher magnetic field substructures, by a factor
of up to $5$ and so we can expect the synchrotron emission to be higher by
factors up to $15.7$, leading to an implied radio luminosity of up to $8.5
\,10^{24} \, {\rm erg/sec/Hz}$, everything else being equal. This luminosity
is in  agreement with the upper limit of the distribution of observed
luminosities, suggesting that in their initial phases the efficiency of
putting kinetic energy into an energetic electron population is indeed of
order $0.1$.  An important next step is to actually model all the selection
effects that dominate the observations (see especially the discussion in
Berhuijsen 1986).

The evolution of a supernova remnant (see, {\elevenit e.g.}, for the
basic physical concepts employed Cox [1972]) is then is divided into four
phases:

First, there is free expansion until the interstellar medium overrun by
the explosion shock is of about the same mass as the mass ejected.  This
happens at a radius of

$$R_e\;=\;1.92 \,{\rm pc}\,({M_{ej} \over M_{\odot}})^{1/3}\,n_o^{-1/3} ,
\eqno\autnum $$

\noindent where $M_{ej}$ is the ejected stellar mass and $n_o$ is again
the environmental density in particles per cc.  Numerical simulations
demonstrate that the expansion is noticeably slowed down long before this
critical radius is reached.

Second, we have a Sedov phase with adiabatic expansion and steady fresh
electron injection.  The observations of radio emitting stars (see paper
CR II) suggest that there is a {\elevenit critical electron injection
velocity} for a shock, below which electron injection is substantially
weakened.  Clearly, the critical velocity in the interstellar medium is
likely to be quite different from that in stellar winds.  We do not wish
to dwell on electron injection physics here (the nonthermal radio emission
of other stars can be a guide here), but adopt somewhat arbitarily as a
critical velocity in the interstellar medium with an Alfv\'enic Machnumber
roughly similar to the massive star wind case, with $1000 \;\rm km/sec$
and obtain then

$$R_{crit}\;=\;9.0 \,{\rm pc}\, (E_{51} / n_o)^{1/3}\, U_{crit,-2.5}^{-2/3}.
\eqno\autnum $$

\noindent An observational comparison of the evidence for the value of the
critical velocity in different sites is an important next step, as for
instance on the basis of the radio and optical observations of the classical
nova GK Per.

Third, when the expansion velocity has decreased below the critical
velocity,  the electron injection ceases (or, to be more accurate, decreases
by many orders of magnitude) and we have further adiabatic expansion, but
with the electron population only modified by adiabatic losses (Shklovsky
1968).   Adiabatic losses of a relativistic particle decrease its energy
by the ratio of the initial and final radius scale.

And finally, fourth, we have the break-up phase due to the various
instabilities in the cooling layer (McCray {\elevenit et al.} 1975,
Chevalier \& Imamura 1982, Smith 1989). Clearly, these instabilities do not
stop the expansion of the dense layer, but lead to a rather chaotic slowdown
of the various shell fragments (see, {\elevenit e.g.}, the beautiful new
colour photos of the Cygnus loop from HST and ROSAT circulating in
the popular press).  The cooling radius is (Cox 1972)

$$R_{cool}\;=\;15.6 \;{\rm pc} \,E_{51}^{3/11} \,(10^{22}L)^{-2/11}
\,n_o^{-5/11}, \eqno\autnum $$

\noindent where $L$ is the cooling coefficient; the cooling curve has been
compared with recent detailed calculations by Schmutzler \& Tscharnuter
(1993) and is still a very good approximation at the temperature range of
interest here.  During this time the energy of any individual relativistic
electron decreases by adiabatic expansion as the ratio of the radii from the
time when injection ceases to the time of break up.  Thus the energy density
of the electron population (from conservation of the adiabatic moment only
and disregarding spatial dilution, see below) decreases by the factor
($p=2.420$)

$$(R_{cool} / R_{crit})^{p-1}\;=\;2.18 \,E_{51}^{-0.09}\,n_o^{-0.17}\,
(10^{22}L)^{-0.26}\,U_{crit,-2.5}^{0.95}. \eqno\autnum $$

\noindent showing a very weak dependence on the environmental density $n_o$.
The dependence on the assumed specific value of the critical velocity is very
nearly linear; hence the numerical factor in this expression would go down
with a smaller value for the critical velocity.  The simple spatial dilution
of the energy density is the same for both protons and electrons and drops
out in the ratio, as long as we are in the Sedov phase (see Shklovsky
[1968], eq. 7.27).  For a tenuous interstellar medium of density about
$0.01$ Hydrogen atoms per cc the factor is increased from $2.18$ to nearly
$4.8$.

This intermediate switch from electron injection with steady
acceleration to a simple adiabatic loss regime determines the net scaling
of the power of the electron population to that of the proton population
in the cosmic rays. The observations suggest that from $1 \;\rm GeV$ the
electron number is only about $0.01$ ({\elevenit e.g.} Wiebel 1992) of the
protons.  From this observed ratio the energy density ratio integrated
over the entire relativistic part of the particle spectrum, protons
relative to electrons, for the spectral index of $-2.42$, is given by $4.3$.
Assuming then, that the ratio of protons and electrons is not influenced by
propagation effects, this suggests an expansion of a factor in radius of
order $3$ between the time when electron injection ceases and the time when
cooling breaks up the shell of the supernova remnant; here we assume that
electrons and protons originally have comparable energy densities of their
relativistic particle populations.  We may have thus identified the origin
of the observed electron/proton ratio in cosmic rays.

The maximum energy particles can possibly reach is given above on spatial
arguments, and depends linearly on the magnetic field.  Near the symmetry
axis we have diffusion parallel to the magnetic field, and, given the
strong turbulence induced by the shock we take there the Bohm limit for the
upstream diffusion coefficient $\kappa_{\mid \mid}\;=\;{1 \over 3}\,{E
\over {Z e B_1}}\,c$.  At the equator we use our radial upstream diffusion
coefficient derived earlier, here for a strong shock, of $\kappa_{\perp}\;
=\;{{r U_1} \over 12}$.  Combining these two upstream diffusion coefficients
(Jokipii 1987) to obtain an acceleration time scale and setting that equal
to the upstream flow time scale of $r/(3 U_1)$ then yields a condition on
the maximum particle energy that can be reached.  This condition is

$${ r \over {3 U_1}}\;=\;{4 \over 3}\,{E \over {Z e B_1}}\,{c \over U_1^2}\,
\mu^2 \, + \, {r \over {3 U_1}}\,(1 - \mu^2) \,{1 \over x}. \eqno\autnum $$

This leads to

$$E_{max}\;=\;{1 \over 4}\,Z e B_1 r\,{U_1 \over c}  \eqno\autnum $$

\noindent in the approximation of $x \simeq 1$, which is true here.  In
stellar winds, on the other hand, $x \simeq 2$, and this approximation
cannot be used (see paper CR I).  A more detailed consideration of the
strength of the magnetic field as a function of latitude does not change
this result.  Since the angular coordinate does not enter this expression,
it is valid for any subsection of the sphere, and thus also for an
inhomogeneous magnetic field.

Using normal interstellar magnetic fields of $5\;\mu \rm G$, and a standard
Sedov explosion with an arbitrary radius then yields a maximum particle
energy.  At breakup, when we suggest the actual mixing of the particle
population into the interstellar medium begins, these maximum
energies are

$$E_{max}({\rm protons})\;=\;2.6 \,10^4 \,F_{CR} \, \rm GeV , \eqno\autnum $$

\noindent and

$$E_{max}({\rm iron})\;=\;7.4 \,10^5 \,F_{CR} \, \rm GeV , \eqno\autnum $$

\noindent where we use

$$F_{CR} \;=\;E_{51}^{0.364}\, (10^{22}L)^{0.091}\, n_o^{-0.273}
. \eqno\autnum$$

The protons ought to dominate over heavier nuclei due to the normal
abundances in the interstellar medium, and higher energies for protons are
clearly reached in the tenuous hot part of the interstellar gas where the
densities are of order $0.01$ particles/cc, and so maximum proton
energies are possible to about $1. \,10^5 \; \rm GeV$.  These numbers are
similar to earlier estimates of the maxium particle energy in a Sedov
expansion phase of a supernova.  Below (see paper CR IV) we obtain from a
fit to the air shower data an estimate for this number of $1.2 \,10^5 \;
\rm GeV$, in very good agreement with these arguments, and suggesting that
the cosmic ray injection is most effective in the tenuous part of the
interstellar medium, an argument which has been made before on quite
different grounds.

There is another observational check, and that is on the electron spectrum
of cosmic rays.  For particle energies above a few GeV, and that corresponds
to synchrotron emission at high radio frequencies normally unobservable in
galaxies because of increasing thermal dust emission, the observed electron
spectrum is about $E^{-3.3 \pm 0.1}$, consistent with a steepening by unity
from injection due to synchrotron losses, in this energy range faster than
leakage losses.  This confirms our injection spectrum derived.  A detailed
discussion of the time scales involved remains to be done.

\bigskip
\line{\elevenit 2.2. Supernova Explosions into a Stellar Wind \hfil}
\vglue 0.4cm

The basic hypothesis for the second site of origin of cosmic rays is, that
we consider explosions into a stellar wind with a Parker spiral topology
(Parker 1958).  We remind the reader that in such a wind in the asymptotic
regime, where the magnetic field decreases with the radial distance $r$
as $1/r$ and the wind velocity is constant, the (tangential) Alfv\'en
velocity is also constant with radial distance.

We also note that the wind in massive stars has a similar energy
integrated over the main sequence life time as the subsequent supernova
explosion.  The wind bubble is large and is itself surrounded by a dense
shell. Hence we can expect the shock of the supernova to disperse this
shell and to mix the energetic particle population produced in the shock
running through the wind directly into the interstellar medium.  Thus,
there is no additional energy dependence introduced here to go from the
spectrum which we calculate below to the injection spectrum of cosmic rays.

Outside the wind acceleration region stellar winds are likely to be
similar to the solar wind, and so we will assume a Parker spiral topology
of the magnetic field ({\elevenit e.g.} Jokipii  {\elevenit et al.} 1977).
This leads then to a tangential drift of

$$V_{d,\theta}\;=\;{2 \over 3}\,c\,r_g/r,  \eqno\autnum $$

\noindent where $r_g$ is the Larmor radius of the particle under
consideration.  The drift is towards the equator for $Z \,B_s \,> \, 0$.
This drift - both gradient and curvature drift -  here is just due to the
unperturbed structure of the stellar wind magnetic field; we will consider
the consequences of additional curvature from turbulence.

\bigskip
\line{\elevenit 2.2.1. The Particle Spectrum below the Knee \hfil}
\smallskip

Consider the structure of a layer shocked by a Supernova explosion into a
stellar wind in the case, that the adiabatic index of the gas is $5/3$ and
the shock is strong.  Then there is an inherent length scale in the system,
namely the thickness of the shocked layer, in the spherical case for a shock
velocity much larger than the wind speed and in the strong shock limit $r/4$.
This leads to very similar expressions for diffusion coefficients and
respective time scales as derived above for the case of an explosion into a
homogeneous medium, just with a factor of $3$ larger length scales.

There is an important consequence of this picture for the diffusion
laterally:  From the residence timescale and the velocity difference across
the shock we find a distance which can be traversed in this time of

$${4 \over 3}\, {r \over c} \, (1 \, - \, {U_2 \over U_1})  \, (U_1 \,-
\,U_2) .  \eqno\autnum $$

Since the convective turbulence in the radial direction also induces motion
in the other two directions, with maximum velocity differences of again
$U_1 \,- \,U_2$, this distance is also the typical lateral length scale.
{}From this scale and again the residence time we can construct an upper
limit to the diffusion coefficient in lateral directions of

$$\kappa_{\theta \theta ,max}\;=\;{4 \over 9} \, (1 \,-{U_2 \over U_1})^3 \,
({U_1 \over c})^2 \, r \,c ,  \eqno\autnum $$

\noindent which is for strong shocks equal to

$$\kappa_{\theta \theta ,max}\;=\;{1 \over 3} \,  ({3 \over 4} \, {U_1 \over
c})^2 \, r \,c .  \eqno\autnum $$

Again in the spirit of the idea, that the smallest dominant scale wins,
this then will begin to dominate as soon as the $\theta$-diffusion
coefficient reaches this maxium at a critical energy.  As long as the
$\theta$-diffusion coefficient is smaller, it will dominate particle
transport in $\theta$ and the upper limit derived here is irrelevant.
When the $\theta$-diffusion coefficient reaches and passes this maximum,
then the particle in its drift will no longer see an increased curvature
due to the convective turbulence due to averaging and the part ($1/3$ of
total for strong shocks) of drift acceleration due to increased curvature
is eliminated.  This then reduces the energy gain, and the spectrum becomes
steeper from that energy on.  The critical particle energy thus implied
will be identified below with the particle energy at the knee of the
observed cosmic ray spectrum, and this reduced energy gain from drifts
provides the steepening of the spectrum beyond the knee.

Again, as above the turbulence increases the curvature of the local medium,
and so increases the drift velocity.  The total drift velocity, combining
now again the curvature $(2/3)$ and gradient $(1/3)$ terms, is then

$$V_{d,\theta}\;=\;{1 \over 3} \,(1 + {U_1 \over {2 U_2}})\, {{c \, r_g}
\over r} , \eqno\autnum $$

\noindent now written for arbitrary shock strength.  It is easily verified
that the factor in front is unity for strong shocks where $U_1/U_2=4$.

The energy gain associated with such a drift is given by the product of
the drift velocity, the residence time, and the electric field.  Upstream
this energy gain is given by

$$\Delta E_1\;=\;{4 \over 3}\,E\,{U_1 \over c}\,f_d\, (1-{U_2 \over U_1}),
\eqno\autnum $$

\noindent where

$$f_d\;=\;{1 \over 3} \,(1 + {U_1 \over {2 U_2}}). \eqno\autnum $$

Thus, $f_d=1$ for strong shocks. The total energy gain is

$${\Delta E \over E}\;=\;{4 \over 3}\,{U_1 \over c}\,f_d\,(1+{U_2 \over
U_1})\,(1-{U_2 \over U_1}). \eqno\autnum $$

The drift energy gain averages over the magnetic field strength during the
gyromotion.  We emphasize that this energy gain is independent of this
average magnetic field, so that even variations of the magnetic field
strength due to convective motions do not change this energy gain.

Considering again a full cycle of a particle going back and forth across a
curved shock produces the following net energy gain:

$${\Delta E \over E}\;=\;{4 \over3} {U_1 \over c} (1 -{U_2 \over U_1})
x \eqno\autnum $$

\noindent where

$$x\;=\;1 + {1 \over 3}\,(1+{U_1 \over {2 U_2}})\,(1+{U_2 \over U_1})
\eqno\autnum $$

\noindent which is $9/4$ for a strong shock when $U_1 / U_2 \;=\;4$.

It is easy to show that the additional energy gain flattens the particle
spectrum.  Consider then adiabatic losses and history of injection as above
leads to a spectral correction relative to a planar shock of

$${{3 U_1} \over {U_1 - U_2}}\;\lbrace {U_2 \over U_1} ({1 \over x} -1)
\;+\; {2 \over x} (b+d) {\kappa_{rr,1} \over {r U_1}} \rbrace.
\eqno\autnum $$

The total spectral change is then for
$U_1 / U_2 \;=\;4$ given by $1/3$, so that the spectrum obtained is

$${\rm Spectrum \,(source)}\;=\;E^{-7/3}. \eqno\autnum $$

After correcting for leakage from the galaxy the spectrum is

$${\rm Spectrum \,(earth)}\;=\;E^{-8/3}  \eqno\autnum $$

\noindent very close to what is observed near earth at particle energies
below the knee.

Such an injection spectrum of $-7/3$ of relativistic particles in strong
and fast shocks propagating through a stellar wind leads to an
unambiguous radio synchrotron emission spectrum of $\nu^{-2/3}$ from
energetic electrons.  Such a radio spectrum has been observed for the
classical nova GK Per (Reynolds \& Chevalier 1984).

Generalizing now for arbitrary wind speed and arbitrary shock strength
we obtain for the thickness of the shocked layer

$${\Delta r \over r}\;=\;{U_2 \over U_1}\;{U_1 \over {V_W+U_1}}.
\eqno\autnum $$

This reduction of the thickness of the layer for a finite wind velocity is
due to the fact that the material which is snowplowed together is not all
gas between zero radius and the current radius $r$, but between zero
radius and $r (1-{V_W / (V_W+U_1)})$, since the gas keeps moving while
the shock moves out towards $r$.  In this case we have

$$f_d\;=\;{1 \over 3}\,(1+{1 \over 6}\,{{r U_1} \over {\kappa_{rr,1}}}
\,{U_1 \over U_2}\,(1-{U_2 \over U_1})).   \eqno\autnum$$

It is easily verified that $f_d=1$ for strong shocks and negligible wind
velocity.  We take here the curvature length scales to be the same on both
sides of the shock, since the curvature is induced by the thickness of the
shock region.

The energy gain associated with the $\theta$-drift is

$${\Delta E \over E}\;=\;{4 \over3} {U_1 \over c} (1 -{U_2 \over U_1})
x \eqno\autnum $$

\noindent where

$$x\;=\;1+3\,{{\kappa_{rr,1}} \over {r U_1}}\,f_d\,
(1+{U_2 \over U_1}) / (1-{U_2 \over U_1}) , \eqno\autnum $$

\noindent which is $9/4$ for negligible wind speeds and a strong shock
when $U_1 / U_2 \;=\;4$; on the other hand, for $V_W / U_1=1$ we have
$x=2.042$ and for the limiting case of large wind speeds compared to
shock speeds $x=1.833$.

The spectral difference to the plane parallel case is thus

$${{3 U_1} \over {U_1 - U_2}}\;\lbrace {U_2 \over U_1} ({1 \over x} -1)
\;+\; {2 \over x} (b+d) {{\kappa_{rr,1}} \over {r U_1}} \, (1+{V_W \over
U_1}) \rbrace. \eqno\autnum $$

We note that

$$\kappa_{rr,1}(1+{V_W\over U_1}) / {r U_1}   \eqno\autnum $$

\noindent is now independent of the wind speed, and
the only effect of the wind which remains is through $x$. For the
sequence of $V_W/U_1\,=\,0.,1.0, \,{\rm and}\, >>1$ we thus obtain
particle spectral index differences, in addition to the index of $7/3$, of
$0.0,\,0.136,\,0.303$, corresponding to synchrotron emission
spectral index of an electron population with the same spectrum, of
$0.667,\,0.735,\,0.818$.  We will use the first two cases as examples
below.  The work of Owocki {\elevenit et al.} (1988) suggests that
typical shocks in winds have a velocity in the wind frame similar to the
wind velocity in the observers frame itself, which implies that, in the
simplified picture here, only spectral indices for the synchrotron emission
between  $0.667$ and $0.735$ are relevant, with an extreme range of spectral
indices  up to $0.818$ for strong shocks.  Obviously, for weaker shocks with
$U_1/U_2 \,<\,4$ the spectrum can be steeper, {\elevenit e.g.} for $U_1/U_2
\,=\,3.5$  we obtain an optically thin spectral index for the synchrotron
emission of $0.734$ for $V_W\, \ll \,U_1$ and $0.815$ for $V_W/U_1 \, = \, 1$.

\bigskip
\line{\elevenit 2.2.2. The Polar Cap \hfil}
\smallskip

We wish to discuss here the bend in the spectrum of cosmic rays at the knee,
near $5 \; 10^6 \; \rm GeV$.

Let us consider the structure of the wind through which the supernova
shock is running. The maximum energy a particle can reach is proportional
to $sin^2 \theta$, since the space available for the gyromotion from a
particular latitude is limited in the direction of the pole by the axis of
symmetry.  Hence, clearly the maximum energy attainable is lowest near the
poles.  Then, consider the pole region itself, where the radial dependence
of the magnetic field is $1/r^2$, and the magnetic field is mostly radial.
We can make two arguments here: Either we put the upstream diffusive scale
$4\;\kappa_{rr,1} /(c \,U_1)$ equal to $r/c$ in the strong shock limit, or
we can put acceleration time and flow time equal to each other.  Both
arguments lead to the same result.  Using here the Bohm limit in the
diffusion coefficient $\kappa_{rr,1} \;=\; {1 \over 3} c  E/{Z e B(r)}$,
since we have a shock configuration near the pole, where the direction of
propagation of the shock is {\elevenit parallel} to the magnetic field -
often referred to as a parallel shock configuration - then leads to a
maximum energy for the particles of

$$E\;=\;{3 \over 4} Z e B(r) r {U_1 \over c}, \eqno\autnum $$

\noindent which is proportional to $1/r$ near the pole, where the magnetic
field is parallel to the direction of shock propagation; the corresponding
gyroradius is then given by ${3 \over 4} {U_1 \over c} r$.  Putting this
equal to the gyroradius of particles that are accelerated further out at
some colatitude $\theta$, where the magnetic field is nearly perpendicular
to the direction of shock propagation, gives the limit where the
latitude-dependent acceleration breaks down.  This then gives the critical
angle as

$$sin \, \theta_{crit} \;=\; {3 \over 4} {U_1 \over c}. \eqno\autnum $$

The angular range of $\theta \, < \, \theta_{crit}$ we refer to as the
{\elevenit polar cap}.  The maximum particle energy at the location of the
critical angle is then given by

$$E_{knee}\; =\;Z e B(r) r ({3 \over 4} {U_1 \over c})^2. \eqno\autnum $$

We identify this energy with the knee feature in the cosmic ray spectrum,
since all latitudes outside the polar cap contribute the same spectrum up
to this energy; from this energy to higher particle energies a smaller part
of the hemisphere contributes and also, the energy gain is reduced, as
argued below.   This is valid in the region where the magnetic field is
nearly perpendicular to the shock, and thus this knee energy is independent
of radius.  The argument described earlier on the limits of the
$\theta$-diffusion leads to the same critical energy.

All this immediately implies that the chemical composition at the knee
changes in such a way, that the gyroradius of the particles at the spectral
break is the same, implying that the different nuclei break off in order
of their charge $Z$, considered as particles of a certain energy (and not as
energy  per nucleon).

Hence it appears plausible to suppose that in the polar cap the effective
diffusion coefficient is quite a bit smaller than further down in latitude,
and so that the particle spectrum is rather close to $E^{-2}$.  On the other
hand, the polar cap is small relative to $4 \pi$ by about $(U_2 / U_1)^2$
and only a spectrum like $E^{-2}$ will make it possible for the polar cap
to contribute appreciably, but only near the knee energy, because the
spectral flux near the knee is increased relative to $1 \; \rm GeV$ by
$(E_{knee} / m_p c^2)^{1/3}$ which approximately compensates for its small
area.  The combination of the polar cap with the rest of the stellar
hemisphere might lead to a situation where up to, say, $10^4 \;\rm GeV$ the
entire hemisphere excluding the polar cap dominates, while from $10^4 \;
\rm GeV$ up to the knee the polar cap begins to contribute appreciably.
Near the knee energy the polar caps might thus contribute equally to the
rest of the $4 \pi$ steradians.  Because of spatial limitations most of
the hemisphere has to dominate again above the knee, although with a
fraction of the hemisphere that decreases with particle energy. The
superposition of such spectra for different chemical elements is tested
in paper CR IV and briefly summarized below.  These tests with airshower
data suggest indeed that near the knee of the overall cosmic ray spectrum
the polar cap contributes noticeably, and produces the sharpness of the
knee feature.

The expression for the particle energy at the knee also implies by the
observed relative sharpness of the break of the spectrum that the actual
values of the combination $B(r) r U_1^2$ must be very nearly the same
for all supernovae that contribute appreciably in this energy range.
Note that $B(r) r$ is evaluated in the Parker regime, and so is related
to the surface magnetic field by $B(r) r \; =\; B_s \; r_s^2 \Omega_s /v_W$,
where the values with index $s$ refer to the surface of the star and $v_W$
is the wind velocity.  Thus the expression

$$B_s \, r_s^2 \, {\Omega_s \over v_W} \, U_1^2 \eqno\autnum $$

\noindent is approximately a universal constant for all stars that explode
as supernova after a Wolf Rayet phase or red giant phase with a strong
wind.  It  may hold for all massive stars of lower mass as well that
explode as supernovae, and might even extend to below the mass the range of
supernovae; this latter possibility is supported by the observation that the
magnetic fields of white dwarfs appear to be higher for higher mass white
dwarfs.  This could be due to an overall evolutionary convergence of
massive stars, which is indeed suggested by evolutionary calculations which,
however, do not take strong magnetic fields or near critical rotation into
account. This relationship suggests as an alternative the speculation that
the mechanical energy of supernovae explosions is the potential energy from
the transient phase when the interior of the exploding star collapses to an
accretion disk, which loses its angular momentum due to torsional Alfv\'en
waves.  Assuming a core rotation of a star shortly before its explosion,
which corresponds to critical rotation on the surface, this speculation
leads with no further dominant parameter to an explosive energy of about
$10^{51} \, \rm erg$  (see paper CR I), corresponding very well to the
typical mechanical energy of supernova explosions.  Conceptually such a
picture has some interesting similarities to protostar formation.
Related ideas have been expressed and discussed by Kardashev (1970),
Bisnovatyi-Kogan  (1970), LeBlanc \&  Wilson (1970), Ostriker \& Gunn
(1971), Amnuel {\elevenit et al.} (1972), Bisnovatyi-Kogan {\elevenit
et al.} (1976), and Kundt (1976), with Bisnovatyi-Kogan (1970) the
closest to the argument here (see paper CR I).  It remains as an
interesting, but  possibly quite difficult next step to calculate the
evolution of massive stars allowing for the possibility of extreme
rotation in the core and strong magnetic fields.

\bigskip
\line{\elevenit 2.2.3. The Particle Spectrum above the Knee \hfil}
\smallskip

Consider the derivation of the spectrum beyond the knee.  Since
the maximum energy particles can attain is a strong function of colatitude,
the spectrum beyond the knee requires a discussion of the latitude
distribution, which we have to derive first.  The latitude distribution is
established by the drift of particles which builds up a gradient which in
turn leads to diffusion down the gradient.  Hence it is clear that drifts
towards the equator lead to higher particle densities near the equator, and
drifts towards the poles lead to higher particle densities there.  Thus the
equilibrium latitude distribution is given by the balancing of the
$\theta$-diffusion and the $\theta$-drift.

The diffusion tensor component $\kappa_{\theta \theta}$
can be derived similar to our heuristic derivation of the radial diffusion
term $\kappa_{r r}$, again by using the smallest dominant scales.  The
characteristic velocity of particles in $\theta$ is given by the erratic
part of the drifting, corresponding to spatial elements of different
magnetic field direction, and this is on average the value of the drift
velocity  $\mid V_{d,\theta}\mid$, possibly modified by the locally
increased values of the magnetic field strength, and the characteristic
distance is the  distance to the symmetry axis $r\,sin\,\theta$;  this
is the smallest dominant scale as soon as the thickness of the shocked
layer is larger than the distance to the symmetry axis, {\elevenit i.e.}
$sin \,\theta\,<U_2 /U_1$.  Thus we can write in this approximation,
{\elevenbf postulate 3},

$$\kappa_{\theta \theta ,1}\;=\;{1 \over 3}\,\mid V_{d,\theta}\mid
\,r\,(1-\mu^2)^{1/2}. \eqno\autnum $$

Here $\mu$ is again the cosine of the colatitude on the sphere we
consider for the shock in the wind.  Interestingly, this can also be
written in the form

$${1 \over 3}\,r_g \,c \, (1-\mu^2)^{1/2} , $$

\noindent where $r_g$ is taken as positive; we also note that $c \,
(1-\mu^2)^{1/2}$ is the maximum drift speed at a given latitude, valid for
the local maximum particle energy.  This suggests that the latitude
diffusion might be usefully thought of as diffusion with a length scale
of the gyroradius, and the particle speed, to within the angular factor
which just cancels out the latitude dependence of the magnetic field
strength in the denominator of the gyroradius.

We assume then for the colatitude dependence a powerlaw
$(1-\mu^2)^{-a}$ and first match the latitude dependence of the diffusion
term and the drift, and then use the numerical coefficients to determine
the exponent in this law. The diffusion term and the drift term have the
same colatitude dependence since the double derivative and the internal
factor of $(1-\mu^2)$ lead to a $(1-\mu^2)^{-a-1}$ for the diffusive term,
while the drift term is just the simple derivative giving the same
expression.  For $(1-\mu^2)\,\ll\,1$ the condition then is $ {2 \over
3}\,a^2\;=\;\pm \,a $.

It is important to remember the sign of these terms.  The diffusive term
is always positive, while the $\theta$-drift term is negative for
$Z \,B_s$ negative.  This means for positive particles and a magnetic
field directed inwards the $\theta$-drift is towards the pole.  In that
case then the exponent $a$ is either zero or $a\,=\,3/2$. Since the drift
itself clearly produces a gradient, the case with $a\,=\,0$ is of no
interest here. It follows that for positive particles and an inwardly
directed magnetic field the latitude distribution is strongly biased
towards the poles, emphasizing in its integral the lower energies, and
thus making the overall spectrum steeper beyond the knee energy.  When
the magnetic field is directed outwards and the particles are positive,
the drift is towards the equator with then a positive gradient with
$(1-\mu^2)^{3/2}$, again in the limit $(1-\mu^2)\,\ll\,1$.  We note that
this exponent $3/2$ is reduced in the case, when the erratic part of the
drift is increased over the steady net drift component.

In our model for the diffusion in $\theta$ we have used the drift
velocity and the distance to the symmetry axis as natural scales in
velocity and in length.  When the $\theta$-drift reaches the maximum
derived earlier, then the latitude drift changes character.  This happens
at a critical energy, which is reached at

$$\kappa_{\theta \theta,1}\;=\;\kappa_{\theta \theta , max} ,
\eqno\autnum $$

\noindent which translates to

$$E_{crit} \;=\;({3 \over 4} \,{U_1 \over c})^2 \, E_{max} \;=\; E_{knee}.
  \eqno\autnum $$

We emphasize that two different basically geometric arguments lead to the
same critical energy, $E_{knee}$.  This means, {\elevenbf postulate 4},
that for

$$E\;>\;E_{knee}  \eqno\autnum $$

\noindent the drift energy gain is down by $2/3$ to the level what the pure
gradient and curvature drift yields, which results in

$$x\;=\;11/6. \eqno\autnum $$

The reduced drift energy gain reduces the value of $x$ below the limiting
value derived earlier, of $9/4$.  This then leads to an overall spectrum of

$$E^{-29/11} , \eqno\autnum $$

\noindent before taking leakage into account, and

$$E^{-98/33}\;\cong\;E^{-3} , \eqno\autnum $$

\noindent with leakage accounted for.  This is what we wanted to derive.
The observational test with airshower data confirms (see paper CR IV and
below), that the particle spectrum beyond the knee is close to the spectrum
derived here; in fact it suggests a slightly steeper spectrum, of close to
$E^{-3.07}$, about $0.1$ steeper than derived here.  This is clearly within
the uncertainties of our analytical approach.

\bigskip
\line{\elevenit 2.2.4. Observational Test: Radio emission \hfil}
\smallskip

There are a number of consistency checks that are possible with the radio
emission of Wolf Rayet stars, radiosupernovae and OB stars (Abbott
{\elevenit et al.} 1986, Bieging {\elevenit et al.} 1989, Cassinelli 1983,
1991, Maheswaran \& Cassinelli 1988, 1992; the references for
radiosupernovae are given below; for details see paper CR II). Additional
checks with X-ray and $\gamma$-ray data are possible as well as with novae.

First of all, it is possible to demonstrate that the classical dynamo
process (Ruzmaikin {\elevenit et al.} 1988) can produce high magnetic
fields inside of massive stars, of order $10^6$ to $10^7$ Gauss, so that
it appears quite plausible that already OB stars which are radiative
outside have appreciable strengths of their surface magnetic fields, of
order $10^3$ to $10^4$ Gauss.  Rotationally induced circulations can
carry the fields on fairly short time scales to the surface for rapidly
rotating stars.  Wolf Rayet stars have as their surface former convective
layers (Langer 1989) and so easily can have as high magnetic fields as
suggested by the cosmic ray argument, also of order $10^3$ to $10^4$ Gauss.
Secondly, it is possible to show that the magnetic fields can provide
additional momentum to the wind of Wolf Rayet stars, and so may help to
explain the strength of these winds (see paper CR II), acting as an
amplifier to the radiative forces.

The radio luminosities of Wolf Rayet stars and of radiosupernovae can
readily be understood with synchrotron emission and free-free absorption
(for free-free absorption Altenhoff {\elevenit et al.} 1960, for a
proper integration of free-free emission Biermann {\elevenit et al.} 1990,
for radiosupernovae Weiler {\elevenit et al.} 1986, 1989, 1990, 1991,
Panagia {\elevenit et al.} 1986, for the supernova 1987A, Turtle
{\elevenit et al.} 1987, Jauncey {\elevenit et al.} 1988, Staveley-Smith
{\elevenit et al.} 1992, for the new supernova in M81 {\elevenit e.g.}
Strom {\elevenit et al.} 1993).  The comparison suggests that there is
a critical shock velocity below which electron injection gets very weak.
For the winds of massive stars this critical velocity is of the order
of $5000 \;\rm km/sec$, interestingly, but probably misleadingly close
to the shock velocity for which the downstream thermal electrons reach
the speed of light.

\bigskip
\line{\elevenit 2.2.5. Observational Test: Air Shower data \hfil}
\smallskip

We wish to test the overall model proposed by asking whether it can
successfully account for the spectrum and chemical composition at particle
energies beyond $10^4 \;\rm GeV$: the special difficulty in this endeavour
is the fact that we know the chemical abundances near TeV particle energies
already, and extrapolating these spectra does not readily give the bump and
knee of the well established overall particle spectrum.

The difficulty in all such attempts originates in the relation between
energy estimation with air showers and the mass of the primary nucleus.
Generally showers initiated by heavy nuclei develop and are
absorbed faster in the atmosphere. Heavy nuclei thus produce showers of
smaller size at the observation levels of all existing experiments.
The effect is stronger for inclined showers, which have to penetrate
through larger atmospheric thickness. We model the shower
size that the primary cosmic ray flux generates in the Akeno detector
(Nagano {\elevenit et al.} 1984), using both vertical showers and slanted
showers; having performed such a fit, we then reconstruct the all particle
spectrum.

Various cosmic ray experiments have given data about the chemical
composition near TeV energies and somewhat beyond, up to $100 \;\rm
TeV$, most notably from the Chicago Group (Grunsfeld {\elevenit et al.}
1988, M\"uller 1989, Swordy {\elevenit et al.} 1990, and M\"uller
{\elevenit et al.} 1991), the JACEE experiment (Parnell {\elevenit
et al.} 1989, Burnett {\elevenit et al.} 1990, and Asakimori
{\elevenit et al.} 1991a, b ), Simon {\elevenit et al.} (1980) and
Engelmann {\elevenit et al.} (1990).  The Oxygen data, where we have
averaged existing data from all four experiments when independent
measurements exist in the same energy range, clearly demonstrate a
power law behaviour with a somewhat flatter spectrum than the low
energy index of about $2.75$. Oxygen has a spectrum of about $2.64$,
corresponding well to the argument made in CR I, that supernova
explosions into existing stellar winds produce not only higher particle
energies, but also spectra flatter than the lower energy particles
from Sedov phase explosions into the interstellar medium.   We use
all these known chemical abundances as a given input, combining the
elements into six groups, and the sources into three different sites
as discussed above.  The six element groups which we use are: a)
Hydrogen, b) Helium, c) Carbon, Nitrogen and Oxygen, d) Neon to
Sulphur, e) Chlorine to Manganese, and f) Iron.  We argue that the
galactic cosmic rays cut off at $10^8 \, Z \; \rm GeV$  and that the
extragalactic component is nearly all protons and Helium nuclei.
In paper CR II (Biermann and Cassinelli 1993) we argue that this is in fact
suggested by a number of stellar observations.

The main uncertain parameters are then the following:

1) The spectral difference between the Sedov phase explosion particles
and the wind explosion particles.

2)  The particle energy $\sim Z$ at the cutoff for the Sedov phase
explosion acceleration.

3)  The relative strength of the polar cap component.

4)  The spectral difference between the energy range below and above the
knee.

5)  The particle energy $\sim Z$ of the knee.

6)  The upper cutoff energy $\sim Z$ for the component beyond the knee.

The theory predicts specific numbers for all these parameters.  The
division of the abundances between the Sedov phase accelerated particles
and the wind explosion accelerated particles can be made using existing
data, especially from JACEE, and thus is no free parameter.

We use {\elevenit directly the shower size spectra} to reconstruct the
cosmic ray spectrum under the assumption of different chemical
compositions. We do this simultaneously for two zenith angles, vertical
($secan \, \theta = 1.0$) and for $secan \, \theta = 1.2$,
{\elevenit i.e.} one moderate  slant angle. At these angles the
statistical errors are small. The shower size is expressed by
$N_e$ - the total number of charged particles at observation level.
The flux of showers with $N_e$ = $10^6$ at $secan \, \theta = 1.2$ is
smaller than the vertical one by a factor of 7. The shapes of the size
spectra are also different.

The detailed modelling procedure is described in Stanev {\elevenit et al.}
(1993, paper CR IV).  The comparison with the data demonstrates that the
fits are well within the acceptable error ranges. The model has the
following parameters:

The cutoff energy (exponential cutoff) for the Sedov phase
accelerated particles is $120 \;\rm TeV$ for protons; the spectral index
of the wind component is $2.66$ below the knee and $3.07$ above the knee;
this steepening occurs at $700 \;\rm TeV$ for protons and is rigidity
dependent. The ratio of the polar cap component to the steeper wind
component is $1$ for heavy nuclei at the bend. Although the polar cap
component does not contribute to shower sizes significantly below or
above the knee, it is essential for reproducing the sharp break at
$N_e = 10^6$.  The reproduction of this shower break, as well as the
observed continuous flattening of the spectra of all heavy nuclei
requires the introduction of a very flat ($E^{-2}$ at injection)
acceleration component, {\elevenit i.e.} the polar cap component.  All
four major parameters are close to their expected values ({\elevenit
predicted} in the earlier papers CR I, CR II, and CR III).

It is clear that within the framework of our picture the knee feature in
the overall spectrum can only be due to the addition of the polar cap
component. Without it the generated size spectrum cannot have a sharp
bend in any model similar to ours, with rigidity dependent features and
composed of different nuclear components.

We note that the shower size distribution is very sensitive to changes in
the parameters. Relatively small changes (by more than $20\%$) of the
cutoff and bend energy, as well as in the composition of the wind
component, affect the calculated size spectra so strongly that they
become inconsistent with data. Together with the direct measurements
in the $1 \;\rm TeV$ region the shower size distributions {\elevenit
constrain} the models to within a small parameter space. Fitting the
shower size distribution at different zenith angles is critical despite
its fairly large error bars.

Several properties stand out:

1)  The {\elevenit correct overall spectrum} is lower in the knee region
and beyond than the conventionally derived spectrum; the factor is about $2$.
The knee itself is not as sharp but the change of the spectral index
takes place at approximately the same energy.

2)  The composition changes rapidly in the knee region, becomes
increasingly heavier, and in our current interpretation is dominated
by the Neon group above the knee.

3)  The proton flux at energy above $10^8 \;\rm GeV$ represents
the extragalactic component.  Because the wind component has a maximum
energy of $10^8 \;\rm GeV$ for protons ($2.6 \, 10^9 \;\rm GeV$ for Fe)
the calculated shower size spectra would become too steep for sizes
above $10^7$ without this extra component. This implies an extragalactic
flux (presumably mostly protons) that is consistent with the independently
derived estimate by Gaisser {\elevenit et al.} (1993) using the Fly's Eye
data; a full comparison with an extragalactic model (Rachen \& Biermann,
1992, 1993:  paper UHE CR I) has been done in Rachen {\elevenit et al.}
(1993, paper UHE CR II).

4)  The comparison demonstrates that our curves {\elevenit fit all
existing data} quite well, including the trend, observed in heavy nuclei,
to exhibit a flattening of the spectrum at the approach to the knee.
The seeming inconsistency with the highest energy direct data for iron
is possibly due to the fact the the JACEE experiment (where these points
come from) has presented its measurements for nuclei with $Z>17$.  It is
difficult to judge what the absolute normalization of the cosmic ray flux
beyond the knee is.  The presentation of the spectrum in the form used
here ($E^{2.75} \,dN/dE$) tends to exaggerate differences in the absolute
normalization.  A relatively small error of the energy determination in
air showers (typical error of $20\%$) changes the normalization of the
absolute flux by a large amount ($65\%$).

We conclude that the {\elevenit most stringent test} yet of the model
proposed in the earlier papers of the series on cosmic ray physics
(CR I, CR II, CR III, UHE CR I,  UHE CR II), which give predicted
spectral shapes for all three sites of origin for cosmic rays is
successful.  Several detailed conclusions can be made:

1)  The abundances of the cosmic rays from explosions into stellar winds
do not require nor imply any admixture from the heavy elements newly
produced in the star exploding.  All the heavy elements already present
in the stellar wind (enriched already to some degree) prior to the
explosion participate in the feeding of the accelerated particle
population, and no other additional source is required.

2)  The nuclei accelerated in the polar cap region are essential for
a successful fit of the shower size spectra. Although the total energy
carried by such nuclei is not more than $1/100$ of the total
wind component, they contribute a major fraction in the knee region.

3)  There is no requirement for other cosmic ray sources, either
from spectral arguments or from abundance arguments; thus pulsars and
compact X-ray binary systems may accelerate lots of particles, but they
need not play a dominant role out in the typical interstellar medium.

4)  The large fraction of heavy nuclei inferred from the Fly's Eye data
(Gaisser {\elevenit et al.} 1993) requires the acceleration of heavy
element nuclei out to at least $3 \, 10^9$ GeV.  This implies (see CR I
and II) that the winds of the stars that explode as supernovae have
magnetic fields at least as strong as $3$ Gauss at a fiducial distance
from the star of $10^{14} \;\rm cm$.  Since this limiting particle
energy is already derived by using the spatial limit given by the
condition that the Larmor radius of the particle fit into the space
available, there is no other way than indeed having these high fields in
the stellar winds of massive stars in the context of our theory  (note
that also Usov \&  Melrose 1992 argue for high magnetic fields, on
different grounds).  The magnetic field strengths required are of order
$10^3$ to $10^4$ Gauss on the surface of the stars, with the exact value
clearly depending on stellar radius and exact wind configuration.  An
important next step is to work out the consequences of these models at
lower energies.

\bigskip
\line{\elevenit 2.3. Starburst and normal Galaxies \hfil}
\smallskip

Starburst galaxies clearly produce a large number of supernovae,
accelerate cosmic rays to energy densities in their interstellar medium
far above that of the interstellar medium in the solar neighborhood, and
also produce break-outs from their gaseous disks to produce winds, which
are visible in a)  optical polarized light and line emission, in b) X-rays
as well as in c) nonthermal radio emission (Kronberg {\elevenit et al.}
1981, 1985, Schaaf  {\elevenit et al.} 1989). Normal galaxies are also
thought to have galactic winds, albeit of less conspicuous power
(Breitschwerdt {\elevenit et al.} 1991, 1993).

We would like to obtain an estimate for the cosmic ray production.  For
this we need to discuss three points, a) the power output of galaxies in
cosmic rays, b) their spectrum, and c) their low energy cutoff.

As our reference we take the well studied starburst galaxy M82
({\elevenit e.g.} Rieke {\elevenit et al.} 1980, Kronberg {\elevenit
et al.} 1981, 1985, Schaaf {\elevenit et al.} 1989).  M82 is the first
galaxy to yield a large number of observable radio supernova remnant
candidates (Kronberg {\elevenit et al.} 1985), the stronger ones being
time variable; this allows a first estimate of the actual supernova rate.
In Kronberg {\elevenit et al.} (1985) we estimate a supernova rate of
about 1 per 3 years, with an uncertainty of a factor of three.  Assuming
that the efficiency of producing cosmic rays is about $10\%$ and the
energy input per supernova is $10^{51} \;\rm ergs$, we thus arrive at
a cosmic ray production luminosity of M82 of $10^{42} \;\rm ergs/sec$;
we have to compare this with the farinfrared luminosity of (Rieke
{\elevenit et al.} 1980) $4 \,10^{10} \,L_{\odot}$.  This gives a ratio
of cosmic ray luminosity to farinfrared luminosity of $0.0069$ with an
uncertainty of a factor of three; we adopt as our reference ratio $0.01$
and keep the uncertainty in mind.  It is an important next step to work
through the large amount of radio data assembled by Kronberg and his
colleagues on M82 to check on all these detailed models.

The spectrum of cosmic rays outside the galaxy converts back to its
source spectrum in the simple leaky box argument; although there are
obvious problems with such arguments (see Biermann 1993), there is
nothing better at present and so we will use this approach here as well.
The acceleration of cosmic rays is discussed above.  For the low energy
cosmic rays this means that their spectrum is about $p^{-2.4}$ at
injection, and again outside the galaxy.

This injection spectrum is for the transrelativistic
regime.  Hence the next question is the low momentum cutoff in this
spectrum.  Here we have information from two sources.  First, when one
repeats the argument of Spitzer \& Tomasko (1968) for interstellar
clouds (Jokipii and Biermann, 1991, as yet unpublished) then one finds
that in order to explain the cosmic ray ionization rate in interstellar
clouds required by the chemistry (Black \& van Dishoeck 1987, Black
{\elevenit et al.} 1990, Black \&  van Dishoeck 1991) low momentum
cutoffs corresponding to kinetic particle energies of $30 \; - \; 100
\;\rm MeV$.  Second, the spallation of cosmic rays produces Be and other
light elements (Gilmore {\elevenit et al.} 1992) also gives an indication
of the low kinetic energy cutoff of the cosmic ray spectrum; again, the
cutoff is in the same kinetic energy range.  The high end of the particle
spectrum is due to explosions into winds, and thus has a rather steep
spectrum, but is dominated by heavy nuclei.

We thus find the following estimates: a cosmic ray luminosity of about
$1 \,\%$ of the farinfrared luminosity, with a spectrum in momentum of
$E^{-2.4}$, and a low kinetic energy cutoff of 30 MeV.  Except for the
spectrum, these estimates are fairly uncertain.

These cosmic ray particles are subject to adiabatic loss as they
come out from a galaxy, and so their energy density as well as their
low and high end particle energies are lowered by a factor which is
likely to be of order at least $10$.  For various plausible models of
galactic evolution this extragalactic cosmic ray population can ionize
and heat the intergalactic medium (Nath \& Biermann 1993), and help to
let us understand the extreme degree of ionization of the intergalactic
matter implied by the Gunn-Peterson test as well as the low Compton $y$
parameter implied by the COBE data (Mather {\elevenit et al.} 1993).
The temperature of the intergalactic medium in such models is about an
order of magnitude higher than based on photoionization models, but still
quite low compared to temperatures which would be required to give a
substantial contribution to the X-ray background.  A further desirable
next step would be a calculation of models for Lyman $\alpha$-clouds
taking low energy cosmic ray particles into account.

\vglue 0.6cm
\line{\elevenbf 3. Radio Galaxy Hot Spots:  Highest Energy Cosmic Rays
\hfil}
\vglue 0.4cm

Any model of the origin of the highest energy cosmic rays has to fulfill
the following conditions: (i) the predicted sources must be able to
accelerate particles to the observed energies up to at least $100 \;
\rm EeV$, (ii) the energy content in relativistic particles and the
number density of the sources must be sufficient to provide the
observed UHE-CR flux, (iii) the observed UHE-CR spectrum must be fitted
in the energy region where the considered contribution dominates over
all others, and (iv) the chemical abundances must match the new Fly's
Eye results (Gaisser {\elevenit et al.}  1993). The investigation of
FR II radio galaxies constitutes a fairly good basis for this purpose:
FR II galaxies are very bright radio objects, their distribution and
evolution is well known up to rather high redshifts (Peacock 1985,
Barthel 1989), and their content of relativistic particles can be well
estimated from their sychrotron emission (Rawlings \& Saunders 1991).
The only problem is to obtain a reliable estimate of the proton to
electron ratio in the energy density of relativistic particles; this
uncertainty introduces  a ``fudge factor'' in our model, which can only
be tested on plausibility but not calculated at present.

\bigskip
\line{\elevenit 3.1. Prediction \hfil}
\smallskip

Clearly, the hot spots in FR II galaxies are not expected to be the only
extragalactic objects that can accelerate particles to ultra-high energies.
Most active galactic nuclei might provide much better conditions for that,
and shock acceleration can also take place in the jets of less luminous
FR I galaxies ({\elevenit e.g.} M87). The reason for the restriction to
the rare class of FR II galaxies is, that highly energetic charged
particles produced deep inside galactic structures will suffer substantial
adiabatic losses on their way out to the extragalactic medium. This
problem applies to all possible cosmic ray sources except the FR II hot
spots, since they are located at the edge of the extended radio lobe of
the galaxy and the particles can readily enter the extragalactic space.
However, protons can escape from galactic cores with very high energies,
since they can undergo isospin flips in $pp$ or $\gamma p$ reactions,
leaving the galaxy as an UHE neutron. At present it is not clear whether
the resulting CR spectrum even extends to the highest energies considered
here (Protheroe \& Szabo 1992, Mannheim 1993a, b); however, it is clear
that airshower data require heavy nuclei to dominate beyond the knee in
contradiction to the model by Protheroe \& Szabo (1992).

Our model is based on the present knowledge about diffusive shock
acceleration, radio galaxy properties, galaxy evolution and a simple model
for intergalactic cosmic ray propagation.

The hot spots in strong extended radio galaxies, denoted as FR II
galaxies following the classification of Fanaroff \& Riley (1974), are
identified as the endpoints of powerful jets ejected by the active nuclei
of the galaxies deep into the extragalactic medium (see, {\elevenit e.g.},
Meisenheimer \&  R\"oser 1989). It has been shown that the radio-to-optical
spectra emitted by those hot spots can readily be explained as synchrotron
radiation from  particles accelerated at a strong shock wave by the first
order Fermi mechanism ({\elevenit e.g.}~Biermann \& Strittmatter 1987,
Meisenheimer {\elevenit et al.}~1989).

The basic argument (Biermann \& Strittmatter 1987) runs as follows:
Protons and other heavier nuclei are accelerated as explained above
(see,  {\elevenit e.g.} Drury 1983) for parallel shocks; these energetic
nuclei have fairly low losses and thus attain large particle energies,
and are thus the first "messenger" to the upstream flow of the coming
shock transition.  The nuclei cause plasma turbulence to develop which
initiates a cascade in wavenumber space.  This cascade is assumed to be,
on the same basis as argued above (Matthaeus \& Zhou 1989, and by analogy
with the solar wind where the turbulence can be measured by spacecraft),
of a Kolmogorov character.  Thus the mean free path for the scattering of
the electrons is fixed by the energetic nuclei.  This mean free path for
the scattering of electrons in turn leads to a maximum particle energy
for electrons, which translates to a maximum synchrotron emission
frequency, which is {\elevenit independent of the strength of the
magnetic field} and thus ought to be similar in both compact sources
as well as in extended hot spot regions; this is exactly the behaviour
found in many red (nonthermal) quasar emission spectra, in BL Lac objects,
in the M87 jet, and in radio hot spots.  Thus, the ubiquitous cutoff
(many papers starting with Rieke {\elevenit et al.} 1976) in the electron
synchrotron emission finds a ready explanation.  There is some variance
in the exact cutoff frequency due to the possibilities (all neglected
except the last point by Biermann \& Strittmatter 1987) due to a) a
slight obliqueness of the shock, b) different transport coefficients
on both sides of the shock, c) weak relativistic boosting between shock
frame and observers frame, d) time dependence and e) the possible
influence of a radiation field, which can modify the observed cutoff
frequency.  There is a possibility to check the assumed Kolmogorov law
with the data of Meisenheimer {\elevenit et al.} (1989):  since particles
accelerated at the shock continuously lose energy as they move downstream
with the flow, the integration of the emission over the downstream region
produces a bend in the emission spectrum at some frequency which is
inversely related to the length of the effective downstream region.
With this effective length available from observation together with the
bend and the cutoff frequency one can limit the possible range of
turbulence spectra of the plasma cascade, and the result suggests support
for the initial assumption of a Kolmogorov spectrum.

Given this agreement, we can then use the theory to derive maximum
particle energies for protons and other nuclei. For the protons a high
energy cutoff of the order of $10^{21}\;\rm eV$ is found if their energy
is mainly limited by synchrotron losses and p-$\gamma$-collisons rather
than by the finite range of the diffusion region or by interactions with
ambient photons.

The simple canonical theory of diffusive shock acceleration
explains the main hot spot features and is therefore a very attractive model
for the acceleration of cosmic ray particles.  Clearly, slightly flatter
spectra as predicted by enhanced acceleration models (Kr\"ulls 1992) are
consistent with most of the results within the error estimates, even
required by the observations of Pic~A~west.

FR II galaxies are extended, steep spectrum radio galaxies, that are
uniquely classified by their double structure due to their luminous
hot spots.  This morpology occurs for radio luminosities above $2\cdot
10^{26}\,\rm W Hz^{-1}$ at $178 \;\rm MHz$, which makes it easy to
distinguish FR II galaxies from other radio galaxies in the radio
luminosity function.  The present state of knowledge is that roughly
$70\%$ of all steep spectrum radio sources in this power range show
FR II structure (Perley 1989), but on the other hand there may also be
some compact flat spectrum sources that can be identified with FR II
galaxies having their jets oriented along the line of sight (Padovani
\& Urry 1992). At high redshifts (above z=1), however, little is known
about the morphology of the sources and it is not clear whether the
Fanaroff-Riley relation still holds; but we will see that this
uncertainty affects the cosmic ray spectrum only below $1 \;\rm EeV$.

The synchrotron radiation emitted by relativistic electrons is the only
(or at least the main) information we get about the content of
relativistic particles in jets. However, the total sychrotron luminosity
$L_s$ of a source allows us to give minimum estimates about their mean
energy density in relativistic particles $U_p$ and $U_e$, protons and
electrons respectively, and in the magnetic field $U_B$. The minimum
energy condition is given by, {\elevenit e.g.}, Pacholczyk (1970).  We
introduce the quantity $k_p = U_p/U_e$. The emission of photons in the
X and $\gamma$ regime from UHE protons in hot spots by synchrotron
radiation and the proton-induced electromagnetic cascades can by expected
(Mannheim \& Biermann 1989, Mannheim {\elevenit et al.}~1991). Recent
GRO detections of blazars, which may be considered as compact
Doppler-boosted emission from jets support this expectation
(Mannheim \& Biermann 1992,  Mannheim 1993a, b) and give evidence
for the presence of energetic protons as employed here. In the future,
if good observations of X-ray and hard $\gamma$ emission from hot spots
can be obtained, it may even be possible to derive definite estimates
on $k_p$.  However, any such estimate has to overcome the difficulty
that we do not know how to estimate the efficiency of converting
jet-power into ultimate energy density in relativistic nuclei in the hot
spots.

For considering FR II galaxies as cosmic ray sources, we are only
interested in the ``proton luminosity'' of a source. From the
synchrotron luminosity one can derive the total jet power, assuming
some value for $k_p$. Rawlings \& Saunders (1991) gave the jet power
for $39$ FR II galaxies, assuming no protons in the jet ($k_p = 0$).

$f$ is the {\elevenit fudge factor \/} in our normalization containing
all the uncertainties briefly mentioned above. We derive $f$ from fitting
our calculated spectra to the data and therefore obtain information on
the relative proton content $k_p$ required.

To calculate the extragalactic source function of cosmic ray protons,
we have to apply our knowledge about jet powers and proton luminosities
to the epoch dependent radio luminosity function (RLF) of FR II galaxies.
The RLF $n_{\rm G}(P_\nu,z)$ gives the number density of radio galaxies
with a specific radio luminosity $ P_\nu$ at a given frequency $\nu$ at
an epoch $z$ respective to the {\elevenit comoving\/} unit volume element.
RLF's can be derived from radio source counts, using the available
information about redshifts and assuming some modelling parameters
(Peacock \& Gull 1981, Windhorst 1984, Peacock 1985).

To connect the proton source function to the RLF, the given relation
between {\elevenit total\/} synchrotron luminosity and kinetic jet
energy is not very useful; we rather have to find a relation between
jet energy and radio luminosity at the frequency for which the RLF is
given. In this work, we use the various luminosity functions given by
Peacock (1985) for different modelling parameters, based on source
counts at $2.7\,$GHz for steep spectrum radio sources.

Protons with Lorentz factors $\gamma_p \simge 10^{10}$ can undergo high
energy reactions even with the very low energy photons of the universal
microwave background radiation (MBR), as pointed out first by Greisen
(1966) and independently by Zatsepin \& Kuzmin (1966). The attenuation
length reduces to merely $10\; \rm Mpc$ for $\gamma_p \simge 10^{11}$
(Stecker 1968), what causes a cutoff in the particle spectra of distant
extragalactic sources, usually denoted as the {\elevenit
Greisen-cutoff\/}. Hillas (1968) demonstrated the influence of the
cosmological evolution on this effect; for very distant sources the
Greisen-cutoff apears at much lower energy, because of the higher
density and temperature of the MBR at earlier epochs.  The most
extensive investigations of the transport of UHE protons in the
evolving MBR were done by Hill \& Schramm (1985) and by Berezinsky
\& Grigor'eva (1988), who both considered the modification of the
spectrum due to the cosmological evolution of the proton sources.
We will use here the method of Berezinsky and Grigor'eva.

In all the discussion sofar we only discussed the evolution of
proton spectra in time; the spatial propagation was always assumed to
be rectilinear. We apply our method to the calculation of proton spectra
from single sources with a clearly defined spatial distance; if there
is any appreciable magnetic field in extragalactic space,
the particles propagate by a combination of drift, convection and
diffusion, so that our results may no longer be valid. We use the
approximation of straight line paths for the energetic particles.  The
limitations and merits of such an approach are discussed at some length
in paper UHE CR I.

The results derived from this modelling lead to spectra with an $E^{-2}$
behaviour at energies well below EeV, then a section with roughly
$E^{-2.75}$ and a bump, and finally a sharp cutoff near $100$ EeV;
the chemical composition is mostly Hydrogen.

\bigskip
\line{\elevenit 3.2 Tests with Airshower data \hfil}
\smallskip

Here we compare the prediction introduced above with
observational data from both the Fly's Eye and the Akeno airshower
detectors.  These experimental data are now available from the
analysis of the chemical composition near EeV energies
(Gaisser et~al.~[1993] for the Fly's Eye experiment in an accurate
analysis, and from Stanev et~al.~[1993] for the Akeno experiment as
a consistency check) and demonstrate that below $1\; \rm EeV$ protons
show a flatter spectrum than the overall spectrum  and thus their relative
proportion increases with energy.  The comparison of the spectral data
and the flux shows that the ultra high energy component of the cosmic
rays can indeed be understood as arising from the hot spots of powerful
radiogalaxies, in flux, spectrum and chemical composition: This result puts
stringent limits on the propagation of high energy cosmic rays and thus
on the properties of the intergalactic magnetic field.

We can use the world data set on the cosmic rays to obtain an averaged
spectrum of all cosmic rays beyond $0.1\;\rm EeV$ as a basis for the
absolute normalization of the new component.

The measurement of the flux of ultra high energy cosmic rays is
difficult because of their small flux (less than $0.05$ particles per
${\rm m}^2\; \rm.ster.year$ above $E=10^8 \;\rm GeV$). Still relatively
large statistics have been collected during the last 20 years by four
experiments, three of which are still active (Akeno (Nagano
et~al.~1984; 1992), Yakutsk (Efimov et~al.~1991) and the Fly's Eye
(Cassiday et~al.~1990; Loh et~al.~1991)). The array at Haverah Park
(Cunningham et~al.~1980) does no longer exist. A recent presentation
and comparison of the results of the four arrays is made by Sokolsky
{\elevenit et al.}  (1992).

Apart from the differences in normalization all data sets show the
same slope ($\gamma$=3.0--3.1) in the region $3 \times 10^{17}$ --
$3 \times 10^{18} \;\rm eV$.  We use the data in this range to normalize
experiments to each other. First an average spectrum is calculated
using the original experimental errors. Then we determine for each
experiment the energy shift $\Delta E$ that brings the data set in
best agreement with the average spectrum. The resulting shifts are
smaller than the suspected systematic errors -- we obtain $\Delta E /E =
-16\%$ for Yakutsk, $-13\%$ for Akeno, $-3\%$ for Haverah Park (which
has the smallest statistical errors) and $+6\%$ for the Fly's Eye. We
apply these energy shifts to the full data sets and obtain a world
data set where the scatter is smaller than the individual error bars.
Finally we bin in logarithmically equal energy bins, combine
individual data points within the same bin and produce the average
flux. This is the flux that we shall use to estimate the strength of
different chemical components further down.

This procedure is not unique and does not eliminate the systematic
errrors in the energy derivation. We claim, however, that it
introduces a standard systematic error, which is nearly the same for
all data sets. The energy shifts required for the averaging are
remarkably small. A better way to achieve a standard energy derivation
from different experiments is a detailed study of the experimental
algorithms (M. Lawrence et~al.~1991) and a comparison and possible
improvement of their theoretical bases.

We have produced a series of model calculations for an input
spectrum of $E^{-2}$ with an intrinsic exponential cutoff at different
energies in the range about $100\;\rm EeV$, and a Hubble constant of
$75\,\rm km\,sec^{-1}\,Mpc^{-1}$.  Such calculations involve an
estimate for the conversion of radio luminosity to jet power (Rawlings
and Saunders 1991), from jet power to energy in energetic particle
populations, and, most importantly, the ratio between the energy
density of the energetic electrons, and the energetic nuclei -- after
all, the observed synchrotron emission traces only the relativistic
electrons.  The combination of these factors is the
``fudge factor'' introduced above, and it is probably dominated by
the unknown nuclei to electron ratio.
Low energy fits to the Fly's Eye data for protons (see below)
imply this factor to be near $3$, and so suggest that the proton to
electron ratio is of order $20$ or less (see UHE CR I and above), a rather
uncertain limit.

We have used two independent methods to estimate the extragalactic
flux from different air shower experiments. First we fit the shower
size distributions published by the Akeno experiment (Nagano
{\elevenit et~al.}~1984) for different zenith angles. Then we use the
results of a recent analysis (Gaisser {\elevenit et~al.}~1993) of the
cosmic ray composition around $10^{18} \;\rm eV$ from the measurements
of the Fly's Eye detector (Cassiday {\elevenit et~al.}~1990).

The Akeno shower array is a traditional detector that measures the
number of charged particles in the shower ($N_e$) that reach the
observation level.  We use data sets for vertical ({\rm
secan}$\,\theta=1.0$, ${\rm atm.~depth} \, =\, 920 \, {\rm g/cm}^2$) and
slightly inclined (${\rm secan}\,\theta=1.2$, ${\rm atm.~depth} \,=\, 1104
{\rm g/cm}^2$) showers. The fitting procedure is described in detail in
paper CR IV and consists of stepping through the energy spectrum of
each chemical component with a small logarithmic step ($10^{0.01}$) and
calculating the size of a number of showers at that energy and primary
mass with the parametrization of Gaisser (1979).  The resulting sizes at
both depths are binned with the appropriate weight, and the sum for all
components is then compared with the experimental data.

In the spectrum and composition model of paper CR IV all galactic nuclear
components have the same spectral index of 3.07 at energies above
$4\times10^{16}\; \rm eV$.  The comparison with the experimental $N_e$
distribution demonstrates that galactic cosmic rays fit well the
shower size distribution up to $N_e = (1.-3.)\times10^7$, but are
insufficient to maintain the calculated spectrum, especially for
inclined showers, in agreement with experimental data for bigger
$N_e$. A better fit requires the introduction of a flatter cosmic ray
component at energy above $10^{16} \;\rm eV$. The agreement is achieved by
introducing a component, consisting of pure Hydrogen, with an energy
spectrum of $(1.4\pm0.3)10^{-7}\times E^{-2} \;\rm
(cm^2\,ster\,s\,GeV)^{-1}$ at energies between $10^{16}$ and
$5\times10^{17}\;\rm eV$ and $(0.47\pm0.09)\times
E^{-2.75}\rm\;(cm^2\,ster\,s\,GeV)^{-1}$ at higher energy, where $E$
is measured in GeV. Because the spectral shape of this component is
entirely different from those of the galactic cosmic rays it is very
likely to represent an emerging extragalactic cosmic ray flux.

For the second estimate we use the recent results from the analysis of
the Fly's Eye measurements of the depth of maximum ($X_{\rm max}$)
distribution in terms of cosmic ray composition. The Fly's Eye is a
different type of detector, which observes directly the longitudinal
development of air showers through the detection of the fluorescent
light from the atmospheric Nitrogen atoms, induced by the shower
charged particles. The amount of fluorescent light is proportional to
the number of charged particles after an account is made for light
scattering and absorption. The data analysis fits individual data
points (taken at various atmospheric depths) to a shower profile and
derives the depth ($X_{\rm max}$) and size ($N_{\rm max}$) of the
shower maximum. $N_{\rm max}$ is proportional to the energy of the
primary nucleus and $X_{\rm max}$ depends on the energy and mass of
the primary nucleus. The sensitivity to the primary mass comes from
the rate of energy dissipation, which is faster in showers initiated
by heavy nuclei. $X_{\rm max}$ of Fe generated showers is about $100 \,
{\rm g/cm}^2$ shallower than that of proton generated shower of the
same energy.

The basic idea of the shower analysis is to simulate a large number of
air showers and compare the results to data.  The error bars include
both the statistical errors from the experimental statistics and the
fitting procedure and the systematic errors from the particle physics
input in the simulation, which also make the errors very asymmetric.

The best curve has an intrinsic cutoff of $100\;\rm EeV$, and a spectrum
which is nearly $E^{-2.75}$ between $1\;\rm EeV$ and $30\;\rm EeV$.
We selected this curve because it fits best the experimental data
simultaneously in the low and the high energy range.  The fit is quite
good, considering the error bars on the data, fitting to about $2\sigma$
or better.

Most of the existing theoretical proposals to explain the origin of
the cosmic ray particles at energies beyond the knee can be excluded
on the basis of the presented data:

\item{1.} Beyond the knee the particles are dominantly heavy nuclei and
not all protons as follows from the proposals by Protheroe \& Szabo
(1992), and by Salamon \& Stecker (1992).
\item{2.} The galactic wind model by Jokipii \& Morfill (1987) would
give chemical abundances much closer to solar abundances
for these cosmic ray particles than the models presented here, as would
the reacceleration model (Ip \& Axford 1992) except for the highest
particle energies.  Again, the observed abundances clearly disagree with
these models.  The essential reason for the difference in the predicted
chemical abundances is, that reacceleration models derive from the low
energy cosmic rays and thus have those abundances, albeit modified, while
the model proposed here derives its nuclei near to and beyond the knee
from the strongly enriched winds of evolved massive stars.

\noindent The difference curve ought to correspond to the galactic
contribution of heavy nuclei; this is first of all confirmed by the
Fly's Eye analysis, which does suggest, that in this particle energy
range there are very few nuclei of Carbon, Nitrogen, and Oxygen, and
many heavier nuclei.  Second, over the particle energy range which is
well measured, {\elevenit i.e.}  below a particle energy of about
$10 \;\rm EeV$, the curve is well described by a powerlaw with an
exponential cutoff near $5\;\rm EeV$, just as suggested by the earlier
cosmic ray arguments (papers CR~I, CR~II, and CR~IV) for the galactic
contribution of cosmic rays.  At energies above $30 \;\rm EeV$ the
subtraction is clearly unreliable, since the difference is between
two large numbers of similar numerical value.

We conclude first that a detailed check of the existing prediction
with the new data now available is successful.  The extragalactic
contribution can be readily modeled in a) particle energy, b) flux and
spectrum, and c) chemical composition.

There is a second important conclusion: Strong shocks in extragalactic
jets and their associated hot spots do produce energetic nuclei.  This
is of interest, since in all arguments about Gamma-ray sources like
the quasar 3C279, Mkn 421 and the like ({\elevenit e.g.} Hartman
{\elevenit et~al.}~1992a, b, c, Punch {\elevenit et~al.}~1992), there
is always the question whether active galactic nuclei accelerate nuclei
at all in their jets.  The successful fit made here suggests strongly,
that this is the case, and that protons dominate the process.  This
gives strong support to hadronic interaction models to explain the
gamma-ray emission from such quasars such as the model we have proposed
(Mannheim \& Biermann 1992; Mannheim 1993a, b; also see following section).

\vglue 0.6cm
\line{\elevenbf 4. Radioloud Quasars \hfil}
\vglue 0.4cm

This section has been written mostly by K. Mannheim.

Extragalactic radio sources are the largest known dissipative
structures (nonthermal objects) known in the universe and as such highly
interesting candidate sources of cosmic rays (see references in
Berezinsky {\elevenit et al.} 1990 and above). There are two ways for
them to inject cosmic ray baryons into the intergalactic medium; (i)
by escape from hot spots often a few hundred kiloparsecs away from
active nucleus of the host galaxy (Biermann 1991, Rachen \& Biermann
1992, and papers UHE CR I and UHE CR II)  and (ii) by neutron escape
from the sub--parsec scale of the jet or the nucleus (Protheroe \&
Szabo 1992).  Both these mechanisms avoid catastrophic adiabatic losses
preventing such escape otherwise.  However, do we have any indications
that baryons are actually present in radio jets?

The leptonic component ($e^\pm$) of nonthermal particles in
jets are the well--studied origin of synchrotron radiation
from radio to X-ray frequencies ({\elevenit e.g.},  see the
conference proceedings edited by Maraschi {\elevenit et al.} 1989 and
Zensus {\elevenit et al.} 1987, Bregman 1990).  Subtracting thermal
emission components like dust infrared radiation (see, {\elevenit e.g.},
Chini {\elevenit et al.} 1989b) and the big blue bump in quasars one
is left with a continuum spectrum of highly varying flux and polarization.
Flat--spectrum radio sources are characterized by jets oriented at small
angles to the line of sight, so that the radiation from the base of the
jet is Doppler--boosted towards the observer. Some of these sources, the
blazars, show very active behaviour of the most compact regions of the jet,
especially with respect to polarization. Probably all flat--spectrum radio
sources appear as blazars during active episodes.  Since a) particle
acceleration takes place in radiojets as is obvious from synchrotron
observations, and b) the direct observations of high energy cosmic rays
(see Gaisser {\elevenit et al.} 1993, and UHE CR II) it is a small step
to argue that this acceleration mechanism not only concerns electrons (and
positrons), but protons (and nuclei) as well.

\bigskip
\line{\elevenit 4.1. The Hadronic Cascade \hfil}
\smallskip

Now, recent gamma ray detections and the radio/X-ray correlation of
extragalactic flat-spectrum radio sources together with the direct
detection of the extragalactic component of cosmic rays (see above)
make the existence of a ultra--relativistic proton population in jets
very probable.  The protons with maximum Lorentzfactors in the range
$10^9-10^{11}$ generate hard photons with energies from keV to TeV via
pion and pair photoproduction and subsequent synchrotron cascade
reprocessing.  The target of the energetic baryons is self--consistently
provided by the synchrotron photon soup from the primary (accelerated)
electrons, while for the secondary pairs cooling always dominates over
acceleration. To simplify things the compact jet with relativistic
electrons and protons giving rise to the radioloud quasar broadband
emission shall be coined the ``proton blazar''.

The simplest approach towards considering protons in addition to electrons
as radiative constituents of radiojets is to make use of the Blandford
\& K\"onigl (1979) model, where the jet is assumed to contain a tangled
magnetic field dominating (in pressure over radiation)
and relativistic particles conically streaming away from
the central engine.  Within this prescription the physical conditions
inside the radiojet plasma can be inferred from continuum observations
at radio through optical frequencies.  It is then straightforward to
calculate secondary emissivities from proton interactions taking into
account that the protons will reach much greater energies than the
electrons, just because they suffer weaker energy losses at the same
energy as electrons.

The pions and pairs created by interactions on the soft target field
produce gamma rays to which the jet is optically thick.  The reason is
that the soft target photons (from the primary accelerated electrons)
are distributed with an inverse powerlaw, so that the higher the gamma ray
energy, the more soft photons satisfy the threshold condition for two
photon pair production.  Thus, a new generation of pairs is produced which
then in turn produces gamma rays again (due to synchrotron losses).
The crucial parameter is the target radiation compactness $l\propto L/r$,
where $L$ denotes the soft luminosity and $r$ the characteristic size
of the emitting region, which can also be calculated from the Blandford
and K\"onigl model yielding typical values $l\approx 10^{-4}-10^{-5}$
for radioloud quasars (extended hot spots have even lower compactnesses).
The energy above which the jet becomes optically
thick with respect to pair creation is then given by

$$E_\gamma^*\approx 10D_{\rm j}l^{-1}\ {\rm MeV}$$

\noindent hence $E_\gamma^*\approx 1-10\ {\rm TeV}$ for typical Doppler
factors of $D_{\rm j}\approx 10$.  This is in accordance with recent GRO
and Whipple (Punch {\elevenit et al.} 1992) observations.
The fact that $l\ll 1$ means that the electromagnetic (synchrotron)
cascade is unsaturated, {\elevenit i.e.} photons at MeV are optically thin.
This results in a spectrum which can be described as the superposition
of only three to four cascade generations
(pair$\rightarrow\gamma$pair$\rightarrow \gamma$...) and the emerging
shape of the spectrum can be quite different depending mainly upon the
ratio $\xi=E_{\rm p,max}/E_\gamma^*$.  When $\xi$ is very great the
spectrum has flux indices $\alpha_X\approx 0.5$ up to the MeV range and
then steepens towards $\alpha_\gamma\approx 1$ until it turns over
steeply in the TeV range (merged cascade generations at gamma ray energies
and the surviving final cascade generation at X-ray energies, {\elevenit cf.}
3C279).  As $\xi$ becomes smaller, the X-/gamma ray spectrum becomes
decomposed into two bumps, one peaking in the MeV range with indices
$\alpha_X\approx 0.5$ and $\alpha_\gamma \approx 1.5$ and the other
peaking in the TeV range with $\alpha_\gamma\approx 0.8$ until it turns over
steeply.  Additional damping of gamma rays may result from strong extended
infrared photons fields as in 3C273.

\bigskip
\line{\elevenit 4.2. Critical Test:  Neutrinos \hfil}
\smallskip

Hadronic interactions lead to considerable neutrino production of
roughly the same luminosity as in the electromagnetic channels.
Using gamma ray observations to normalize neutrino fluxes one
can readily calculate expected count rates in neutrino detectors.
It is crucial to consider that while the gamma ray spectrum reflects
cascading, the neutrino spectrum reflects the initial injection.
Due to the inverse powerlaw nature of the target agent this means
that the neutrino flux density spectrum from $p\, \gamma$ interactions
is flat.
Since the energy loss time scale for $p \, p$ collisions is constant,
in contrast to the $p \, \gamma$ losses described above, the emissivity of
secondaries from $p \, p$ collisions is much steeper, {\elevenit i.e.}
reflects the power law of the protons.
Therefore the neutrinos from $p \, p$ collisions are important at lower
neutrinos energies than the neutrinos from $p \, \gamma$ collisions,
{\elevenit i.e.} below $\approx 10^{-2}E_{\rm \nu,max}$.

The possibility of detecting neutrinos
from flat--spectrum radio quasars with the Fly's Eye
experiment is investigated in Mannheim {\elevenit et al.} (1992).  There
is  a significant chance that the HiRes (Cassiday {\elevenit et al.} 1989)
might be able to detect proton blazars.
The neutrino flux crosses the atmospheric background flux at roughly
$E_\nu\approx 1 \,~{\rm PeV}$ and has a spectrum $F_\nu\propto
E_\nu^{-1}$ below approximately the same energy and is flat above up
to roughly $10^9 \;\rm ~GeV$.

Neutrino detection at lower energies
is feasible with  experimental
designs currently under construction like AMANDA, DUMAND II
or BAIKAL (Stenger {\elevenit et al.} 1991, Halzen 1991) in the TeV
range and by advanced analysis of horizontal atmospheric showers up
to the PeV range (Halzen \& Zas 1992).
Although extragalactic neutrino beams would be a unique tool for
elementary particle physics to study weak interactions in a kinematical
regime far beyond planned laboratory designs, it remains to be shown
wether it is possible to detect the neutrinos from radioloud quasars
below PeV because of the flatness of the flux.

\bigskip
\line{\elevenit 4.3. Further Consequences \hfil}
\smallskip

Hadronic interactions exhibit still another fingerprint:
In every second inelastic proton-photon event the leading nucleon
emerging from the interaction fireball is a neutron. Since neutrons
are not magnetically confined anymore, they can escape the acceleration
region in the jet (Biermann \& Strittmatter 1987).  The optical depth
for neutrons transverse to the jet is less than unity, so that the
ultra--relativistic neutrons can leave the source without turning back
into confined protons (provided that the distance of the jet from the
site of the thermal UV nucleus is great enough to prevent damping by these
photons). Since the frame in which the scattering centers isotropize the
accelerated protons rests in the bulk flow of the jet, the emitted
neutrons as seen from an observer stationary with respect to the host
galaxy are streaming along a cone with the blazar in its apex. As a
corollary it follows that {\elevenit Baryonic blazar beams contain
energetic photons, neutrinos and neutrons with comparable luminosities.}

It remains to be shown, wether the neutrons have observable consequences
({\elevenit cf.} Kirk \& Mastichiadis 1989).

While it has been suggested by Sikora {\elevenit et al.} (1989) and
Begelman {\elevenit et al.} (1991) ({\elevenit cf.}  MacDonald {\elevenit
et al.} [1991] for further consequences of energetic neutrino
production.) that neutrons from a hypothetical accretion disk in an
AGN can explain  observed gas outflows, the mechanism proposed works
best for moderately energetic neutrons.  To convert the power of the
relativistic neutrons into kinetic power of thermal plasma authors assume
effective coupling of the two media via excitation of Alfv\'en--waves by
$\beta$--decay protons.

However, it is by no means clear how the microphysics shall interplay
to lock the particles isotropically to the plasma when their energy
is very high ({\elevenit e.g.}, Berezinsky {\elevenit et al.}
1990). It must be remembered that the luminosity of
neutrons from the jet peaks at the highest neutron energy
($\approx 10^9$~GeV).
Without such isotropic locking there are no adiabatic losses.
Moreover, the distance neutrons travel before suffering
$\beta$--decay is given by

$$R_n=\gamma_nc\Delta\tau_n\approx
\left(\gamma_n\over 10^8\right)\thinspace
{\rm kpc}\eqno\autnum$$

Proton blazars generating most neutron luminosity at
$\gamma_{\rm n}\ge 10^9$ are therefore clearly injectors
of cosmic rays, because the neutrons decay at $R_{\rm n}\ge 10$~kpc
outside the main central galaxy.  Only for $\gamma_{\rm n}< 10^9$ neutrons
decay well within the host galaxy.  The greatest number of these neutrons
come from $p\, p$ collisions in the jet. Their luminosity is
much less than that of neutrons from $p\, \gamma$ collisions.  It must
also be remembered that the great luminosity of the proton blazar
appears only in the direction of the jet.  After isotropization the
true luminosity available for conversion into the kinetic power of
a wind is reduced by the factor $D_{\rm j}^{-4}\approx 10^{-4}$.

Other models based on $p\, p$-interaction in a thick accretion disk
(Bednarek 1993), on electron/positron beams (Henri {\elevenit et al.}
1993), and on synchrotron and inverse Compton processes (Maraschi
{\elevenit et al.} 1992, Dermer \& Schlickeiser 1992)  have been
presented.  Hadronic interactions distinguish themselves most strongly
through the neutrino emission from the more classical electron/positron
arguments; also, the observed high energy cosmic ray particles strongly
suggest that protons and heavier nuclei get abundantly accelerated in
the cosmos (see earlier sections in this review).   The basic hadronic
interaction processes are being rediscussed in some detail by Berezinsky
\& Gazizov (1993a, b).

To summarize our model, what can be learned from adding relativistic protons
to the usual relativistic electrons observed in compact radio jets?

First of all, if the protons manage to be accelerated up to the
high energies where their energy loss rate is equal to the electron
energy loss rate, they generically induce high energy emission from
X--rays to gamma rays.  The spectrum is complex, since it results
from the superposition of several cascade generations.  These various
cascade generations allow a variety of source spectra ({\elevenit e.g.}
3C279 vs. 3C273).  High redshift sources suffer from interaction with a
possible infrared background (Stecker {\elevenit et al.} 1992), and so
their observable photon energy range is limited.

At present it is unknown, wether
there is also a generic proton/electron ratio and hence
a generic ratio $L(>X)/L(<UV)$.   The EGRET detections
of flat--spectrum sources seem to indicate that either there is
significant scatter in the distribution of the proton/electron
ratio or that the
proton acceleration (taking longer than the electron acceleration)
is sometimes interrupted before reaching the maximum proton energies.

Secondly, the presence of protons leads to neutron production which
has remarkable astrophysical consequences.  At ultra--high energies
the neutrons escape from the host galaxy without adiabatic
losses injecting cosmic ray
protons.  At lower energies they can accelerate gas surrounding
the radio jet within the host galaxy (and its halo).

Finally, with protons brought into the game, the flat--spectrum
radio sources generate a {\elevenit diffuse background of high energy
neutrinos} which is detectable with planned experiments.

Important next steps are here first to try to
understand the rapid time variability observed in such sources (Qian
{\elevenit et al.} 1991, Quirrenbach {\elevenit et al.} 1991, 1992),
and second to calculate the induced $\gamma$-ray background, both due
to the interaction of cosmic ray particles interacting with the microwave
background, as well as the superposition of all flat radio spectrum quasars.

\vglue 0.6cm
\line{\elevenbf 5. The Galactic Center and Radioweak Quasars \hfil}
\vglue 0.4cm

\bigskip
\line{\elevenit 5.1. The Galactic Center \hfil}
\smallskip

The recent radio observation of the Galactic center source Sgr A* with
mm-VLBI by Krichbaum {\elevenit et al.} (1993) for the first time
resolved the source spatially; these observations suggest an elongated
one-sided morphology quite similar to the jet observed in radio quasars.
Fitting the flux density and the radio spectrum of this central source
with the simple classical model for extragalactic radio sources
(Blandford \& K\"onigl 1979), we can deduce limits for the important
ratio of jet power to disk accretion luminosity (Falcke  {\elevenit
et al.} 1992, 1993).  It turns out that the data lead to a lower limit
for this ratio of order $0.1$.  This suggests that even for such a weak
source akin to an AGN, the jet power is of similar order of magnitude
as the observed accretion disk luminosity.  As an important next step
it will be interesting to see whether this conclusion holds up for
radioweak quasars in general; below we will speculatively assume that
it does.

The feeding of the central accretion disk close to the black hole from
larger scales further out, out to about $100$ pc can be thought of as
an overall nonsteady accretion flow ({\elevenit cf.} Weizs\"acker
1943, 1951, L\"ust 1952) with a substantial effective viscosity in
the interstellar gas (Linden {\elevenit et al.} 1993, Biermann
{\elevenit et al.} 1993).  The accretion rate for the central disk is
very low (Falcke {\elevenit et al.} 1993), while on the larger scale
the cloud motions can be traced out very well with such an accretion
disk model, which has avery much higher accretion rate, more in
correspondence with the star formation rate in that area.

If there is a powerful jet, possibly emanating from the inner edge of the
accretion disk, then nuclei might be accelerated to high particle energies in
shocks at the boundary or inside.  The effect of energetic particles on the
accretion disk again, and also on molecular clouds will be considered below.

\bigskip
\line{\elevenit 5.2. Radioweak Quasars \hfil}
\smallskip

\bigskip
\line{\elevenit 5.2.1. The Hadronic Cascade in the Accretion Disk \hfil}
\smallskip

The active nuclei of galaxies (AGN), ranging in luminosity
from Seyfert galaxies to quasars, are the most powerful individual
sources of radiation in the Universe. To explain the power emitted by
such objects, one generally assumes the existence of a central engine
in which the gravitational energy of matter falling into a supermassive
black hole gets converted into radiation.  Even though so far only
electromagnetic radiation has been observed, it is generally assumed
that other particles are accelerated and emitted as well to explain
the tight relationship between the non-thermal and thermal components
in the IR, UV and X-ray spectra (Lightman \& White 1988, Pounds
{\elevenit et al.} 1989, Fabian {\elevenit et al.} 1990, Clavel
{\elevenit et al.} 1992, Chini {\elevenit et al.} 1989a). Of special
interest are neutrinos, since they can travel cosmological distances
without losing the information on the direction they originated from.
Large underwater detectors (Alatin {\elevenit et al.} 1991, Stenger 1991)
or detectors in the antarctic ice cap (Barwick {\elevenit et al.} 1991),
are used as neutrino telescopes. Recent calculations have shown that the
flux of neutrinos originating in an AGN could be detected by such
experiments (Stecker {\elevenit et al.} 1991, 1992a, 1992b, Szabo and
Protheroe 1992,  Mannheim {\elevenit et al.} 1992, Biermann 1992,
Mannheim 1992).  Data from proton decay experiments (Meyer 1991) and
airshower arrays Halzen \& Zas (1992) is already sensitive enough to
constrain such models significantly.

In the following we are interested in the production of neutrinos in
radio-weak AGN. In the ``standard'' model for AGN one assumes
that the central black hole is surrounded by an accretion disk of
infalling matter. Besides that, one expects to find bipolar outflow of
gas and plasma perpendicular to the disk (jet). Jets can be seen in
different objects with disk accretion and in many
AGN (Bridle \& Perley 1984, Meisenheimer \& R\"oser 1991).
One expects shocks in the plasma of the jets (Mannheim 1992) which could
accelerate protons through first order Fermi acceleration to energies
ranging from~$10^6\,{\rm GeV}$ to~$10^9\,{\rm GeV}$,
depending on the distance to the black hole, with a powerlaw spectrum of
$E^{-2}$ (Biermann \& Strittmatter 1987).  The observation of
$\gamma$-rays with the same kind of spectrum (Punch {\elevenit et al.}
1992) indicates that indeed there is shock-acceleration outside the core
of the AGN, so that the photons can escape.

Niemeyer (1991) showed that the far infrared (FIR)
emission of AGN can be explained by {\elevenit assuming} that accelerated
particles diffuse back from a central location above the accretion disk
and heat dust clouds beyond the outer region of the accretion disk (see
the next subsection). In the inner part of the disk, protons hitting the
disk will initiate hadronic cascades through interactions with the gas
in the disk.  This could feed into an electromagnetic cascade and thereby
generate the observed X-ray and gamma emission (Biermann 1992).  Clearly,
this is a {\elevenit speculative} picture at present, but it is suggested
to have merit on the basis of the analoguous situation in the center of our
Galaxy, discussed briefly above.

To describe the accretion disk, we use the model by Shakura \&
Sunyaev (1973).  For our purposes, we will concentrate on the inner region,
which is the radiation dominated part of the disk (neglecting the thin,
innermost ring around the black hole).

Due to the observed short timescales of the variability of the X-ray and
UV~components of the AGN spectrum, this part of the electromagnetic
radiation must predominantly originate in a small, central region
around the central black hole (Stecker {\elevenit et al.} 1991, 1992a,
b) with a  typical mass of $10^8 M_\odot$ for luminous AGN.  We expect
neutrino production to take place in the same region, so we need to know
the incoming particle flux in this region. From Niemeyer \& Biermann
(1993 and below) we know that the FIR spectra are very well fitted by
assuming that the protons originate from a source in the jet extending
outward from $z_0 = O(3\cdot10^{15}\,{\rm cm})\approx 100 R_S$.

For a source that close to the center of the disk, the discussion of
the maximum proton energy in (Biermann \& Strittmatter 1987) has to
be extended to take the UV photons from the disk into account.  The
photon spectrum has a pronounced bump in the UV region (Stecker
{\elevenit et al.} 1991, 1992a, b, Clavel {\elevenit et al.} 1992).
Therefore, the reaction $p\gamma\to  N\pi$ is only effective for protons
with energies above the threshold $E_{\rm th}\approx 8\cdot10^6\,
{\rm GeV}$ for the interaction with UV photons. This interaction channel
limits the maximum proton energy.

If we take $\dot M/M_{\rm edd} \approx 0.1$ and
$\alpha \approx 0.1$, the column density perpendicular through the disk
is high enough compared to the mean free path for $p\, p$- and $n\,
p$-collisions of $O(50 \,{\rm g}/\,{\rm cm}^2)$ for a hadronic cascade
to develop. The magnetic field confines the protons to the disk, which
leads to a further increase of the effective column density seen by the
protons. Even though neutrons which are produced in the cascade are
not confined by the magnetic field, the amount of gas present is
sufficient to prevent them from escaping; they interact before they
can decay.

The hadronic cascade in the disk is much simpler than cascades in the
earth's atmosphere (Gaisser 1990, Gaisser \& Yodh 1980, Gaisser
{\elevenit et al.} 1978), since the density of the accretion disk is
so low that all unstable particles decay rather than interact. Therefore
the cascade consists of a nucleonic part which feeds into the mesonic and
electro-magnetic channels; no pion-nucleon reactions occur.

The electromagnetic output of the hadronic cascade is reprocessed in
the inner disk. Pair cascades, inverse Compton scattering, and
reflection on cold material change the shape of the initial
$E^{-2}$-spectrum and provide a steep turnover around
$E_\gamma \approx m_e \approx 511\,{\rm keV}$ (Zdziarski \& Coppi 1991,
Zdziarski {\elevenit et al.} 1990, Ghisellini 1987, Svensson 1987,
Zdziarski  {\elevenit et al.} 1991, Done {\elevenit et al.} 1990).
This may produce the observed X-ray and gamma emission of the AGN which
we will calculate in detail elsewhere.

The neutrino spectrum from a single source mirrors the
$E^{-2}$-spectrum of the protons, only that it is shifted down by a
factor of $\approx 0.05$. The upper cutoff of the
neutrino spectrum depends on the details of the shock acceleration
process in the jet, since that determines the maximum energy reached by
the protons (Biermann 1992).

To be able to use the relation between electromagnetic output and
neutrino production to predict the diffuse neutrino background, we
have to estimate how much of the observed, diffuse X-ray and $\gamma$-ray
background results from hadronic cascades in AGN.

There is growing evidence from the ROSAT all sky survey that an
appreciable fraction of the background is due to active galactic
nuclei, possibly a dominant proportion (Hasinger {\elevenit et al.} 1991).

To be conservative, we allow for a factor of three maximum
between the total hard X-ray emission from an AGN galaxy (starburst
and active nucleus together) and the contribution strictly from the
hadronic cascade.

The observations of the hard X-ray background at those photon energies
minimally influenced by reprocessing (Fabian {\elevenit et al.} 1990),
{\elevenit i.e.}, at energies where the original $E^{-2}$~powerlaw is
still visible, give a possible range for the unreprocessed energy
total of~$1.0 \cdot 10^5$ to~$1.4 \cdot 10^5 \,{\rm eV} \,{\rm cm}^{-2}
\,{\rm s}^{-1} \,{\rm sr}^{-1}$. Using the above mentioned conservative
estimate, we arrive at a lower limit of~$3 \cdot 10^4  \,{\rm eV} \,
{\rm cm}^{-2} \,{\rm s}^{-1}\,{\rm sr}^{-1}$
for the contribution of hadronic cascades
to the X-ray and $\gamma$-ray background.

Since the cosmological redshift is the same for
neutrinos and photons, we can scale the background neutrino luminosity
using the fraction of the X-ray and $\gamma$-ray background
derived above. We get

$$  N(E_\nu) = 1.7 \cdot 10^{-12} \left(E_\nu \over \,{\rm TeV}\right)^{-2}
               \,{\rm cm}^{-2} \,{\rm s}^{-1} \,{\rm sr}^{-1} \,
{\rm GeV}^{-1} \eqno\autnum$$

\noindent as a conservative limit for the sum of all neutrino species.
About~$2/3$ are muon neutrinos, the remaining~$1/3$ are electron neutrinos.
This prediction is a factor of $2.5$ lower than the experimental limit
set by the Frejus experiment (Meyer 1991).  For each family, the
number splits evenly into neutrinos and anti-neutrinos.
We expect the background spectrum to have a softer cutoff at high
energies than a single source,
since the maximum neutrino energy and the redshift vary between different
AGN. Therefore, the cutoff in the superposition of all spectra
will be smoothed out.

The main point in our model is the production of neutrinos as the
result of a hadronic cascade initiated by $p\, p$-interactions. Compared
to the $p\, \gamma$-channel, which has a threshold of
{$E_{\rm th}\approx 8\cdot10^6\,{\rm GeV}$} for photoproduction
on UV~photons, all protons above a few hundred~$\,{\rm MeV}$
contribute in $p\, p$-interactions. As a consequence, we expect neutrino
emission even from sources with a low cutoff in the primary proton spectrum.
This way, we include contributions from a larger class of sources both
to the neutrino background and to the diffuse X-ray background.
Similar to the result of other authors (Stecker {\elevenit et al.}
1992a, b,  Szabo \& Protheroe 1992) , our model displays the tendency
to produce neutrinos strikingly close to existing limits (Meyer 1991,
Halzen \& Zas 1992).

Modifications of this $p\, p$-model can lead to a reduced neutrino flux
while reprocessing of the electromagnetic component ensures an
unchanged X-ray emission. The most drastic modification --- reducing
the maximum proton energy and correspondingly the maximum neutrino
energy --- can ultimately decrease the observable extragalactic
neutrino flux by moving the cutoff below the cross-over with the
steeper spectrum of atmospheric neutrinos. More realistically,
steepening of the AGN proton distribution at a sufficiently low break
energy is an alternative to reduce the predicted event rate in a
neutrino detector. In both cases, neutrino production via
$p\gamma$-reactions becomes ineffective.

On the other hand, a proton spectrum flatter than~$E^{-2}$ leads to
the dominance of $p\, \gamma$-reactions at high energies. This, again
keeping the X-ray background unchanged, enhances the neutrino flux far
above the $\,{\rm TeV}$ range while reducing the flux below. Details
depend on the modelling of the target photon field ({\elevenit i.e.},
contributions from the disk, corona, jet, {\elevenit etc.}).

A likely consequence is the effect of the energetic particles (see,
{\elevenit e.g.} also, Ferland \& Mushotzky 1984, Kazanas \& Ellison
1986) and photons on stars in their neighborhood of the central engine,
{\elevenit i.e.} the rotating central black hole, the inner accretion
disk and the inner jet.  The stellar distribution is modified by
disruption of those stars that get too close to the central black hole,
by stellar collisions, and possibly also by agglomeration and
interaction with the disk.  Stellar atmospheres exposed to a strong
flux of energetic particles may expand and thus start a stellar wind,
which in turn gets ionized by the ultraviolet radiation field, a
process which can be described as an HII-region turned inside out
(Scoville \& Norman 1988, Kazanas 1989); a similar effect can occur
when energetic neutrinos penetrate the stars and change their structure
from the inside, bloating the stars and thus moving lower mass main
sequence stars into the red giant region (MacDonald {\elevenit et al.}
1991).  An additional process which may confuse the observations is
the likehood of star formation in the dense molecular clouds which are
observed to exist rather close to the central engine, {\elevenit e.g.}
by absorption (Lawrence \& Elvis 1982, Mushotzky 1982, Antonucci and
Miller 1985, Barthel 1989) or by maser emission ({\elevenit e.g.}
Claussen \& Lo 1986)).  The broad emission line region may thus be the
ensemble of slow stellar winds embedded in a strong ultraviolet radiation
field.

An important point in our speculative model is that the acceleration of
the protons and other nuclei takes place above the disk.
This way, the acceleration takes place in an environment more suitable
than an accretion disk.

\bigskip
\line{\elevenit 5.2.2. The farinfrared emission of radioweak quasars
\hfil}
\smallskip

The FIR emission from active galactic nuclei was believed for a long
time to be due to nonthermal processes ({\elevenit e.g.} Edelson 1986).
The observations of a few nearby Seyfert galaxies (Hildebrandt
{\elevenit et al.} 1977, Telesco \& Harper 1980, Rieke \& Low 1975,
Roche {\elevenit et al.}  1984, Neugebauer {\elevenit et al.} 1979,
Miley {\elevenit et al.} 1984)  gave the first hint, that dust emission
might be important as an additional source.  Then, the very sensitive
mm-observations at the 30m IRAM telescope (Chini {\elevenit et al.}
1989a, b) combined with the IRAS quasar survey (Neugebauer {\elevenit
et al.} 1986) turned the trend; the dust emission from radioweak quasars
is now well accepted to be generally due to warm dust (Sanders
{\elevenit et al.} 1989, A. Lawrence {\elevenit et al.} 1991,
Barvainis 1990).

There are theoretical possibilities to reproduce many, but not all of
the observed spectra with synchrotron emission due to two populations of
relativistic electrons via self-absorption and emission (de Kool
{\elevenit et al.} 1989, Schlickeiser {\elevenit et al.} 1991).  However,
the  observations of very large amounts of molecular gas in radioweak
quasars (Barvainis {\elevenit et al.} 1989, Alloin {\elevenit et al.}
1992) demonstrated that the radioweak quasars contain certainly
sufficient gas and dust mass to produce all the emission observed in
the FIR. Hence, our understanding of radioweak quasars now includes an
extended region of dust clouds, presumably in a disk configuration.
It is less clear what the energy source of this dust is.  The difficulty
lies in the fact that the FIR emission of radioweak quasars is the
strongest of any of the observed wavelength bands usually, and
therefore any primary source must be even stronger, allowing for some
waste and inefficieny.  Another difficulty is, that we can infer from
Plancks law that the radial scale from which the farinfrared emission
is arising, ranges over a fairly large radial scale out to distances of
order $100 \; \rm pc$.

A natural hypothesis, which is actually supported
by the discovery of large amounts of molecular gas, is that a
circumnuclear starburst also provides the FIR emission seen.  It
appears as a very reasonable hypothesis that stars are being formed
in these molecular clouds surrounding the central engine; a number of
nearby Seyfert galaxies and  galaxies with extreme FIR luminosities
clearly demonstrate the activity of such starbursts (a now well
established example is the Seyfert galaxy NGC1068, Wilson {\elevenit
et al.} 1992). However, starbursts have a very clear correlation
between their FIR and radio emission, and also have typical radio
spectra:  Normal galaxies have nonthermal radio emission spectra with
flux density $S_{\nu}\,\sim\,\nu^{-0.8 \pm 0.1}$, sometimes modified
by some thermal free-free emission, or by free-free absorption in a
clumpy interstellar medium, as in the famous starburst galaxy M82
(Kronberg {\elevenit et al.} 1981, 1985).  The ratio of FIR to radio
emission is much larger in many cases for radioweak quasars than for
starburst galaxies, clearly requiring an additional energy source (Chini
{\elevenit et al.} 1989a, 1989b).  The task, however, remains and is an
important next step, whether extremely intense regions of star
formation also suppress the radio emission relative to the farinfrared
emission; resolution of this question may come from infrared line
emission observations.  Finally, in some cases the spatial structure of
the radioemission has been determined, and is found to be analoguous
to radio jets of low power ({\elevenit e.g.}, Lacy {\elevenit et al.}
1992), again inconsistent with starbursts as the source of the dominant
radioemission. Other possibilities are discussed in Niemeyer \&
Biermann (1993), like the X-ray heating model of Sanders {\elevenit
et al.} (1989); in such a model rather extreme warping of the layer of
dust clouds is required to geometrically allow irradiation of all the
dense gas.  Here we introduce a hypothesis that circumvents geometric
difficulties, but uses a putative jet and its particle acceleration:

We propose that the dusty clouds are heated by
relativistic particles. These particles are thought to originate
from a source on the axis of rotational symmetry, possibly a jet
or knots in a jet and its interaction regions with the environment.
The difference between radioloud and radioweak quasars in our picture
is not that the jets in radioweak quasars have a low kinetic luminosity
(Lacy {\elevenit et al.} 1992), but that their radio  emission is very
much weaker, despite their large power.  This is analoguous to the weak
nonthermal radio emission of Wolf Rayet stars as compared to the large
nonthermal radioemission of radiosupernovae, which differs probably only
in the velocity of the shockwave which causes the particle injection and
acceleration (see Biermann \& Cassinelli 1993).  Thus, in our speculative
picture radioweak and radioloud quasars both have jets of strong power,
but in one case the jet power is dissipated fairly deep inside the host
galaxy, and in the other case the dissipation is very far outside the
host galaxy.  The source is assumed to be located above the disk and to
transform a fair fraction of the entire source accretion power into
relativistic particles, analogous to the argument made above for the
Galactic center source. Such a concept is consistent with the model for
strong jets discussed in the literature.  The particles easily scatter
in the magnetic plasma that permeates the halo  above the disk all the
way from the central engine.  And since dust clouds of the column
densities typical for central regions of galaxies readily provide sinks
for energetic particles, they can be heated.

We calculated (Niemeyer \& Biermann 1993) the FIR spectra that can
be produced in this model, and compare them to observations.  The
basic model parameter is the radial behaviour of the diffusion
coefficient that governs the transport of the energetic particles from the
putative central source out to the dust clouds in the disk.  The essential
test is then whether any reasonable model for this diffusion gives a
physical explanation for the observed spectra.

Independent of any particular model for the diffusion
coefficient, we assume here for the diffusion coefficient a general
powerlaw dependence on spherical radius $r$.  It can be shown that the
spectrum itself is independent of the numerical value of the diffusion
coefficient, since we have assumed a stationary state (see Niemeyer \&
Biermann 1993).  However, for these basic assumptions to be justified,
that lead to a stationary state, the diffusion coefficient has to be
quite large.
We consider also the possibility that the source is extended along the
symmetry axis $z$; this case can be considered as a limiting addition of
many point sources.  The physical concept is given by many knots in a jet.

Dust particles are of mixed chemical composition with different grain
size distribution.  Mathis {\elevenit et al.} (1977)
developed a dust model for the composition, absorption and emission of
interstellar dust.  They succeeded to approximate the dust grain size
distribution with a powerlaw.  This powerlaw can be
interpreted as a quasistationary fragmentation (Biermann \& Harwit 1980).

We do not discuss in detail here how the energy is
transformed from an impinging flux of energetic particles into an
effective gas and dust heating.  There are two channels which are
clearly important.

a) The $p \, p$ or nucleus - nucleus collisions which
transform particle energy into pions, which in
turn decay and finally deposit their energy in electrons and positrons
as well as $\gamma$-photons of a range of energies.  These
electrons/positrons and $\gamma$s in turn thermalize their energy by
further encounters with atoms and their shells (see, {\elevenit e.g.},
Spitzer \& Tomasko 1968).  In this picture about $1/3$ of the primary proton
and nuclei energy density is dissipated, the rest goes into neutrinos and
energetic photons.

b)  The second process is the
excitation of Alfv\'en waves by a gradient of the particle
distribution, which is necessarily formed when many particles are
absorbed in $p \, p$ and corresponding collisions (Skilling 1975a,
1975b, 1975c, Skilling \& Strong 1976).  This wave excitation in turn
can lead to a very strong wavefield, which dissipates readily in a
neutral-ion plasma like inside dense dust clouds; it would dissipate
considerably less in a purely ionized medium, such as the innermost
accretion disk.  In molecular clouds of a low degree of ionization this
diminishes the resulting $\gamma$-emission due to an appreciable optical
thickness in the hadronic encounters.  An important next step is to work
out the consequences for the molecular clouds in the Galactic center,
where there are unresolved questions regarding the flux of cosmic rays
and their confinement in the galactic magnetic field ({\elevenit e.g.}
Chi \& Wolfendale 1993), the heating of clouds and the amount of
molecular gas.  In this case, a large proportion of the primary energy
density of the protons and nuclei can be dissipated.

Here we simply assume that a fixed fraction of the energy of the
impinging particles is deposited as heat. We determine the dust
temperature by assuming that emission and absorption balance.

Since the incoming flux of energy depends on radius in the disk geometry
considered, the dust temperature is also dependent on radius $r$.
We calculate the FIR spectrum of a finite disk in the wavelength range
$10$ und $1300 \,{\rm\mu m}$. The most important characteristica to be
interpreted are the rapid rise from the mm-wavelengths to the FIR
$L_{\nu}\sim \nu^\alpha$ with $\alpha \ge 2.5$, and, on the other
hand, the slow decrease with frequency beyond the farinfared with a
spectral index $\alpha\simeq -1$.

All cases considered of the radial behaviour of the diffusion
coefficient show the steep increase from the mm-range towards the
FIR, as required by the observations. The emission through the IR
towards the NIR decreases very slowly. The general trend of the
calculated spectra show a good representation of what is seen in the
observations.  In the mm to FIR wavelength range the local spectral
index $\alpha$ is always $\ge 2.5$, independent of the specific
dependence of the diffusion coefficient on radius.  On the other hand,
the spectra in the IR to NIR range are indeed strongly dependent on
the source distribution function and on the specific law for the radial
dependence of the diffusion coefficient chosen. The decline through
the IR gets stronger with an increasing radial dependence of the
diffusion coefficient.

It is important to note that there is no minimum distance the source
has to have. This is because with decreasing minimum distance more
particles are deposited in the inner part of the disk, but the radial
dependence of the temperature remains the same in the outer part of
the disk.  The maximal IR--luminosity decreases with decreasing minimal
distance $z_{\rm min}$. The relation can be described by a logarithm
scale factor: If the minimal distance $z_{\rm min}$ decreases by a
factor $X$, the IR--luminosity decreases by a factor $Y=\ln X$ for a
given source power.

The calculations show that for sufficiently high values of the column
density the optical depth at high infrared frequencies is significantly
larger than unity. Hence the limiting case of high optical depth
describes the spectrum, and the numerical value of the optical depth
itself drops out. Correspondingly we treat the case of optical depth
much smaller than unity at low frequencies for all values of column
densities considered.  In this case the actual value of the optical
depth does matter, since the emission is directly proportional to it.
Specifically, the dependence of the absorption cross section on
frequency can be approximated in the FIR/mm-range by a power law.  For
extremely low values of the column density the optical
depth becomes smaller than unity at all frequencies, and the local
maxima and minima of the absorption cross section then determine the
shape of the emission spectrum directly and produce the various maxima
and minima that we described.  Should some observed spectra show such
features, then we could derive estimates for the column density.

Nearly all observed spectra can be fitted with a line source with an
intensity $z^{-1}$.  Therefore a basic model is a line source, with an
intensity $z^{-1}$ starting at some indetermined
distance above the the center of symmetry - as long as this distance
is not larger than of order $0.1$ pc - and a diffusion coefficient
which is $D \sim r$ or $\sim r^2$.

The source energy goes about equally into the central region, where the
disk is too hot for dust, and the outer region, where infrared emission is
caused.  Of the fraction of energy which goes into the outer region, we
have used $1/3$ to heat the dust and dense gas specifically, but, depending
on the microphysics assumed (see the discussion above), this fraction
could approach unity.

A line source with source strength proportional to $z^{-1}$ can be
understood as due to the particle population energy density $\sim \,
z^{-2}$, a lateral gradient scale $\sim \, z$, a transport coefficient
$\sim \, z$ also, and a circumference $\sim z$; again we use here a
scaling argument for the transport coefficient.

To summarize we find that for reasonable parameters we can indeed
reproduce the observed farinfrared spectra of radioweak quasars.
We find that for a diffusion coefficient in the region above the disk,
which scales with radius $r$, and a line source strength $\sim \,
z^{-1}$, with a total source  luminosity of $6$ to $10$ times the
observed infrared luminosity, we can reproduce and interpret the
spectra of radioweak quasars from the mm to  the near infrared region.

There are obvious consequences of our model for the
X-ray, gamma- and neutrino emission from the inner disk, as well as
likely consequences for the heating of interstellar clouds close to a
central engine (by wave excitation and dissipation), which remain to
be worked out.  The analogy to the Galactic center also suggests that the
accretion disk luminosity crudely scales with the jet power, and with the
FIR emission, but maybe larger by a factor in the range 1 to 10.
Such correlations will provide a severe test for the proposed speculative
model.

\vglue 0.6cm
\line{\elevenbf 6. Conclusions and Outlook \hfil}
\vglue 0.4cm

Energetic particles can have an observable effect in many astronomical
objects, from heating and ionization of interstellar material to TeV
$\gamma$-rays from Active Galactic Nuclei (AGN).  Key to all such
arguments and an important test for our physical insight is the
acceleration of the observed cosmic ray particles, their energy range,
spectrum and chemical composition.  There has been recent progress in
this area and in this review we have described some of it.

\bigskip
\line{\elevenit 6.1. Galactic Cosmic Rays \hfil}
\smallskip

We propose a fully developed physical concept to account for
the cosmic ray particle energies, spectrum and chemical abundances.  We
suggest, that the cosmic rays observed near earth arise from three sites:

1)  Supernova explosions into the nearly homogeneous interstellar medium;
these cosmic rays reach particle energies of about $100 \,Z \; \rm TeV$,
and have an injection spectrum of near $E^{-2.4}$, where $Z$ is the
charge of the nucleus considered.

2)  Supernova explosions into a former stellar wind produce a slightly
flatter spectrum of near $E^{-2.3}$ at injection below a bend near
$700 \, Z \;\rm TeV$, beyond which the spectrum steepens to near
$E^{-2.7}$ up to a maximum energy of about $100 \, Z \;\rm PeV$.
This means that heavy nuclei dominate at large particle energies.
Since here the acceleration is in a shock that runs through the wind of
an evolved star, the heavy elements are enriched.   The heavy
element contribution may even be important at low energies.  Air shower
data have been successfully fitted with this model in the energy range
from 10 TeV to near EeV.

3)  The hot spots of powerful radio galaxies provide the third site
of origin.  The predictions from this model prediction have also been
fitted successfully to airshower data from 0.1 EeV to 100 EeV.

We emphasize that the proposal rests on a plausible but nevertheless
speculative assumption about the nature of the transport of energetic
particles in perpendicular shock waves, namely that there are large fast
convective motions across the shock interface; this notion is, however,
supported by radio polarization observations. The model has predictive
power.  The predictions for both the galactic and the extragalactic
cosmic rays underwent their most severe tests with the airshower data
quite successfully, with all the parameters fitted rather close to their
predicted values, and with little "wiggle room".

The leakage of low energy cosmic rays from normal and starburst galaxies
at high redshift may ionize and heat the intergalactic medium, meeting
all known constraints.

\bigskip
\line{\elevenit 6.2. Extragalactic Cosmic Rays \hfil}
\smallskip

The third site of cosmic ray production is the hot spots of powerful radio
galaxies.  The full tests against the new airshower data from Fly's Eye have
been made, demonstrating that this concept successfully explains particle
energy, spectrum and chemical composition in the particle energy range from
01. EeV to 100 EeV.  One consequence is that the mean free path for high
energy cosmic rays in the intergalactic medium has to be of order the
bubble size inferred from galaxy distribution studies, and thus of the
same order of magnitude as the distances between powerful radiogalaxies.

\bigskip
\line{\elevenit 6.3. Radioloud quasars \hfil}
\smallskip

The new GRO observations of such flat spectrum radio quasars demonstrate,
that they have higher luminosities in the $\gamma$-ray range than at any
other wavelength, with one object even showing emission near TeV photon
energies.

We review the hadronic interaction picture to explain these
observations, tracing the $\gamma$-ray emission to the consequences of
hadronic interactions of nuclei accelerated in shocks in relativistic
jets.  Our model is capable of explaining a variety of hard X-ray to
$\gamma$-ray spectra such as observed by Comptel and Egret.  The
spectra deduced have some interesting analogies to the spectra of
$\gamma$-ray bursts, reviewed in chapter 3 of this book.

\bigskip
\line{\elevenit 6.4. The Galactic Center and Radioweak quasars \hfil}
\smallskip

The Galactic center has been demonstrated by recent mm-VLBI observations
to contain an elongated morphology in its compact radio emission
suggestively similar to jets in radioloud quasars; in simple models
this jet can be shown to carry nearly as much power as is in the
accretion disk luminosity.   For radioweak quasars it has only
recently become clear (Chini  {\elevenit et al.} 1989a) that their
farinfrared emission is due to dust,  and is thus thermal radiation.
The origin of this energy is not clear;  several possibilities exist,
and one of these possibilities is the hypothesis, that the active
nucleus provides a large proportion of its power in the form of energetic
particles - arising from shocks in a putative jet analoguous to our
interpretation of the Galactic center - which in turn heat the molecular
clouds which contain the dust.  Various consequences of this hypothesis
have been described, including  some consequences for the emission from
the inner accretion disk (X-rays and neutrinos) as well as the farinfrared
emission spectra of quasars.

\bigskip
\line{\elevenit 6.5. The 12 next steps \hfil}
\smallskip

The comprehensive physical concept proposed here for AGN and Galactic
sites of origin of cosmic rays stands and its essential parameters are
fixed.  There remain many important steps to be done and so the proposal
can be understood as a program; we will list here those steps that seem
feasible within the near future:

\item{1.} The secondary to primary ratio of the nuclei in the cosmic
rays and their relation to the structure and evolution of molecular
clouds.
\item{2.} A simulation of the selection effects governing our
observations of normal Sedov-type supernova remnants.
\item{3.} A comparative study of different acceleration sites for
energetic electrons, like supernova remnants, radiosupernovae, novae,
stars in order to induce a physical concept for the critical velocity of
electron injection.  Obviously, the injection of protons and heavier
nuclei ought to be studied in parallel.
\item{4.} A check on why the knee feature of the
cosmic ray spectrum is so sharp.  This sharpness suggests either a) a
complete convergence of the properties of massive stars, including
rotation and magnetic fields, or b) a speculative picture like the one
suggested in paper CR I that traces the origin of the mechanical energy of
supernovae to the potential energy at that depth where the innermost
section of the star forms an accretion disk for a few seconds, before
magnetic fields transport its angular momentum away and most of the disk
collapses further.
\item{5.} The abundances of the chemical elements
in the cosmic rays at low energy.
\item{6.} A study of the supernova remnant candidates in the starburst
galaxy M82.
\item{7.} The effect of low energy cosmic rays from normal and
starburst galaxies at high redshift on Lyman $\alpha$-clouds.
\item{8.} A calculation of the $\gamma$-ray background due to
cosmic ray interaction.
\item{9.} The rapid variability of blazars, as inferred from the same shock
acceleration and hadronic interaction concept, which successfully lets us
understand the $\gamma$-ray emission, its variability and spectrum.
\item{10.} The connection betwen radioweak quasars and our Galactic center
in structure and energetic photon emission.
\item{11.} The contribution to the farinfrared emission of radioweak
quasars from the circumnuclear starburst.
\item{12.} The heating of giant molecular clouds in our Galactic center,
their $\gamma$-ray emission and the cosmic ray flux near to the Galactic
center.

Obviously, the most important task of all is to check the notion of the fast
convective parallel transport at shocks, both observationally and
computationally.

\vglue 0.6cm
\line{\elevenbf Acknowledgements \hfil}
\vglue 0.4cm

I wish to thank all my present and former graduate students as well my
postdocs who have contributed a large part of the work reported here: M.
Diewald, H. Falcke, S.v. Linden, M. Niemeyer, A. Poll, J. Rachen, H.
Scherer, and Drs. W. Kr\"ulls, K. Mannheim, and T. Schmutzler as well as
Drs. B. Nath, L. Nellen, and R. Schaaf.  The author also wishes to thank
his recent collaborators Drs. J.P. Cassinelli, T.K. Gaisser, J.R. Jokipii,
T. Stanev, and R.G. Strom for their scientific partnership, extensive
discussions of cosmic ray, supernova and stellar physics as well as
generous hospitality at their institutes.  An  essential part of this
work was carried out during a five-month sabbatical in 1991 of the author
at Steward Observatory at the University of Arizona, Tucson, Arizona, USA.
The author wishes to thank Steward Observatory, its director, Dr. P.A.
Strittmatter, and all the Tucson colleagues for their generous hospitality
during this time and during many other visits. Final corrections of this
manuscript were done during a stay in the spring of 1993 at the Weizmann
Institute of Science, Rehovot, Israel; the author wishes to thank Dr. M.
Milgrom and all Israeli colleagues for the generous hospitality extended
there to his family and himself.  High Energy Physics with the author has
been and still is supported by several grants from the Deutsche
Forschungsgemeinschaft (DFG Bi 191/6,7,9), the Bundesministerium f\"ur
Forschung und Technologie (DARA FKZ 50 OR 9202) and a NATO travel grant
(CRG 9100072).

\vglue
0.6cm \line{\elevenbf References \hfil}
\vglue 0.4cm
\medskip

\item{1.}  D.C. Abbott, J.H. Bieging, E. Churchwell, A.V. Torres,
     {\elevenit Astrophys.J}. {\elevenbf 303} (1986) 239.

\item{2.}  S.~D. Alatin {\elevenit et~al.}, Preprint (Institute for
     High Energy Physics, Zeuthen) (1991) PHE 91-17.

\item{3.} H. Alfv\'en,
{\elevenit Phys. Rev. 2nd ser.} {\elevenbf 55} (1939) 425.

\item{4.}  D. Alloin, R. Barvainis, M.A. Gordon, R.R.J. Antonucci,
     {\elevenit Astron. \& Astroph.} {\elevenbf 265} (1992) 429.

\item{5.}  W. Altenhoff, P.G. Mezger, H. Wendker, G. Westerhout,
     {\elevenit Publ.Obs. Bonn} {\elevenbf No.59} (1960) 48.

\item{6.} P.R. Amnuel, O.Kh. Guseinov, F.K. Kasumov,
{\elevenit Sov. Astr.A.J.} {\elevenbf 16} (1973) 932
({\elevenit Astr.Zh.} {\elevenbf 49} (1972) 1139).

\item{7.} R.R.J. Antonucci, J.S. Miller,
{\elevenit Astrophys.J.} {\elevenbf 297} (1985) 621.

\item{8.} Asakimori {\elevenit et~al.}, in Proceedings of 22nd ICRC,
(Dublin Institute for advanced studies), Dublin, Ireland, (1991a) vol. 2, p.57.

\item{9.} Asakimori {\elevenit et~al.}, in Proceedings of 22nd ICRC,
(Dublin Institute for advanced studies), Dublin, Ireland, (1991b) vol. 2, p.97.

\item{10.} W. Baade, F. Zwicky,
{\elevenit Proc. Nat. Acad. Science} {\elevenbf 20} (1934) 259.

\item{11.} P.D. Barthel,
{\elevenit Astrophys.J.} {\elevenbf 336} (1989) 606.

\item{12.} R. Barvainis,
{\elevenit Astrophys.J.} {\elevenbf 353} (1990) 419.

\item{13.} R. Barvainis, D. Alloin, R. Antonucci,
{\elevenit Astrophys.J. Letters} {\elevenbf 337} (1989) L69.

\item{14.} S. Barwick, {\elevenit et~al.},  in {\elevenit Third International
Workshop on Neutrino Telescopes}, (Venice, Italy, 1991).

\item{15.}  W. Bednarek,
{\elevenit Astrophys.J. Letters} {\elevenbf 402} (1993) L29.

\item{16.}  M.C. Begelman, M. de Kool, M. Sikora,
{\elevenit Astrophys.J.} {\elevenbf 382} (1991) 416.

\item{17.} E.G. Berezhko, G.F. Krymskii,
{\elevenit Usp. Fiz. Nauk} {\elevenbf 154} (1988) 49
({\elevenit Sov. Phys. Usp.} {\elevenbf 31} (1988) 27).

\item{18.} V.S. Berezinsky, S.I. Grigor'eva,
{\elevenit Astron. \& Astroph.} {\elevenbf 199} (1988) 1.

\item{19.}  V.S. Berezinskii, S.V. Bulanov, V.A. Dogiel, V.L. Ginzburg (ed.),
and V.S. Ptus\-kin, {\elevenit Astrophysics of Cosmic Rays}, (North
Holland, Amsterdam, 1990).

\item{20.}  V.S. Berezinsky, A.Z. Gasizov, {\elevenit Gran Sasso preprint}
(1993) No. LNGS-93-52.

\item{21.}  V.S. Berezinsky, A.Z. Gasizov, {\elevenit Gran Sasso preprint}
(1993) No. LNGS-93-53.

\item{22.} E.M. Berkhuijsen,
{\elevenit Astron.\&Astroph.} {\elevenbf 166} (1986) 257.

\item{23.}  D.L. Bertsch, {\elevenit et~al.},
{\elevenit Astrophys.J. Letters} {\elevenbf 405} (1993) L21.

\item{24.} J.H. Bieging, D.C. Abbott, E.B. Churchwell,
{\elevenit Astrophys.J.} {\elevenbf 340} (1989) 518.

\item{25.} P.L. Biermann, M. Harwit,
{\elevenit Astrophys.J. Letters} {\elevenbf 241} (1980) L105.

\item{26.}  P.L. Biermann, P.A. Strittmatter,
{\elevenit Astrophys.J.} {\elevenbf 322} (1987) 643.

\item{27.}  P.L. Biermann, in {\elevenit Currents in Astrophysics and
Cosmology}, eds. G. Fazio and R. Silberberg,  (Cambridge Univ. Press ,
Cambridge, UK, scheduled to appear 1993, in press since 1990).

\item{28.} P.L. Biermann, R. Chini, A. Greybe-G\"otz, G. Haslam, E. Kreysa,
P.G. Mezger,
{\elevenit Astron.\& Astroph. Letters} {\elevenbf 227} (1990) L21.

\item{29.} P.L. Biermann, in Nagano and Takahara eds. (1991), p. 301.

\item{30.} P.L. Biermann, in {\elevenit High Energy Neutrino
Astrophysics}, eds. V.J. Sten\-ger, J.G. Learned, S. Pakvasa, X. Tata,
(World Scientific, Singapore, 1992), p. 86.

\item{31.} P.L. Biermann,
{\elevenit Astron.\&Astroph.} {\elevenbf 271} (1993)  649 (CR I).

\item{32.} P.L. Biermann, J.P. Cassinelli,
{\elevenit Astron.\&Astroph.} {\elevenbf }  (1993) (in press, CR II).

\item{33.} P.L. Biermann, R.G. Strom,
{\elevenit Astron.\&Astroph.} {\elevenbf } (1993)  (in press, CR III).

\item{34.} P.L. Biermann, W.J. Duschl, S.v. Linden,
{\elevenit Astron.\&Astroph.}  {\elevenbf } (1993) (in press).

\item{35.} G.S. Bisnovatyi-Kogan,
{\elevenit Sov.Astr.A.J.} {\elevenbf 14} (1971) 652
({\elevenit Astron.Zh.} {\elevenbf 47} (1970) 813).

\item{36.}  G.S. Bisnovatyi-Kogan, Yu.P. Popov, A.A. Samochin,
{\elevenit Astroph.Sp.Sc.} {\elevenbf 41} (1976) 287.

\item{37.} J.H. Black, E.F.v. Dishoeck,
{\elevenit Astrophys.J.} {\elevenbf 322} (1987) 412.

\item{38.}  J.H. Black {\elevenit et~al.},
{\elevenit Astrophys.J.} {\elevenbf 358} (1990) 459.

\item{39.} J.H. Black, E.F.v. Dishoeck,
{\elevenit Astrophys.J. Letters} {\elevenbf 369} (1991) L9.

\item{40.} R.D. Blandford, A. K\"onigl,
{\elevenit Astrophys. J.} {\elevenbf 232} (1979) 34.

\item{41.}  R. Blandford, R. Narayan, R.W. Romani,
{\elevenit Astrophys.J. Letters} {\elevenbf 301} (1986) L53.

\item{42.} R. Blandford, D. Eichler,
{\elevenit Physics Reports} {\elevenbf 154} (1987) 1.

\item{43.} G.R. Blumenthal,
{\elevenit Phys. Rev. D} {\elevenbf 1} (1970) 1596.

\item{44.} T.J.Bogdan, H.J. V\"olk,
{\elevenit Astron.\& Astroph.} {\elevenbf 122} (1983) 129.

\item{45.} R. Braun, W.M. Goss, J.L. Caswell, R.S. Roger,
{\elevenit Astron.\& Astroph.} {\elevenbf 162} (1986) 259.

\item{46.} R. Braun, S.F. Gull, R.A. Perley,
{\elevenit Nature} {\elevenbf 327} (1987) 395.

\item{47.} J.N. Bregman,
{\elevenit Astron. Astrophys. Rev.} {\elevenbf 2} (1990) 125.

\item{48.}  D. Breitschwerdt, J.F. McKenzie, H.J. V\"olk,
{\elevenit Astron. \& Astroph.} {\elevenbf 245} (1991) 79.

\item{49.}  D. Breitschwerdt, J.F. McKenzie, H.J. V\"olk,
{\elevenit Astron. \& Astroph.} {\elevenbf 269} (1993) 54.

\item{50.} A.~H. Bridle and R.~A. Perley,
{\elevenit Ann. Rev. Astron. Astrophys.} {\elevenbf 22} (1984) 319.

\item{51.} S. Britzen, Diploma Thesis, (University of Bonn, 1993).

\item{52.} T.H. Burnett, {\elevenit et~al.},
{\elevenit Astrophys.J. Letters} {\elevenbf 349} (1990) L25.

\item{53.} G.L. Cassiday, {\elevenit et~al.}, in {\elevenit Proc. of the
Workshop on Particle Astrophysics:  Forefront Experimental Issues}, ed. E.B.
Norman, (World Scientific, Singapore, 1989), p. 259.

\item{54.} G.L. Cassiday, {\elevenit et~al.},
{\elevenit Astrophys. J.} {\elevenbf 356} (1990) 669.

\item{55.} J.P. Cassinelli, in {\elevenit Wolf-Rayet stars:
Observation, Physics, Evolution}, eds. C.W.H. de Loore, A.J. Willis, (Reidel,
Dordrecht, Netherlands, 1983), p. 173.

\item{56.} J.P. Cassinelli, in {\elevenit Wolf-Rayet Stars and
Interrelations with Other Massive Stars in Galaxies}, IAU Sympos. No. 143,
eds. K.A van der Hucht and B. Hidayat, (Kluwer, Dordrecht,
Netherlands, 1991), p. 289.

\item{57.} R.A. Chevalier,
{\elevenit Ann.Rev.Astron.\&Astroph.} {\elevenbf 15} (1977) 175.

\item{58.}  R.A. Chevalier, J.N. Imamura,
{\elevenit Astrophys.J.} {\elevenbf 261} (1982) 543.

\item{59.} X. Chi, A.W. Wolfendale,
{\elevenit Nature} {\elevenbf 362} (1993) 610.

\item{60.}  R. Chini, E. Kreysa, P.L. Biermann,
{\elevenit Astron.\&Astroph.} {\elevenbf 219} (1989a)  87.

\item{61.}  R. Chini, P.L. Biermann, E. Kreysa, H.-P. Gem\"und,
{\elevenit Astron.\&Astroph. Letters} {\elevenbf 221} (1989b) L3

\item{62.} M.J. Claussen, K.-Y. Lo,
{\elevenit Astrophys.J.} {\elevenbf 308} (1986) 592.

\item{63.} J. Clavel, {\elevenit et~al.},
{\elevenit Astrophys.J.} {\elevenbf 393} (1992) 113.

\item{64.} G. Cocconi,
{\elevenit Nuovo Cimento} {\elevenbf 3} (1956) 1433.

\item{65.} D.P. Cox,
{\elevenit Astrophys.J.} {\elevenbf 178} (1972) 159.

\item{66.} G. Cunningham, {\elevenit et~al.},
{\elevenit Astrophys.J. Letters} {\elevenbf 236} (1980) L71.

\item{67.} C.D. Dermer, R. Schlickeiser,
{\elevenit Science} {\elevenbf 257} (1992) 1642.

\item{68.} J.R. Dickel, R. Sault, R.G. Arendt, Y. Matsui, K.T. Korista,
{\elevenit Astrophys.J.} {\elevenbf 330} (1988) 254.

\item{69.}  J.R. Dickel, W.J.M.van Breugel, R.G. Strom,
{\elevenit Astron.J.} {\elevenbf 101} (1991) 2151.

\item{70.}  B.L. Dingus, {\elevenit et~al.},
{\elevenit IAU Circular} (1993), No. 5690.

\item{71.} C. Done, G. Ghisellini, and A.~C. Fabian,
{\elevenit Mon. Not. Roy. Astr. Soc.} {\elevenbf 245} (1990) 1.

\item{72.}  G.S. Downs, A.R. Thompson,
{\elevenit Astron.J.} {\elevenbf 77} (1972) 120.

\item{73.} L.O'C. Drury,
{\elevenit Rep.Prog.Phys.} {\elevenbf 46} (1983) 973.

\item{74.} L.O'C. Drury, W.J. Markiewicz, H.J. V\"olk,
{\elevenit Astron. \& Astroph.} {\elevenbf 225} (1989) 179.

\item{75.} R.A. Edelson,
{\elevenit Astrophys.J. Letters} {\elevenbf 309} (1986) L69.

\item{76.} N.N. Efimov, {\elevenit et~al.}, in
Nagano and Takahara eds. (1991), p.~20.

\item{77.}  J.J. Engelmann, {\elevenit et~al.},
{\elevenit Astron.\& Astroph.} {\elevenbf 233} (1990) 96.

\item{78.} A.C. Fabian, I.M. George, S. Miyoshi, and M.J. Rees,
{\elevenit Monthly Not. Roy. astr. Soc.} {\elevenbf 242} (1990) 14p.

\item{79.} H. Falcke, P.L. Biermann, W.J. Duschl, P.G. Mezger,
{\elevenit Astron.\&Astroph.}  {\elevenbf 270} (1993) 102.

\item{80.}  H. Falcke, K. Mannheim, P.L. Biermann,
{\elevenit Astron.\& Astroph.} {\elevenbf } (1992) (submitted).

\item{81.} S.A.E.G. Falle, in {\elevenit Neutron Stars and their
Birth Events}, ed. W. Kundt, (Kluwer, Dordrecht, 1990), p. 303.

\item{82.} B.L. Fanaroff, J.M. Riley,
{\elevenit Monthly Not. Roy. Astron. Soc.} {\elevenbf 167} (1974) 31P.

\item{83.} G.J. Ferland, R.F. Mushotzky,
{\elevenit Astrophys.J.} {\elevenbf 286} (1984) 42.

\item{84.} E. Fermi,
{\elevenit Phys. Rev. 2nd ser.} {\elevenbf 75} (1949) 1169.

\item{85.} E. Fermi,
{\elevenit Astrophys.J.} {\elevenbf 119} (1954) 1.

\item{86.}  C.E. Fichtel, {\elevenit et~al.},
{\elevenit IAU Circular} (1992), No. 5460.

\item{87.} M.A. Forman, J.R. Jokipii, A.J. Owens,
{\elevenit Astrophys.J.} {\elevenbf 192} (1974) 535.

\item{88.} T.~K. Gaisser, R.~J. Protheroe, K.~E. Turver, and McComb,
{\elevenit Rev. Mod. Phys.} {\elevenbf 50} (1978) 859.

\item{89.} T.K. Gaisser, in {\elevenit Proc. Cosmic Ray Workshop}, Salt Lake
City, Utah, 1979).

\item{90.} T.~K. Gaisser and G.~B. Yodh,
{\elevenit Ann. Rev. Nucl. Part. Sci.} {\elevenbf 30} (1980) 475.

\item{91.} T.K. Gaisser, U.P. Sukhatme, G.B. Yodh,
{\elevenit Phys.Rev. D} {\elevenbf 36} (1987) 1350.

\item{92.} T.~K. Gaisser, {\elevenit Cosmic Rays and Particle Physics}
(Cambridge University Press, Cambridge, UK, 1990).

\item{93.} Gaisser, T.K., Stanev, T. {\elevenit et al.},
{\elevenit Phys.Rev D} {\elevenbf 47} (1993) (in press).

\item{94.}  D.J. Galloway, M.R.E. Proctor,
{\elevenit Nature} {\elevenbf 356} (1992) 691.

\item{95.} M. Garcia-Munoz, G.M. Mason, J.A. Simpson,
{\elevenit Astrophys.J.} {\elevenbf 217} (1977) 859.

\item{96.} G. Ghisellini,
{\elevenit Monthly Not. Roy. Astr. Soc.} {\elevenbf 224} (1987) 1.

\item{97.} G. Gilmore, {\elevenit et~al.},
{\elevenit Nature} {\elevenbf 357} (1992) 379.

\item{98.} V.L. Ginzburg,
{\elevenit Dokl. Akad. Nauk SSSR} {\elevenbf 92} (1953) 1133
({\elevenit NSF-Transl.} {\elevenbf 230}.)

\item{99.} G. Golla, M.Sc. Thesis (University Bonn, 1989).

\item{100.} K. Greisen,
{\elevenit Phys. Rev. Lett.} {\elevenbf 16} (1966) 748.

\item{101.} J.M. Grunsfeld, {\elevenit et~al.},
{\elevenit Astrophys.J. Letters} {\elevenbf 327} (1988) L31.

\item{102.}  S.F. Gull,
{\elevenit Monthly Not.Roy.Astr.Soc.} {\elevenbf 161} (1973) 47.

\item{103.}  C.R. Gwinn, J.M. Moran, M.J. Reid, M.H. Schneps,
{\elevenit Astro\-phys.J.} {\elevenbf 330} \break (1988a) 817.

\item{104.}  C.R. Gwinn, J.M. Cordes, N. Bartel, A. Wolszczan, R.L. Mutel,
{\elevenit Astrophys.J. Letters} {\elevenbf 334} (1988b) L13.

\item{105.} F. Halzen, in Nagano and Takahara eds. (1991), p. 91.

\item{106.} F. Halzen and E. Zas,
{\elevenit Phys. Lett.} {\elevenbf 289B} (1992) 184.

\item{107.} R.C. Hartman, {\elevenit et~al.},
{\elevenit Astrophys.J. Letters} {\elevenbf 385} (1992a) L1.

\item{108.} R.C. Hartman, {\elevenit et~al.},
{\elevenit IAU Circular}, (1992b), No. 5477.

\item{109.} R.C. Hartman, {\elevenit et~al.},
{\elevenit IAU Circular}, (1992c), No. 5519.

\item{110.} G. Hasinger, M. Schmidt, J. Tr{\"u}mper,
{\elevenit Astron. \& Astroph.} {\elevenbf 246} (1991) L2.

\item{111.} S. Hayakawa,
{\elevenit Progr. Theoret. Physics} {\elevenbf 15} (1956) 111.

\item{112.}  G. Henri, G. Pelletier, J. Roland,
{\elevenit Astrophys.J. Letters} {\elevenbf 404} (1993) L41.

\item{113.}  W. Hermsen, {\elevenit et~al.},
{\elevenit MPE-Comptel Preprint} (1992).

\item{114.} V.F. Hess,
{\elevenit Phys. Z.} {\elevenbf 13} (1912) 1084.

\item{115.} R.H. Hildebrandt, {\elevenit et~al.},
{\elevenit Astrophys.J.} {\elevenbf 216} (1977) 698.

\item{116.} C.T. Hill, D.N. Schramm,
{\elevenit Phys. Rev. D} {\elevenbf 31} (1985) 564.

\item{117.} A.M. Hillas,
{\elevenit Can. J. Phys.} {\elevenbf 46} (1968) S623.

\item{118.} A.M. Hillas,
{\elevenit Ann. Rev. Astron.\& Astroph.} {\elevenbf 22} (1984) 425.

\item{119.} S.D. Hunter, {\elevenit et~al.},
{\elevenit IAU Circular}, (1992), No. 5594.

\item{120.}W.-H. Ip, W.I. Axford, in {\elevenit Particle
acceleration in cosmic plasmas}, eds. G.P. Zank, T.K. Gaisser, (AIP Conf.Proc.
No. 264, 1992), p.~400.

\item{121.} D.L. Jauncey, {\elevenit et~al.},
{\elevenit Nature} {\elevenbf 334} (1988) 412.

\item{122.}  J.R. Jokipii, E.H. Levy, W.B. Hubbard,
{\elevenit Astrophys.J.} {\elevenbf 213} (1977) 861.

\item{123.}  J.R. Jokipii,
{\elevenit Astrophys.J.} {\elevenbf 313} (1987) 842.

\item{124.}  J.R. Jokipii, G. Morfill,
{\elevenit Astrophys.J.} {\elevenbf 312} (1987) 170.

\item{125.} J.R. Jokipii, in {\elevenit Radiowave scattering in the
Interstellar Medium}, eds. J.M. Cordes, B.J. Rickett, D.C. Backer, AIP
Conference Proc. No. 174, (1988), p. 48.

\item{126.} F.C. Jones, D.C. Ellison,
{\elevenit Space Science Reviews} {\elevenbf 58} (1991) 259.

\item{127.} K. Kamper, S.van den Bergh,
{\elevenit Astrophys.J.Suppl.} {\elevenbf 32} (1976) 351.

\item{128.} G. Kanbach, {\elevenit et~al.},
{\elevenit IAU Circular}, (1992), No. 5431.

\item{129.}  N.S. Kardashev,
{\elevenit Sov.Astr.A.J.} {\elevenbf 14} (1970) 375
({\elevenit Astron.Zh.} {\elevenbf 47} (1970) 465).

\item{130.} D. Kazanas, D.C. Ellison,
{\elevenit Astrophys.J.} {\elevenbf 304} (1986) 178.

\item{131.} D. Kazanas,
{\elevenit Astrophys.J.} {\elevenbf 347} (1989) 74.

\item{132.} J.G. Kirk, A. Mastichiadis,
{\elevenit Astron. \& Astroph.} {\elevenbf 213} (1989) 75.

\item{133.} W. Kohlh\"orster,
{\elevenit Phys. Z.} {\elevenbf 14} (1913) 1153.

\item{134.} M.de Kool, M.C. Begelman,
{\elevenit Nature} {\elevenbf 338} (1989) 484.

\item{135.} T.P. Krichbaum, {\elevenit et al.},
{\elevenit Astron. \& Astroph.} {\elevenbf 237} (1990) 3.

\item{136.}  T. Krichbaum, {\elevenit et~al.},
{\elevenit Astron. \& Astroph.} {\elevenbf } (1993) (in press).

\item{137.} P.P. Kronberg, P.L. Biermann, F.R. Schwab,
{\elevenit Astrophys.J.} {\elevenbf 246} (1981) 751.

\item{138.}  P.P. Kronberg, P.L. Biermann, F.R. Schwab
{\elevenit Astrophys.J.} {\elevenbf 291} (1985) 693.

\item{139.} W.M. Kr\"ulls,
{\elevenit Astron.\&Astroph.} {\elevenbf 260} (1992) 49.

\item{140.}  G.F. Krymskii, S.I. Petukhov,
{\elevenit Sov.Astr.A.J.Letters} {\elevenbf 6} (1980) 124
({\elevenit Pis. Astron.Zh.} {\elevenbf 6} (1980) 227).

\item{141.} W. Kundt,
{\elevenit Nature} {\elevenbf 261} (1976) 673.

\item{142.} M. Lacy, S. Rawlings, G.J. Hill,
{\elevenit Monthly Not. Roy. Astron. Soc.} {\elevenbf 258} (1992) 828.

\item{143.} P.O. Lagage, C.J. Cesarsky,
{\elevenit Astron.\& Astroph.} {\elevenbf 118} (1983) 223.

\item{144.} N. Langer,
{\elevenit Astron.\& Astroph.} {\elevenbf 210} (1989) 93.

\item{145.}  R.B. Larson,
{\elevenit Monthly Not.Roy.Astr.Soc.} {\elevenbf 186} (1979) 479.

\item{146.}  R.B. Larson,
{\elevenit Monthly Not.Roy.Astr.Soc.} {\elevenbf 194} (1981) 809.

\item{147.} A. Lawrence, M. Elvis,
{\elevenit Astrophys.J.} {\elevenbf 256} (1982) 410.

\item{148.} A. Lawrence, {\elevenit et~al.},
{\elevenit Monthly Not. Roy. Astron. Soc.} {\elevenbf 248} (1991) 91.

\item{149.} M.A. Lawrence, R.J.O. Reid, A.A. Watson,
{\elevenit J. Phys. G: Nucl. Part. Phys.} {\elevenbf 17} (1991) 733.

\item{150.} J.M. LeBlanc, J.R. Wilson,
{\elevenit Astrophys.J.} {\elevenbf 161} (1970) 541.

\item{151.} A.~P. Lightman and T.~R. White,
{\elevenit Astrophys.J.} {\elevenbf 335} (1988)  57.

\item{152.} S.v. Linden, W.J. Duschl, P.L. Biermann,
{\elevenit Astron. \& Astroph.} {\elevenbf 269}  (1993) 169.

\item{153.} E.C. Loh, {
\elevenit et~al.}, in Nagano and Takahara (1991), p.
345.

\item{154.} R. L\"ust,
{\elevenit Z. f. Naturf.} {\elevenbf 7a} (1952) 87.

\item{155.} F. Macchetto, {\elevenit et al.},
{\elevenit Astrophys.J. Letters} {\elevenbf 373} (1991) L55.

\item{156.}  J. MacDonald, T. Stanev, P.L. Biermann,
{\elevenit Astrophys.J.} {\elevenbf 378} (1991) 30.

\item{157.}  M. Maheswaran, J.P. Cassinelli,
{\elevenit Astrophys.J.} {\elevenbf 335} (1988) 931.

\item{158.}  M. Maheswaran, J.P. Cassinelli,
{\elevenit Astrophys.J.} {\elevenbf 386} (1992) 695.

\item{159.}  K. Mannheim, P.L. Biermann,
{\elevenit Astron. \& Astroph.} {\elevenbf 221} (1989) 211.

\item{160.} K. Mannheim, W.M. Kr\"ulls, P.L. Biermann,
{\elevenit Astron. \& Astroph.} {\elevenbf 251} (1991) 723.

\item{161.} K. Mannheim, P.L. Biermann,
{\elevenit Astron. \& Astroph. Letters} {\elevenbf 253} (1992) L21.

\item{162.} K. Mannheim, T. Stanev, P.L. Biermann,
{\elevenit Astron. \& Astroph. Letters} {\elevenbf 260} (1992)  L1.

\item{163.} K. Mannheim,  in {\elevenit High Energy Neutrino Astrophysics},
     Workshop Hawaii 1992, edited by V.~J. Stenger, J.~G. Learned, S. Pakvasa,
     and X. Tata (World Scientific Publishing Co., Singapore, 1992), p.\ 105.

\item{164.} K. Mannheim,
{\elevenit Astron. \& Astroph.} {\elevenbf 269} (1993a) 67.

\item{165.} K. Mannheim,
{\elevenit Physical Review D} {\elevenbf } (1993b) (submitted).

\item{166.}  L. Maraschi, T. Maccacaro, M.-H. Ulrich (eds.), {\elevenit
Proc. of the Workshop on BL Lac Objects}, Lecture Notes in Physics,
vol. {\elevenbf 334}, (Springer Verlag, Berlin, 1989).

\item{167.}  L. Maraschi, G. Ghisellini, A. Celotti,
{\elevenit Astrophys.J. Letters} {\elevenbf 397} (1992) L5.

\item{168.}  W.J. Markiewicz, L.O'C. Drury, H.J. V\"olk,
{\elevenit  Astron. \& Astroph.} {\elevenbf 236} (1990) 487.

\item{169.}  J.C. Mather, {\elevenit et~al.}, (1993), COBE-preprint 93-01.

\item{170.} J.S. Mathis, W. Rumpl, K.H. Nordsieck,
{\elevenit Astrophys.J.} {\elevenbf 217} (1977) 425.

\item{171.}  W.H. Matthaeus, Y. Zhou,
{\elevenit Phys.Fluids B} {\elevenbf 1} (1989) 1929.

\item{172.}  R. McCray, R.F. Stein, M. Kafatos,
{\elevenit Astrophys.J.} {\elevenbf 196} (1975) 565.

\item{173.}K. Meisenheimer, H.J. R\"oser (eds.), {\elevenit Hot
Spots in Extragalactic Radio \break Sources}, Proc. Ringberg Castle, 1988;
(Springer, Heidelberg, 1989).

\item{174.} K. Meisenheimer, {\elevenit et~al.},
{\elevenit Astron. \& Astroph.} {\elevenbf 219} (1989) 63.

\item{175.} K. Meisenheimer, H.J. R\"oser (eds.),
{\elevenit Jets in Extra\-gal\-actic Rad\-io \break Sour\-ces}, (Sprin\-ger
Ver\-lag, Heidel\-berg, to be published), proceedings Ringberg Castle meeting
1991.

\item{176.} H. Meyer, Preprint, University of Wuppertal (unpublished, 1992).

\item{177.}  P.F. Michelson, {\elevenit et~al.},
{\elevenit IAU Circular}, (1992), No. 5470.

\item{178.} G.K. Miley, {\elevenit et~al.},
{\elevenit  Astrophys.J. Letters} {\elevenbf 278} (1984) L79.

\item{179.}  D.K. Milne,
{\elevenit Austral.J. Phys.} {\elevenbf 24} (1971) 757.

\item{180.}  D.K. Milne,
{\elevenit Austral.J. Phys.} {\elevenbf  40} (1987) 771.

\item{181.} C. von Montigny {\elevenit et~al.},
{\elevenit MPE - Preprint} (1992) No. 233.

\item{182.} D. M\"uller,
{\elevenit Adv.Space Res.} {\elevenbf 9} (1989) (12)31.

\item{183.} D. M\"uller, {\elevenit et~al.},
{\elevenit Astrophys.J.} {\elevenbf 374} (1991) 356.

\item{184.} R.F. Mushotzky,
{\elevenit Astrophys.J.} {\elevenbf 256} (1982) 92.

\item{185.} M. Nagano, {\elevenit et~al.},
{\elevenit Journal of Physics G: Nucl. Part. Phys.} {\elevenbf 10} (1984) 1295.

\item{186.} M. Nagano,F. Takahara (eds.), 1991, {\elevenit Astrophysical
Aspects of the Most Energetic Cosmic Rays}; Proc.~ICRR Internat. Sympos., Kofu,
1990; (World Scientific, Singapore, 1991).

\item{187.} M. Nagano, {\elevenit et al.},
{\elevenit J. Phys. G: Nucl. Part. Phys.} {\elevenbf 18} (1992) 423.

\item{188.} R. Narayan, in {\elevenit Radiowave scattering in the
Interstellar Medium}, eds. J.M. Cordes, B.J. Rickett, D.C. Backer, AIP
Conference Proc. No. 174, (1988), p. 17.

\item{189.} B.B. Nath, P.L. Biermann,
{\elevenit Monthly Not. Roy. Astron. Soc.} {\elevenbf } (1993) (submitted).

\item{190.} L. Nellen, K. Mannheim, P.L. Biermann,
{\elevenit Physical Review D} {\elevenbf } (1993) (in press).

\item{191.} G. Neugebauer, J.B. Oke, E.E. Becklin, K. Matthews,
{\elevenit Astrophys.J.} {\elevenbf 230} (1979) 79.

\item{192.} G. Neugebauer, G.K. Miley, B.T. Soifer, P.E. Clegg,
{\elevenit Astrophys.J.} {\elevenbf 308} (1986) 815.

\item{193.} M. Niemeyer, M.Sc. Thesis, (Universit{\"a}t Bonn, 1991).

\item{194.} M. Niemeyer, P.L. Biermann,
{\elevenit Astron. \& Astroph.} {\elevenbf } (1993) (submitted).

\item{195.}  T.G. Northrop, {\elevenit The Adiabatic Motion of Charged
Particles} (Interscience Publishers, New York, 1963).

\item{196.}  J. Ormes, P. Freier,
{\elevenit Astrophys.J.} {\elevenbf 222} (1978) 471.

\item{197.}  J.P. Ostriker, J.E. Gunn,
{\elevenit Astrophys.J. Letters} {\elevenbf 164} (1971) L95.

\item{198.} S.P. Owocki, J.I. Castor, G.B. Rybicki,
{\elevenit Astrophys.J.} {\elevenbf 335} (1988) 914.

\item{199.} A.G. Pacholczyk A.G., {\elevenit Radio Astrophysics}
(Freeman and Company, San Francisco, CA, 1970).

\item{200.} P. Padovani, C.M. Urry,
{\elevenit Astrophys.J.} {\elevenbf 387} (1992) 449.

\item{201.} N. Panagia, R.A. Sramek, K.W. Weiler,
{\elevenit Astrophys.J. Letters} {\elevenbf 300} (1986) L55.

\item{202.} E.N. Parker,
{\elevenit Astrophys.J.} {\elevenbf 128} (1958) 664.

\item{203.} T. Parnell, T. {\elevenit et~al.},
{\elevenit Adv.Space Res.} {\elevenbf 9} (1989) (12)45.

\item{204.} J.A. Peacock, S.F. Gull,
{\elevenit Monthly Not. Roy. Astron. Soc.} {\elevenbf 196} (1981) 611.

\item{205.} J.A. Peacock,
{\elevenit Monthly Not. Roy. Astron. Soc.} {\elevenbf 217} (1985) 601.

\item{206.} R.A. Perley R.A., in Meisenheimer \& R\"oser (1989), p. 1.

\item{207.} K.A. Pounds, K. Nandra, G.C. Steward, K. Leighly,
{\elevenit Monthly Not. Roy. Astron. Soc.} {\elevenbf 240} (1989) 769.

\item{208.}  L. Prandtl,
{\elevenit Zeitschrift angew. Math. und Mech.} {\elevenbf 5} (1925) 136.

\item{209.}  V.L. Prishchep, V.S. Ptuskin,
{\elevenit Sov.Astr.A.J.} {\elevenbf 25} (1981) 446
({\elevenit Astron.Zh.} {\elevenbf 58} (1981) 779).

\item{210.} R.J. Protheroe, A.P. Szabo,
{\elevenit Phys. Rev. Lett.} {\elevenbf 69} (1992) 2885.

\item{211.} C.W. Punch, {\elevenit et~al.},
{\elevenit Nature} {\elevenbf 358} (1992) 477.

\item{212.} S.J. Qian, {\elevenit et~al.},
{\elevenit Astron. \& Astroph.} {\elevenbf 241} (1991) 15.

\item{213.} A. Quirrenbach, A. Witzel, {\elevenit et~al.},
{\elevenit Astrophys.J. Letters} {\elevenbf 372} (1991) L71.

\item{214.} A. Quirrenbach, {\elevenit et~al.},
{\elevenit Astron. \& Astroph.} {\elevenbf 258} (1992) 279.

\item{215.} J. Rachen, Master of Science Thesis, (University of Bonn,
1992).

Proc.

\item{216.} J.P. Rachen, P.L. Biermann, in {\elevenit Particle acceleration in
cosmic plasmas}, G.P. Zank, T.K. Gaisser, eds., (AIP Conf. No. 264, 1992),
p. 393.

\item{217.} J.P. Rachen, P.L. Biermann,
{\elevenit Astron. \& Astroph.} {\elevenbf 272} (1993) 161 (UHE CR I).

\item{218.} J.P. Rachen, T. Stanev, P.L. Biermann,
{\elevenit Astron. \& Astroph.} {\elevenbf }  (1993) (in press, UHE CR II).

\item{219.} S. Rawlings, R. Saunders,
{\elevenit Nature} {\elevenbf 349} (1991) 138.

\item{220.} S.P. Reynolds, R.A. Chevalier,
{\elevenit Astrophys.J. Letters} {\elevenbf 281} (1984) L33.

\item{221.} S.P. Reynolds, D.M. Gilmore,
{\elevenit Astron.J.} {\elevenbf 92} (1986) 1138.

\item{222.}  B.J. Rickett,
{\elevenit Ann.Rev.Astron. \& Astroph.} {\elevenbf 28} (1990) 561.

\item{223.} G.H. Rieke, {\elevenit et~al.},
{\elevenit Nature} {\elevenbf 260} (1976) 754.

\item{224.} G.H. Rieke, F.J. Low,
{\elevenit Astrophys.J. Letters} {\elevenbf 199} (1975) L13.

\item{225.} G.H. Rieke, {\elevenit et~al.},
{\elevenit Astrophys.J.} {\elevenbf 238} (1980) 24.

\item{226.} P.F. Roche, D.K. Aitken, M.M. Phillips, B. Whitmore,
{\elevenit Monthly Not. Roy. Astron. Soc.} {\elevenbf 207} (1984) 35.

\item{227.} A. Ruzmaikin, D. Sokoloff, A. Shukurov,
{\elevenit Nature} {\elevenbf 336} (1988) 341.

\item{228.} M.H. Salamon, F.W. Stecker,
{\elevenit Preprint} {\elevenbf } (1992).

\item{229.} D.B. Sanders, E.S. Phinney, G. Neugebauer, B.T. Soifer, K.
Matthews,
{\elevenit Astrophys.J.} {\elevenbf 347} (1989) 29.

\item{230.}  R. Schaaf, W. Pietsch, P.L. Biermann, P.P. Kronberg, T.
Schmutzler,
{\elevenit Astrophys.J.} {\elevenbf 336} (1989) 722.

\item{231.} R. Schlickeiser, P.L. Biermann, A. Crusius-W\"atzel,
{\elevenit Astron. \& Astroph.} {\elevenbf 247} (1991) 283.

\item{232.} T. Schmutzler, W. Tscharnuter,
{\elevenit Astron. \& Astroph.} {\elevenbf } (1993) (in press).

\item{233.} N. Scoville, C. Norman,
{\elevenit Astrophys.J.} {\elevenbf 332} (1988) 163.

\item{234.} E.R. Seaquist, {\elevenit et al.},
{\elevenit Astrophys. J.} {\elevenbf 344} (1989) 805.

\item{235.} N.~I. Shakura and R.~A. Sunyaev,
{\elevenit Astron. \& Astroph.} {\elevenbf 24} (1973) 337.

\item{236.} I.S. Shklovskii,
{\elevenit Dokl. Akad. Nauk SSSR} {\elevenbf 91} (1953) 475
({\elevenit Lib. of Congress Transl.} {\elevenbf RT-1495}).

\item{237.} I.S. Shklovsky, {\elevenit Supernovae},
(Wiley-Interscience, London, 1968).

\item{238.} M. Sikora, M.C. Begelman, B. Rudak,
{\elevenit Astrophys.J. Letters} {\elevenbf 341} (1989) L33.

\item{239.}  R. Silberberg, C.H. Tsao, M.M. Shapiro, P.L. Biermann,
{\elevenit Astrophys.J.} {\elevenbf 363} (1990) 265.

\item{240.} Simon {\elevenit et~al.},
{\elevenit Astrophys.J.} {\elevenbf 239} (1980) 712.

\item{241.} J. Skilling,
{\elevenit Monthly Not. Roy. Astr. Soc.} {\elevenbf 172} (1975a) 557.

\item{242.} J. Skilling,
{\elevenit Monthly Not. Roy. Astr. Soc.} {\elevenbf 173} (1975b) 245.

\item{243.} J. Skilling,
{\elevenit Monthly Not. Roy. Astr. Soc.} {\elevenbf 173} (1975c) 255.

\item{244.} J. Skilling, A.W. Strong,
{\elevenit Astron. \& Astroph.} {\elevenbf 53} (1976) 253.

\item{245.}  M.D. Smith,
{\elevenit Monthly Not. Roy. Astr. Soc.} {\elevenbf 238} (1989) 235.

\item{246.} P. Sokolsky, {\elevenit et~al.},
{\elevenit Phys. Reports} {\elevenbf 217} (1992).

\item{247.}  S.R. Spangler, C.R. Gwinn,
{\elevenit Astrophys.J. Letters} {\elevenbf 353} (1990) L29.

\item{248.} L. Spitzer Jr., M.G. Tomasko,
{\elevenit Astrophys.J.} {\elevenbf 152} (1968) 971.

\item{249.}  T. Stanev, in {\elevenit Particle acceleration in cosmic
plasmas}, G.P. Zank, T.K. \break Gaisser, eds., (AIP Conferenve Proc. No. 264,
1992), p. 379.

\item{250.} T. Stanev, P.L. Biermann, T.K. Gaisser
{\elevenit Astron. \& Astroph.} {\elevenbf } (1993) (in press, CR IV).

\item{251.} L. Staveley-Smith, R.N. Manchester, {\elevenit et~al.},
{\elevenit Nature} {\elevenbf 355} (1992) 147.

\item{252.} R.G. Strom, {\elevenit et al.},
{\elevenit IAU Circular} {\elevenbf 5762} (1993).

\item{253.} A.~P. Szabo and R.~J. Protheroe,  in {
\elevenit High energy
neutrino
     astrophysics}, Workshop Hawaii 1992, edited by V.~J. Stenger, J.~G.
Learned,
     S. Pakvasa, and X. Tata (World Scientific Publishing Co., Singapore,
1992),
     p.\ 24.

\item{254.} F.W. Stecker,
{\elevenit Phys. Rev. Lett.} {\elevenbf 21} (1968) 1016.

\item{255.} F.~W. Stecker, C. Done, M.~H. Salamon, and P. Sommers,
{\elevenit Phys. Rev. Lett.} {\elevenbf 66} (1991) 2697.

\item{256.} F.~W. Stecker, C. Done, M.~H. Salamon, and P. Sommers,  in
{\elevenit High energy neutrino astrophysics}, Workshop Hawaii 1992, edited
by V.~J. Stenger, J.~G. Learned, S. Pakvasa, and X. Tata (World Scientific
Publishing Co., Singapore, 1992a), p.\ 1.

\item{257.} F.W. Stecker, O.C. De Jager, M.H. Salamon,
{\elevenit Astrophys.J. Letters} {\elevenbf 390} \break (1992b) L49.

\item{258.} V.~J. Stenger,  Workshop on {\elevenit Trends in Astroparticle
     Physics}, Aachen, Germany, 1991.

\item{259.} R. Svensson,
{\elevenit Monthly Not. Roy. Astr. Soc.} {\elevenbf 227} (1987) 403.

\item{260.}  S.P. Swordy, D. M\"uller, P. Meyer, J. L'Heureux, J.M.
Grunsfeld,
{\elevenit Astrophys.J.} {\elevenbf 349} (1990) 625.

\item{261.} C.M. Telesco, D.A. Harper,
{\elevenit Astrophys.J.} {\elevenbf 235} (1980) 392.

\item{262.} A.J. Turtle, {\elevenit et~al.},
{\elevenit Nature} {\elevenbf 327} (1987) 38.

\item{263.} V.V. Usov, D.B. Melrose,
{\elevenit Astrophys.J.} {\elevenbf 395} (1992) 575.

\item{264.}  H.J. V\"olk, P.L. Biermann,
{\elevenit Astrophys.J. Letters} {\elevenbf 333} (1988) L65.

\item{265.} A. Wandel, {\elevenit et al.},
{\elevenit Astrophys.J.} {\elevenbf 316} (1987) 676.

\item{266.} A. Wandel,
{\elevenit Astron. \& Astroph.} {\elevenbf 200} (1988) 279.

\item{267.}  T.C. Weekes, {\elevenit et~al.},
{\elevenit IAU Circular}, (1992), No. 5522

\item{268.} K.W. Weiler, R.A. Sramek, N. Panagia, J.M.van der Hulst, M.
Salvati,
{\elevenit Astrophys.J.} {\elevenbf 301} (1986) 790.

\item{269.} K.W. Weiler, N. Panagia, R.A. Sramek, J.M.van der Hulst, M.S.
Roberts, L. Nguyen,
{\elevenit Astrophys.J.} {\elevenbf 336} (1989) 421.

\item{270.} K.W. Weiler, N. Panagia, R.A. Sramek,
{\elevenit Astrophys.J.} {\elevenbf 364} (1990) 611.

\item{271.} K.W. Weiler, S.D.van Dyk, N. Panagia, R.A. Sramek, J.L. Discenna,
{\elevenit Astrophys.J.} {\elevenbf 380} (1991) 161.

\item{272.} C.F.v. Weizs\"acker,
{\elevenit Z.f. Astrophys.} {\elevenbf 22} (1943) 319.

\item{273.} C.F.v. Weizs\"acker,
{\elevenit Astrophys.J.} {\elevenbf 114} (1951) 165.

\item{274.} B. Wiebel, unpublished HEGRA-Note, (University of
Wuppertal, 1992).

\item{275.} A.~S. Wilson, M. Elvis, A. Lawrence, J. Bland-Hawthorn,
{\elevenit Astrophys.J. Letters} {\elevenbf 391} (1992)  L75.

\item{276.} R. Windhorst, Ph.D.~thesis, (Sterrewacht Leiden,
     Netherlands, 1984).

\item{277.} P.F. Winkler, R.P. Kirshner,
{\elevenit Astrophys.J.} {\elevenbf 299} (1985) 981.

\item{278.} P.F. Winkler, J.H. Tuttle, R.P. Kirshner, M.J. Irwin,
in {\elevenit Supernova Remnants and the Interstellar Medium}, eds. R.S. Roger,
T.L. Landecker, (Cambridge University Press, Cambridge,1988), p. 65.

\item{279.}  A. Witzel, {\elevenit et~al.},
{\elevenit Astron. \& Astroph.} {\elevenbf 206} (1988) 245.

\item{280.} G.T. Zatsepin, V.A. Kuzmin,
{\elevenit JETPh Lett.} {\elevenbf 4} (1966) 78.

\item{281.} A. Zdziarski, {\elevenit et~al.},
{\elevenit Astrophys.J. Letters} {\elevenbf 363} (1990) L1.

\item{282.} A.A. Zdziarski and P.~S. Coppi,
{\elevenit Astrophys.J.} {\elevenbf 376} (1991) 480.

\item{283.} A.~A. Zdziarski, P.~T. \.{Z}ycki, R. Svensson, and E. Boldt,
Preprint (1991).

\item{284.}  J.A. Zensus, T.J. Pearson (eds.), {\elevenit Proc. of
     a Workshop on superluminal radio sources}, (Cambridge Univ. Press, 1987)

\bye